\documentclass[12pt, letterpaper]{JHEP3}
\pdfoutput=1
\usepackage{amsmath}
\usepackage{amssymb}
\usepackage{cite}
\usepackage{epsfig}
\usepackage{graphicx}

\newcommand{\comment}[1]{}
\newcommand{\abs}[1]{\left|#1\right|}

\title{Predictions from Star Formation in the Multiverse}

\author{Raphael Bousso and Stefan Leichenauer\\
  Center for Theoretical Physics, Department of Physics\\
  University of California, Berkeley, CA 94720-7300, U.S.A.\\
  {\em and}\\
  Lawrence Berkeley National Laboratory, Berkeley, CA 94720-8162,
  U.S.A.}

\abstract{We compute trivariate probability distributions in the
  landscape, scanning simultaneously over the cosmological constant,
  the primordial density contrast, and spatial curvature.  We consider
  two different measures for regulating the divergences of eternal
  inflation, and three different models for observers.  In one model,
  observers are assumed to arise in proportion to the entropy produced
  by stars; in the others, they arise at a fixed time (5 or 10 billion
  years) after star formation.  The star formation rate, which
  underlies all our observer models, depends sensitively on the three
  scanning parameters.  We employ a recently developed model of star
  formation in the multiverse, a considerable refinement over previous
  treatments of the astrophysical and cosmological properties of
  different pocket universes.  For each combination of observer model
  and measure, we display all single and bivariate probability
  distributions, both with the remaining parameter(s) held fixed, and
  marginalized.  Our results depend only weakly on the observer model
  but more strongly on the measure.  Using the causal diamond measure,
  the observed parameter values (or bounds) lie within the central
  $2\sigma$ of nearly all probability distributions we compute, and
  always within $3\sigma$.  This success is encouraging and rather
  nontrivial, considering the large size and dimension of the
  parameter space.  The causal patch measure gives similar results as
  long as curvature is negligible.  If curvature dominates, the causal
  patch leads to a novel runaway: it prefers a negative value of the
  cosmological constant, with the smallest magnitude available in the
  landscape.}

\begin{document}

\section{Introduction}

String theory appears to give rise to a large vacuum landscape,
containing perhaps ten to the hundreds of metastable vacua with three
large spatial dimensions~\cite{BP,KKLT}.  (See, e.g.,
Ref.~\cite{Sche06} for a discussion of earlier work.) Parameters that
appear fundamental at low energies can vary among these vacua and must
be predicted statistically.  The probability for a particular value is
proportional to the expected number of times it is observed.  In
particular, the cosmological constant, $\Lambda$, will vary.  Thus,
the landscape of string theory provides a theoretical foundation for
Weinberg's~\cite{Wei87} famous (and correct) prediction of a small but
nonzero value of $\Lambda$~\cite{BP,Pol06,Bou07}.  In the string
landscape, however, not only $\Lambda$ but also many other parameters
are expected to scan.  This means that there are many additional
opportunities to falsify the theory.

It is legitimate to consider only a subset of the landscape, defined
by one variable parameter (such as $\Lambda$), with all other
parameters fixed to their observed values.  If our observations are
highly atypical among the set of observations made in this restricted
class of vacua, then the theory is ruled out.  If they are typical,
then the theory has passed a first test.  We can then move on to test
the theory further, by predicting a second parameter---for example,
the primordial density contrast, $Q$.  If this succeeds, we have yet
another chance to falsify the theory by computing a joint probability
distribution over both parameters: now we might find that our universe
is very unlikely compared to one in which both parameters differ.  If
the theory is still not ruled out, we can consider a third parameter
(in the present paper, the amount of spatial curvature), and compute
yet more probability distributions.  Each new probability distribution
we compute is another chance for the theory to fail.  

In this paper, we present the first detailed computation of a
trivariate probability distribution in the landscape.  We display all
single-variable and bivariate distributions that can be extracted from
it.

The computation of such probability distributions is complicated by a
number of challenges.  What are observers?  Given a model for
observers, can we actually compute how many observations will be made
as a function of the scanning parameters?  In this paper, we consider
three models for observers, all of which require computing the rate at
which stars form, as a function of time.  We have recently developed a
numerical tool for computing the star formation rate in vacua with
different values of $\Lambda$, $Q$, and spatial
curvature~\cite{BouLei08}.  Here, we apply our star formation model to
the challenge of estimating the rate of observations made in this
three-parameter space of vacua.  As far as we know, this is the first
time that the cosmological and astrophysical evolution of other vacua
has been modeled at such level of detail.

Another challenge is the measure problem.  Long-lived vacua with
positive cosmological constant, which are abundant in the string
landscape, lead to eternal inflation~\cite{BP}. Globally, spatially
infinite bubbles of each type of vacuum are produced over and over.
Everything that can happen will happen infinitely many times. To
compute relative probabilities, such as the ratio of the numbers of
times two different parameter values are observed, this divergence has
to be regulated.

Recent years have seen considerable progress on the measure
problem. Several proposals have been ruled out because they conflict
violently with observation~\cite{LinLin96,Gut00a,Gut00b,Gut04,
  Teg05,BouFre07,BouFre06b,Pag06, Pag06b,Lin07,Gut07}. Interestingly,
several measures that manage to evade the most drastic problems appear
to be closely related.  They differ at most by subexponential
geometric factors~\cite{Bou09}.  Indeed, some of them have been shown
to be precisely equivalent~\cite{BouFre08b, BouYan09}, despite having
superficially a very different form.  This apparent convergence is
encouraging. It is all the more important to thoroughly test extant
proposals, and to discriminate between them, by computing probability
distributions and comparing them to observation.

Here, we consider two closely related but inequivalent members of the
surviving group of measures: the causal diamond
cut-off~\cite{Bou06,BouHar07} and the causal patch
cut-off~\cite{Bou06}. A particularly interesting discovery has been
that these two measures provide a novel catastrophic boundary on
parameter space, beyond which observations are suppressed---not for
dynamical reasons, like galaxy formation, but geometrically.  For
example, for some values of a scanning parameter, the cut-off region
may have exponentially small comoving volume, and for this reason
alone will contain a very small number of observers.  This geometric
effect provides a stronger upper bound on $\Lambda$ than the
disruption of structure~\cite{BouHar07}. (This result is reproduced
here as a special case.)  It also provides a stronger constraint on
the ratio of dark matter to baryonic matter~\cite{Fre08}.  In both
cases, geometric suppression has significantly improved agreement
between theory and observation.  The results presented here will
reflect the effects of geometric suppression in a larger parameter
space, and we will highlight these effects in the discussion of our
results (Sec.~\ref{sec-discussion}).

\paragraph{Scope and method} We consider three cosmological
parameters: the cosmological constant, $\Lambda$; the primordial
density contrast, $Q\equiv \frac{\delta\rho}{\rho}$; and the spatial
curvature. We parametrize spatial curvature by the logarithmic
quantity $\Delta N$, which can be thought of as the number of
inflationary $e$-foldings minus the minimum number required to explain
the observed flatness of our universe. We scan over the
three-dimensional parameter space
\begin{eqnarray}
  10^{-3}\Lambda_0< & \abs{\Lambda} & <10^5\Lambda_0\\
  10^{-1}Q_0< & Q & <10^2Q_0\\
  -3.5 < & \Delta N & <\infty ~,
\label{eq-range}
\end{eqnarray}
where $\Lambda_0$ and $Q_0$ are the observed values.  For each
combination $(\Lambda,Q,\Delta N)$, we compute a history of structure
formation and of star formation in the corresponding universe.  We use
our own model of star formation~\cite{BouLei08}, which was designed to
handle variations of these parameters over several decades.  The upper
limit on $Q$ is motivated by a change of regime found in
Ref.~\cite{BouLei08}: For $Q<10^2Q_0$, most of the star formation
happens well after recombination, where we can trust our model; for
larger values, we cannot not.

We obtain single- and multivariate probability distributions by
computing the expected number of times each parameter combination
$(\Lambda,Q,\Delta N)$ is observed in the multiverse.  We consider
three different models for observers. One model assumes that the rate
of observation tracks the rate of entropy production by
stars~\cite{Bou06,BouHar07}.  The other two are based on the
assumption that the rate of observations follows the rate at which
stars are produced, with a delay of five or ten billion years.

Our computation is numerical.  Even an elementary treatment of the
physics of structure formation and star formation involves a complex
interplay of different phenomena.  In our own universe, several of
these processes, such as structure formation, radiative galaxy
cooling, Compton cooling of galaxies, galaxy mergers, observer
evolution, and vacuum domination, happen roughly on the same time
scale, a billion years to within about one order of magnitude.  (The
lack of any known symmetry that could explain this multiple
coincidence is itself evidence for a multiverse~\cite{BouHal09}.)  The
parameter range we consider includes values in which curvature, too,
comes to dominate at a time comparable to the previously mentioned
scales.  Coincidences of scales preclude a separation into
well-defined analytic regimes, necessitating a numerical computation.

Coincidences of scales arise not just for our own universe, but
persist on certain hypersurfaces of the parameter space we consider.
The time of structure formation scales as $Q^{-3/2}$; the radiative
cooling time as $Q^{-3}$; the time of vacuum domination as
$\Lambda^{-1/2}$; and the time of curvature domination as
$\exp(3\Delta N)$.  So, for example, in universes whose parameters lie
near a certain hypersurface of constant $Q^3/\Lambda$, the beginning
of structure formation and its disruption by vacuum domination will
not be well separated.  In the neighborhood of such surfaces,
analytical arguments are very imprecise, and numerical treatment is
essential.

Numerical computation is somewhat complementary to analytical
arguments.  Our code becomes unstable when certain separations of
scales become too large.  This limits the parameter range we can
consider numerically.  Strictly speaking, our results pertain only to
the subset of the landscape defined by the above range of parameters.
But for the same reason---a good separation of scales---we can often
extrapolate analytically to a larger range.  Near some boundaries of
our parameter range, the probability density is negligible, and
analytic arguments tell us that it will continue to decrease.  In
these cases, we can safely neglect the missing part of the probability
distribution.  We can also do so if the probability density is
increasing towards the boundary, but there is a catastrophic change of
regime at the boundary that sharply suppresses the number of
observations in universes beyond the boundary.  (For example, if we
increase $Q$ and $\Lambda$ while holding $Q^3/\Lambda$ fixed,
eventually vacuum domination will occur before recombination.  Since
star formation can begin only after recombination, when dark matter
halos are exponentially dilute, such universes have negligible
probability of being observed.)  However, near some boundaries, the
probability distribution is increasing and there is no change of
regime at or near the boundary.  In this case, the probability
distribution may be dominated by regions outside the parameter range
we consider numerically.  In general, we can use analytic arguments to
understand its behavior in this regime.  An example is the runaway
towards small values of $|\Lambda|$ that we find with the causal patch
measure.

\paragraph{Results} Our results are fully displayed in
Sec.~\ref{sec-results}.  Its six subsections correspond to the six
combinations of measure and observer model we consider.  For each
model, we show about 30 plots corresponding to different combinations
of parameters that are varied, held fixed, or integrated out.  We
discuss our results in Sec.~\ref{sec-discussion}, where we highlight
several interesting features.  We provide a qualitative understanding
of these features, and we explain how probability distributions depend
on the measure and the observer model. Most of our results do not
depend strongly on how observers are modeled.\footnote{The case where
  both $Q$ and $\Lambda$ vary is an exception. When observers are
  modeled by a time delay, larger $Q$ does not lead to a preference
  for larger $\Lambda$; with entropy production, it does. In neither
  case, however, do we find that our values of $Q$ and $\Lambda$ are
  very unlikely.} However, they do depend on the measure, allowing us
to discriminate between the causal diamond and the causal patch.  Let us briefly describe our most important findings.

We find that the {\em causal diamond\/} measure is good agreement with
observation for all parameter combinations, independently of details
of the landscape (see Figs. \ref{posDia}--\ref{posnegDia10Gyr}). The observed values are
within\footnote{In some plots they appear just outside of $2\sigma$,
  but by a margin that is negligible compared to the uncertainties in
  our computation both of probabilities and of the confidence
  contours.}  $2\sigma$ in all plots, except if $\Lambda$ is scanned
over both positive and negative values, and $Q$ is simultaneously
scanned; in this case, they lie within $2\sigma$ or $3\sigma$.  This
is in large part because the negative range, $\Lambda<0$, is between
12 and 25 times more probable than the positive range, depending on
the observer model.

The {\em causal patch\/} measure, on the other hand, yields a
nonintegrable probability distribution near $|\Lambda|=0$ in the
absence of a cut-off, i.e., of a smallest possible value of
$|\Lambda|$.  This runaway is explained analytically in
Sec.~\ref{sec-lambdacurvonly}.\footnote{This result, as well as the
  preference for large curvature mentioned below, was anticipated in
  unpublished analytical arguments by Ben Freivogel.}  The onset of
this limiting behavior is at $|\Lambda|\sim t_{\rm c}^{-2}$, where
$t_{\rm c}$ is the time at which curvature comes to dominate.  In
particular, the runaway does not occur at all in the absence of
spatial curvature ($t_{\rm c}\to\infty$).  

The strength of the runaway depends on the sign of the cosmological
constant.  For $\Lambda<0$, the probability density grows like
$|\Lambda|^{-1}$ as $|\Lambda|\to 0$, i.e., it grows exponentially in
the display variable $\log_{10}(|\Lambda|/\Lambda_0)$.  The rapid
growth is evident in several plots in Figs.~\ref{negPatch},
\ref{negPatch5Gyr}, and~\ref{negPatch10Gyr}.  For $\Lambda>0$, the
probability density is independent of $\log_{10}(|\Lambda|/\Lambda_0)$
for $\Lambda\ll t_{\rm c}^{-2}$.  Because of our limited parameter
range, this milder runaway is not readily apparent in the relevant
plots in Figs.~\ref{posPatch}, \ref{posPatch5Gyr},
and~\ref{posPatch10Gyr}, but it can be predicted analytically.

Thus, if spatial curvature is large enough to dominate before
$\Lambda$ does, the causal patch predicts a negative cosmological
constant whose magnitude is the {\em smallest\/} among all anthropic
vacua in the landscape.  Whether this runaway is a problem or a
success depends on the (unknown) size of the string landscape. It
would certainly be a problem if the landscape is so large that it
contains anthropic vacua with cosmological constant much smaller than
$\Lambda_0\sim 10^{-123}$ in magnitude. In this case the causal patch
measure would predict at high confidence level that we should find
ourselves in a vacuum with $-\Lambda_0\ll \Lambda <0$, and so would be
ruled out. It might be a success, on the other hand, if the observed
$\Lambda$ corresponds to one of the smallest values available among
the finite number of anthropic vacua in the landscape. The size of the
landscape would be directly responsible for the observed scale
$10^{-123}$, with the density of its discretuum providing an
``ur-hierarchy'' from which other hierarchies can be
derived~\cite{Bou06,BouHal09}.  Even in this case the causal patch
prefers negative values of the cosmological constant (and somewhat
larger curvature than the observed upper bound), but only by a factor
of order 10, not strongly enough to be ruled out by observation.

At fixed values of $\Lambda$, the causal patch leads to a stronger
preference for curvature than the causal diamond.  This is explained
analytically in Sec.~\ref{sec-lambdacurvonly}.  The pressure is
particularly strong for $\Lambda<0$, where the probability density
grows very rapidly, like $\exp(-9\Delta N)$, towards small values of
$\Delta N$.  This is not a true runaway problem, because there is a
catastrophic boundary from the disruption of structure formation that
will suppress the probability for sufficiently small values of $\Delta
N$.  However, after $\Delta N$ is marginalized, this effect would
contribute additional weight to vacua with negative cosmological
constant even if the runaway towards $\Lambda=0$ was suppressed by a
lower bound $\Lambda_{\rm min}\sim \Lambda_0$ on the magnitude of the
cosmological constant from the discretuum.

Thus, we find evidence that the causal patch does not yield
probability distributions compatible with observation, unless (1) we
are very close to the smallest value of $|\Lambda|$ in the discretuum
($\Lambda_{\rm min}\sim \Lambda_0$), or (2) the prior probability
distribution differs from what we have assumed (for example, by
suppressing curvature so strongly that all anthropic vacua can be
treated as spatially flat, $t_{\rm c}\gtrsim \Lambda_{\rm
  min}^{-1/2}$; this would be the case if all inflationary models in
the landscape have a very large number of e-foldings).

Another possibility is worth mentioning.  The causal patch measure
(with particularly simple initial conditions) was recently shown to be
equivalent~\cite{BouYan09} to the light-cone time cut-off~\cite{Bou09}
on the multiverse.  The latter is motivated~\cite{GarVil08} by analogy
with the holographic UV-IR connection of the AdS/CFT correspondence.
The boundary structure in the future of eternally inflating regions
differs sharply from that in the future of regions with $\Lambda\leq
0$.  Since the analogy with AdS/CFT is most compelling in regions with
positive cosmological constant, it is natural to consider the
possibility that the causal patch measure may give correct relative
probabilities only for observations in such regions.  This restriction
would eliminate the worst of the above problems, which pertain mainly
to negative values of $\Lambda$.  (It would also eliminte the
divergence for $\Lambda=0$~\cite{BouFre06,MerAda08}.)  There remains a
weak (logarithmic) runaway towards $\Lambda=0$ from above
($\Lambda>0$), but this would not be a problem if $-\log\Lambda_{\rm
  min}\sim O(100)$, a plausible value for the string
landscape~\cite{BP,DenDou04b}.

\paragraph{Relation to recent work} Our work can be regarded as a
substantial extension and refinement of Ref.~\cite{BouHar07}, where
the probability distribution over positive values of $\Lambda$ was
estimated from entropy production in the causal diamond (the ``causal
entropic principle''). Here we consider a larger number and range of
parameters, two different measures, and three different models for
observers. Whereas in Ref.~\cite{BouHar07} the effects of the single
parameter $\Lambda$ on the star formation history were negligible in
the most important range of the probability distribution, here we are
forced to compute the entire star formation history numerically for
each value of $(\Lambda,Q,\Delta N)$.

Other interesting extensions of Ref.~\cite{BouHar07} include
Refs.~\cite{CliFre07,BozAlb09,PhiAlb09}. Cline {\em et
  al.}~\cite{CliFre07} compute a bivariate probability distribution
over (positive) $\Lambda$ and $Q$; and Bozek {\em et
  al.}~\cite{BozAlb09} compute a bivariate distribution over
(positive) $\Lambda$ and spatial curvature. In principle, these
portions of Refs.~\cite{CliFre07,BozAlb09} could be regarded as
special cases of the present work, with $\Lambda>0$ and either $\Delta
N$ or $Q$ held fixed, infinities regulated by the causal diamond
measure, and observers modeled in terms of the entropy produced by
dust heated by stars. However, our results differ because we model
star formation and dust temperature differently.

Both~\cite{CliFre07} and~\cite{BozAlb09} employ the analytic star
formation model of Hernquist and Springel (HS)~\cite{HerSpr02}. This model
was designed to closely fit data and numerical simulations of the
first 13.7 Gyr of our own universe. The HS model exhibits some
unphysical features when extrapolated to later times or different
values of $(\Lambda,Q,\Delta N)$. For example, because it does not
take into account the finiteness of the baryon supply in a halo, the
HS model predicts unlimited star formation at a constant rate after
structure formation is disrupted by a positive cosmological constant
or by negative spatial curvature. Our own star formation
model~\cite{BouLei08} includes only the most important physical
effects governing star formation, and so provides only a rough (though
surprisingly good) fit of the observed star formation history.
However, our model includes not just those effects which govern star
formation during the first 13.7 Gyr of our own universe, but is
designed to apply in a wide range of $(\Lambda,Q,\Delta N)$, and at
all times after recombination. Differences between our results and
those of Refs.~\cite{CliFre07,BozAlb09} can be traced mainly to how we
model star formation. A more subtle difference from
Ref.~\cite{CliFre07} arises from our treatment of the dust temperature
dependence on the virial density. A trivial difference from
Ref.~\cite{BozAlb09} is the choice of prior probability distribution
for the parameter $\Delta N$. More detail is given in
Sec.~\ref{sec-prior}.

Salem~\cite{Sal09} computes a probability distribution over positive
and negative values of $\Lambda$, with all other parameters fixed,
using the causal patch measure. Observers are modeled as arising at a
fixed time delay after the formation of {\em galaxies\/} that are
similar to the Milky Way in a specific sense~\cite{Sal09}. The special
case in this paper most similar to Ref.~\cite{Sal09} is our
computation of a probability distribution over positive and negative
$\Lambda$ with $Q$ and $\Delta N$ fixed, using the causal patch
measure and modeling observers by a 10 Gyr time delay after star
formation.  Despite the different observer model, our results for this
case agree very well with Salem's.  We find that the observed value of
$\Lambda$ is nearly three standard deviations above the mean of the
predicted distribution: 99.7\% of observers see a smaller value than
ours, and most of them see a negative value. In fact, our observed
value of $\Lambda$ is outside $2\sigma$ no matter how we model
observers, as long as the causal patch is used. (The causal diamond is
in better agreement with observation.)

\section{Making predictions in the landscape}
\label{sec-framework}

In this section, we will explain how we compute probabilities in the
multiverse.  We will explore two different measures, described in
Sec.~\ref{sec-measures}.  In Sec.~\ref{sec-prior}, we will discuss
prior probabilities and cosmological selection effects, and in
Sec.~\ref{sec-observers} we will describe three ways of modeling
observers.  In Sec.~\ref{sec-summary}, we will explain how these
ingredients are combined to obtain a probability distribution over the
parameters $(\Lambda,Q,\Delta N)$.

\subsection{Two choices of measure}
\label{sec-measures}

Before we can compute anything, we need to remove the divergences that
arise in an eternally inflating universe. We will consider two
slightly different measures:

\begin{figure}[t]
\begin{center}
\includegraphics[width=6in]{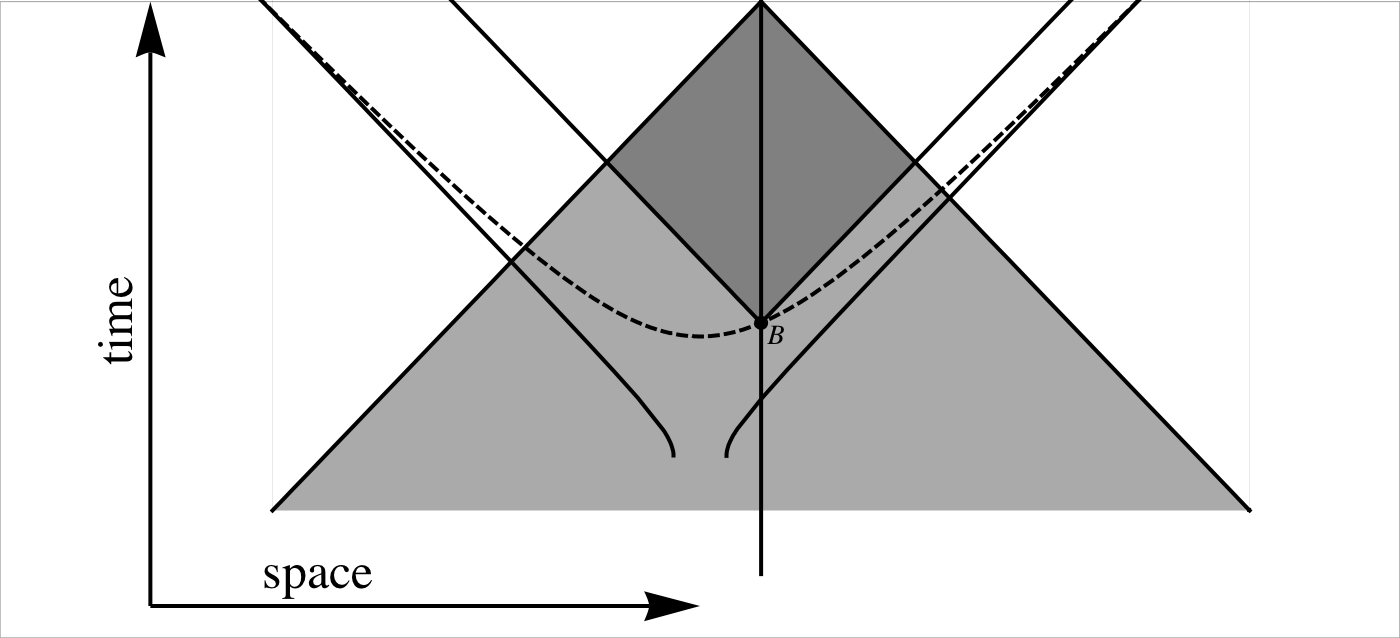}
\caption{The causal patch (shaded triangle) is the past of the
  future endpoint of a geodesic (vertical line) in
  the multiverse.  The causal diamond (dark shaded) is the intersection of the
  causal patch with the future of the point $B$, where the geodesic
  intersects a surface of reheating (dashed).}
\label{fig-cut-offs}
\end{center}
\end{figure}

\subsubsection{Causal patch cut-off}

The causal patch is the past of a point on the future boundary of the
spacetime (Fig.~\ref{fig-cut-offs}).  Consider a pocket universe
described by the Friedmann-Robertson-Walker metric
\begin{equation}
ds^2=-dt^2+a^2(t) \left[ d\chi^2+f^2(\chi) d\Omega^2\right]~.
\end{equation}
We consider only open universes (which include flat universes as a
special case), so $f(\chi)=\sinh\chi$.  The causal patch is the set of
points with $\chi<\chi_{\rm patch}$, where
\begin{equation} 
\chi_{\rm patch}(t) = \int_t^{t_{\rm max}} \frac{dt}{a(t)}~, 
\label{eq-patch} 
\end{equation}
and $t_{\rm max}$ is the time of the crunch.  For a long-lived
de~Sitter vacuum, we can take $t_{\rm max}\approx\infty$.

In any long-lived de~Sitter vacuum ($\Lambda>0$), the patch coincides
with the interior of the event horizon, because a late-time decay into
a terminal vacuum (with $\Lambda\leq 0$) does not affect the size of
the event horizon at early times.  In vacua with $\Lambda<0$, the
causal patch is the past of a point on the future singularity (the
``big crunch'').  We will not consider $\Lambda=0$ vacua in this
paper.  The causal patch has divergent four-volume in such
vacua~\cite{BouFre06,MerAda08}.

The causal patch cut-off was motivated by the resolution of the
quantum xeroxing paradox in black
holes~\cite{SusTho93,BouFre06a}. Recently, the measure was shown to be
exactly equivalent to the light-cone time
cut-off~\cite{Bou09,BouYan09}, which was motivated by an analogy with
the AdS/CFT correspondence~\cite{GarVil08,Bou09}. The analogy is most
compelling in eternally inflating regions of the multiverse (``eternal
domains'').  From this viewpoint, it is conceivable that the regime of
validity of the causal patch cut-off is limited to vacua with
$\Lambda>0$. Our results will offer some phenomenological evidence for
this possibility, in that we will find that the measure is least
successful in vacua with $\Lambda<0$.

\subsubsection{Causal diamond cut-off}

Consider a geodesic in the multiverse.  The causal diamond is the
intersection of the causal past of the future endpoint of the geodesic
(the causal patch) with the causal future of some earlier point $B$.
We will follow Ref.~\cite{BouHar07}, where $B$ was taken to be the
point where the geodesic intersects the surface of reheating
(Fig.~\ref{fig-cut-offs}).   Thus, the causal
diamond is the set of points with $\chi<\chi_{\rm dia}$, where
\begin{eqnarray} \nonumber
  \chi_{\rm dia}(t) & = & \min\{\chi_{\rm patch}(t), \eta(t)\}\\
  & = &
  \frac{\eta_{\rm max}}{2} - \abs{\frac{\eta_{\rm max}}{2}-\eta(t)}~,
\label{eq-diamond}
\end{eqnarray} 
where
\begin{equation}
\eta(t) = \int_{t_{\rm rh}}^t \frac{dt}{a(t)}~,
\end{equation}
and $\eta_{\rm max} = \eta(t_{\rm max})$.

Because of the additional restriction to the future of $B$, the
diamond cannot be larger than the patch.  With our choice of $B$, the
diamond will be smaller than the patch approximately until
$\Lambda$-domination, and it will coincide with the patch after
$\Lambda$-domination.

Our choice of $B$ is motivated by the absence of matter prior to
reheating.  However, the concept of a reheating surface is not
completely sharp.  Nevertheless, the causal diamond may be an
approximation to a more generally defined cut-off; a candidate will be
discussed in future work.  (In
Ref.~\cite{Bou06}, the point $B$ was taken to be the starting point of
the geodesic on some initial spacelike hypersurface.  Then most pocket
universes will lie entirely in the future of $B$.  Except for very
unnatural initial conditions, the region excluded from the diamond but
present in the patch will be an empty de~Sitter region with large
cosmological constant.  Thus, with this choice, the causal diamond
gives the same probabilities as the causal patch.)

\subsection{Prior distribution and cosmological selection}
\label{sec-prior}

The probability distribution over an observable parameter $x$ can be
operationally defined as the relative abundance of the various
outcomes of all measurements of this parameter in the whole
universe.\footnote{See, e.g.,
  Refs.~\cite{HarSre07,BouFre07,GarVil07,Pag09} for discussions of
  this claim.}  It will be useful for us to think of this probability
distribution as a convolution of the following three distributions:
\begin{itemize} 
\item Prior distribution.  The relative abundance of different values
  of the parameter $x$ among vacua in the theory landscape
\item Cosmological selection effects.   The relative abundance of the
  different vacua in the universe will differ from the prior
  distribution because of selection effects of
  cosmological dynamics and/or initial conditions
\item Anthropic selection effects: whether, and how frequently, some
  value of $x$ is observed may depend on $x$
\end{itemize} 
Once a measure has been chosen (see the previous subsection), all
three distributions listed above can be computed. Let us discuss the
first two in turn; we will devote a separate subsection to the third.

\paragraph{Prior distribution} Because the cosmological constant is
effectively a random variable and $\Lambda=0$ is not a special point,
the prior distribution of $\Lambda$ can be approximated as flat in the
anthropically relevant regime ($\Lambda\ll 1$):
\begin{equation}
\frac{d\tilde p}{d\Lambda}\propto 1~,
\end{equation}
which translates into a prior proportional to $\Lambda$ for
$\log_{10}\Lambda$, our choice of display parameter.

We know much less about the prior distributions of spatial curvature,
$\Delta N$, and the primordial density contrast, $Q$, in the string
landscape.  There are certain prior distributions which seem
implausible, such as a strong preference for large hierarchies (e.g.,
for small $\log_{10} Q$ or large $\Delta N$), but this still leaves
considerable uncertainty.  For definiteness, $Q$ will be assumed to
have a prior which is flat in $\log_{10} Q$, which we view as the most
optimistic choice among reasonable alternatives:
\begin{equation}
\frac{d\tilde p}{d\log_{10} Q}\propto 1~.
\end{equation}
For curvature, Ref.~\cite{FreKle05} estimated
\begin{equation}
\frac{d\tilde p}{dN}\propto \frac{1}{N^4}~.
\label{eq-nprior}
\end{equation}
We shall use this prior distribution together with the assumption that $\Delta N =0$ corresponds to $N=60$. 

Despite the large uncertainties about the priors, our results will be robust in the following sense: In cases where we find a tension between the prediction of a measure and observation, this tension could only be removed by choosing a highly implausible prior on $Q$ or $\Delta N$.

\paragraph{(No) cosmological selection} Relative to the very high
energy scales that determine the decay channels and decay rates of
metastable landscape vacua, the length of inflation and the mechanism
for generating density perturbations can plausibly be assumed to arise
at relatively low energies, and thus, to be uncorrelated with the
production rate of various vacua.  This also holds for the
cosmological constant, since we are interested only in an anthropic
range of values, $\Lambda\ll 1$.  These values can only be measured at
very low energy density, and so cannot be correlated with the
nucleation rate of vacua.  Therefore, we will ignore cosmological
selection effects.\footnote{We assume, however, that there is no
  ``staggering problem''~\cite{SchVil06,Sch06}, in which cosmological
  selection effects lead to such unequal probabilities for different
  vacua as to effectively eliminate most of the landscape and render
  it unable to solve the cosmological constant problem.  The presence
  of this problem depends on details of the landscape, and in the case
  of the two local measures considered here, on initial conditions.
  It is absent in plausible toy models~\cite{OluSch07,Sch08}.}

\subsection{Three ways of modeling observers}
\label{sec-observers}

Finally, we must compute the expected number of instances of observing
different values of the parameters $(\Lambda,Q,\Delta N)$ in the
cut-off region.  In general, these values will be correlated with the
presence and number of observers, so we must compute the number of
observers as a function of $(\Lambda,Q,\Delta N)$.  In principle,
there is no reason why such a computation could not be performed in a
sufficiently powerful theory, by a sufficiently able theorist.  In
practice, we struggle to define ``observer'' or ``observation'' in
complete generality.  In this paper, we will consider three models for
observers.  In the first two, we focus on observers ``like us'', which
arise near a star, a certain number of years (5 Gyr in the first
model, 10 Gyr in the second) after the formation of stars.  The third
model uses entropy production as a more general proxy for
observers~\cite{Bou06,BouHar07}.\footnote{In combination with the
  causal diamond cut-off this has been called the ``Causal Entropic
  Principle''.  However, we should stress that the question of
  modeling observers is, at least naively, orthogonal to the measure
  problem.  Entropy production could be used as an observer proxy
  essentially in any measure.}  We will now describe these models in
more detail.

\subsubsection{Observers = stars + time delay of 5 or 10 Gyr}

We are at liberty to restrict our attention to any class of observers
that includes us, for example, observers that emerge near stars.  This
probably ignores some observers in the landscape.  There may be
regions without stars that nevertheless contain observers powered by
some other source of free energy.  However, we can still ask whether
our observations are typical of observers in the class we have
defined; a theory in which they turn out to be highly atypical can be
ruled out.  
% (For the same reason, it is legitimate to restrict attention to a
% three-parameter subset of the landscape that includes our vacuum, as
% we do in this paper.)

We need to know both the spatial and temporal location of observations
in order to compute what they will observe.  Concretely, let us assume
that each star, on average, gives rise to some number of observers
after a fixed ``evolutionary'' delay time $t_{\rm delay}$.  Then the
number of observations made at the time $t$, per unit
comoving volume and unit time, is
\begin{equation}
  \frac{d^2 n_{\rm obs}}{dV_{\rm c}\, dt}(t) \propto 
  \dot{\rho}_\star (t-t_{\rm delay})~,
  \label{eq-starobs}
\end{equation}
where $\dot{\rho}_\star(t)\equiv d^2 m_\star/dV_{\rm c}\,dt$ is the
star formation rate, i.e., the amount of stellar mass produced per
unit time and per unit matter mass.  Because we assume a fixed
initial mass function for stars, this equation holds
independently of how the number of observers may depend on the mass of
the star, so we will not need to make any particular assumption about
this distribution.

In this paper, we explore two choices: $t_{\rm delay}=5\,{\rm Gyr}$,
and $t_{\rm delay}=10\,{\rm Gyr}$.  The first choice corresponds to
the evolutionary timescale of life on earth.  It defines a class of
observers that are like us in the sense that they exist at equal time
delay after the birth of their respective star.  In this case, stars
with lifetimes less than 5 Gyr do not contribute to observations and
should be excluded in Eq.~(\ref{eq-starobs}).  However, by the remark
at the end of the previous paragraph, this only affects the constant
of proportionality in Eq.~(\ref{eq-starobs}), but not the normalized
probability distributions.

The second choice defines a slightly different class of observers,
which are like us in the sense that they exist 10 Gyr after {\em
  most\/} stars in the universe are produced.  (In our universe, the
peak of the star formation rate was about $10\,{\rm Gyr}$ ago.)

\subsubsection{Observers = entropy production}

The second law of thermodynamics guarantees that the spontaneous
formation of an ordered subsystem (like a frozen pond, or a galaxy)
will be compensated by increased entropy in the remaining system.  In
practice, this increase tends to overcompensate vastly, and the
overall entropy increases.  This motivated one of us to
propose~\cite{Bou06} that the emergence of complex structures such as
observers is correlated with the production of entropy.  The simplest
ansatz is that the rate of observation is proportional (on average) to
the rate of entropy production:
\begin{equation}
  \frac{d^2 n_{\rm obs}}{dV_{\rm c}\, dt}(t) \propto 
  \frac{d^2 S}{dV_{\rm c}\, dt}(t)~.
  \label{eq-entobs}
\end{equation}

In Ref.~\cite{BouHar07}, it was shown that most of the entropy
produced inside the causal diamond in our universe comes from dust
heated by stars. In the absence of complex molecules, let alone stars
and galaxies, the entropy production would be much lower. The simple
criterion of entropy production thus turns out to capture several
conditions often assumed explicitly to be necessary for
life. Moreover, it succeeds very well in postdicting our rather
unusual location: If life is correlated with entropy production, then
most life forms will find themselves when and where most entropy
production takes place: near stars, during the era while stars are
burning.  Indeed, the probability distribution over positive values of
$\Lambda$ computed from the causal entropic principle (weighting by
entropy production in the causal diamond) proved to be in excellent
agreement with observation~\cite{BouHar07}.

Here we refine and extend the validity of that prescription.  We will
use the star formation rates calculated according to
Ref.~\cite{BouLei08} in order to properly account for the changes in
star formation as cosmological parameters are varied.  In addition, we
will account for the dependence of the dust temperature on the virial
density.

In contrast to the causal diamond, overall entropy production in the
causal patch is dominated by the entropy produced at reheating.  This
is evidently not a good proxy for observers.  In the context of the
causal patch cut-off, we will model observers specifically in terms of
the entropy produced by dust heated by stars (or, as above, by a time
delay).

\subsection{Summary}
\label{sec-summary}

The rate of observation, per unit comoving volume and unit time, in a
universe with parameters $(\Lambda, Q, \Delta N)$ is given by
\begin{equation}
\frac{d^2n_{\rm obs}}{dV_{\rm c}\,dt}(t;\,\Lambda,Q,\Delta N)\propto \begin{cases}
\frac{d^2 S}{dV_{\rm c}dt}(t;\,\Lambda,Q,\Delta N) 
& \text{(entropy production)}\\
\dot\rho_\star(t-5\,{\rm Gyr};\Lambda,Q,\Delta N) 
& \text{(star formation plus 5 Gyr)}\\
\dot\rho_\star(t-10\,{\rm Gyr};\Lambda,Q,\Delta N) 
& \text{(star formation plus 10 Gyr)}
\end{cases}
\label{eq-3mod}
\end{equation}
depending on which model for observers we use; see
Sec.~\ref{sec-observers}.  In the second (third) case, we set the rate
of observation to zero for $t<5$ Gyr ($t<10$ Gyr).

The total number of observations in a universe with parameters
$(\Lambda, Q, \Delta N)$ is given by integrating the above rate over
the time and comoving volume contained in the causal patch or causal
diamond:
\begin{equation} 
  n_{\rm obs}(\Lambda, Q, \Delta N)=
  \int_0^\infty dt~ V_{\rm c}(t)~
  \frac{d^2n_{\rm obs}}{dV_{\rm c}\,dt}(t;\,\Lambda,Q,\Delta N)~,
\label{eq-nobs}
\end{equation} 
where the comoving volume at time $t$ is given by
\begin{equation}
V_{\rm c}(t)=\begin{cases}
\int_0^{\chi_{\rm patch}(t)} d\chi~ 4\pi\sinh^2\chi 
& \text{(causal patch measure)}\\
\int_0^{\chi_{\rm diamond}(t)} d\chi~ 4\pi\sinh^2\chi 
& \text{(causal diamond measure)}
\end{cases}
\label{eq-vc}
\end{equation}
and $\chi_{\rm patch}$ and $\chi_{\rm diamond}$ are given in
Eqs.~(\ref{eq-patch}) and (\ref{eq-diamond}).  

The probability distribution over $(\Lambda, Q, \Delta N)$ is obtained
by multiplying the prior probability for a universe with $(\Lambda, Q,
\Delta N)$, discussed in Sec.~\ref{sec-prior}, by the number of
observations made in such a universe:
\begin{equation}
  \frac{d^3p}{d\log\Lambda\,d\log_{10} Q\,d(\Delta N)}= 
  \frac{\Lambda}{(60+\Delta N)^4}~ n_{\rm obs}(\Lambda, Q, \Delta N)~.
\end{equation}
With three choices of observer model and two choices of measure, we
thus consider a total of six different models for computing
probabilities in the multiverse.

\section{Star formation and entropy production
  in the multiverse}

In this section we describe in detail how we compute the quantities
appearing in Eq.~(\ref{eq-3mod}).  In Sec.~\ref{sec-stars} we review
our star formation model~\cite{BouLei08}.  In Sec.~\ref{sec-entropy},
we estimate the rate of entropy production by the dust heated by
stars, following~\cite{BouHar07,BouHarTA}.

\subsection{Star formation}
\label{sec-stars}

To compute the rate of star formation per unit time and unit comoving
volume in a universe with parameters $(\Lambda,Q,\Delta N)$,
\begin{equation}
  \dot{\rho}_\star(t;\,\Lambda,Q,\Delta N)
  \equiv \frac{d^2 m_\star}{dV_{\rm c}\, dt}~,
\end{equation}
we use the model we developed in Ref.~\cite{BouLei08}.  The following
summary will skip many details, and the reader is encouraged to
consult Ref.~\cite{BouLei08} for a more thorough discussion.  There
are three steps to star formation: (1) density perturbations grow and
collapse to form dark matter halos; (2) baryons trapped in the halo
cool and condense; (3) stars form from the cooled gas.

Cosmological perturbations can be specified by a time-dependent power
spectrum, $\mathcal{P}$, which is a function of the wavenumber of the
perturbation, $k$.  The r.m.s.\ fluctuation amplitude, $\sigma$,
within a sphere of radius $R$, is defined by smoothing the power
spectrum with respect to an appropriate window function:
\begin{equation}
  \sigma^2 = \frac{1}{2\pi^2}\int_0^\infty 
  \left(\frac{3\sin(kR)-3\,kR\cos(kR)}{(kR)^3}\right)^2
  \mathcal{P}(k)k^2\,dk~.
\end{equation}
The radius $R$ can be exchanged for the mass $M$ of the perturbation,
using $M = 4\pi\rho_{\rm m}R^3/3$.  Generally $\sigma$ factorizes as
\begin{equation}
  \sigma(M,t) = Q\,s(M)\,G(t)~.
\end{equation}
$Q$ sets the overall scale of density perturbations, and is one of the
parameters we vary. The scale dependence $s(M)$ is held fixed; we use
the fitting formula provided in Ref.~\cite{TARW}:
\begin{equation}
  s(M) = \left[ 
    \left(9.1\mu^{-2/3}\right)^{-0.27} 
    + 
    \left(50.5\log_{10} \left(834 + \mu^{-1/3}\right)-92\right)^{-0.27}
  \right]^{-1/0.27}
\end{equation}
with $\mu=M/M_{\rm eq}$, where
\begin{equation}
M_{\rm eq} = 1.18\times 10^{17} m_\odot 
\end{equation}
is roughly the mass contained inside the horizon at matter-radiation
equality.  The linear growth function $G(t)$ satisfies
\begin{equation}
\frac{d^2G}{dt^2}+2H\frac{dG}{dt}=4\pi G_{\rm N}\rho_{\rm m}G~
  \label{eq-g}
\end{equation}
with the initial conditions $G=5/2$ and $\dot{G} = 3H/2$ at $t=t_{\rm eq}$.
For each value of $\Lambda$ and $\Delta N$, we numerically compute the
scale factor $a(t)$ and, from Eq.~(\ref{eq-g}), $G(t)$.

Density perturbations grow and collapse to form structure.  The
Press-Schechter function, $F$, gives the total fraction of mass
collapsed into structures of mass $<M$~\cite{PressSchechter}:
\begin{equation}
F(<M,t) = {\rm Erf}\left(\frac{1.68}{\sqrt{2} \sigma(M,t)}\right)~.
\label{eq-PSFraction}
\end{equation}
We can compute the mass density of a collapsed halo (called the virial
density, $\rho_{\rm vir}$) using the spherical top-hat collapse model.
The virial density does not depend on the mass of the object, but only
on the time of collapse.

After collapse, the baryonic component of a halo must cool and undergo
further condensation before stars can form.  We require that this
cooling process happen sufficiently quickly.  The most efficient
cooling mechanisms require ionized gas, so only those halos with a
virial temperature above $10^4~{\rm K}$ can cool further.  This
translates into a time-dependent lower mass limit for star-forming
halos.  Also, we require that halos cool on a timescale faster than
their own gravitational timescale, $t_{\rm grav} = (G_{\rm N}\rho_{\rm
  vir})^{-1/2}$.  This is motivated by observations indicating that
cooling-limited galaxy formation is ineffective, providing a
time-dependent upper mass limit on star-forming halos.

A halo in the allowed mass range is assigned an individual star
formation rate based on its mass and time of virialization:
\begin{equation}
\label{eq-SingleHaloSFR}
\frac{dm_\star^{\rm single}}{dt}(M, t_{\rm vir})=
\frac{1}{6}\sqrt{\frac{32}{3\pi}} 
\frac{M}{t_{\rm grav}(t_{\rm vir})}~.
\end{equation}
We use the extended Press-Schechter formalism~\cite{LaceyCole} to sum
over the formation times of all halos of mass $M$ in existence at time
$t$:
\begin{equation}\label{eq-AvgHaloSFR}
  \frac{dm^{\rm avg}_\star}{dt}(M,t)= 
 \frac{1}{6}\sqrt{\frac{32}{3\pi}} \int_{t_{\rm min}}^{t_{\rm max}} \;
  \left[ \frac{M}{t_{\rm grav}(t_{\rm vir})}\,
    \frac{\partial P}{\partial t_{\rm vir}}(t_{\rm vir},M,t)\right]\,
  dt_{\rm vir}~.
\end{equation}
The function $P$ is the probability that a halo of mass $M$ at time
$t$ virialized before $t_{\rm vir}$, and is derived in
Ref.~\cite{LaceyCole}:
\begin{equation}\label{eq:EPSFunction}
P(<t_{\rm vir},M,t) = \int_{M/2}^M \frac{M}{M_1}\frac{d\beta}{dM_1}(M_1,t_{\rm vir},M,t)  \,dM_1~,
\end{equation}
where
\begin{equation}
\beta(M_1,t_1,M_2,t_2)={\rm Erfc}\left( \frac{1.68}{Q\sqrt{2 (s(M_1)^2 - s(M_2)^2)}}\left(\frac{1}{G(t_1)}-\frac{1}{G(t_2)}\right)    \right)~.
\end{equation}
$t_{\rm min}$ and $t_{\rm max}$ are specific functions of $M$ and $t$
designed to restrict the range of integration to only those halos
which are capable of cooling and have not yet used up all of their
cold gas supply~\cite{BouLei08}.

Finally, the star formation rate itself is given by summing over all
values of halo mass, weighted by the Press-Schechter distribution
function:
\begin{equation}
\dot{\rho}_\star(t;\,\Lambda,Q,\Delta N)= 
\sqrt{\frac{8}{27\pi}}\int  dM 
\int_{t_{\rm min}}^{t_{\rm max}} dt_{\rm vir}\, 
\frac{\partial F}{\partial M}(M,t)  
\frac{1}{t_{\rm grav}(t_{\rm vir})} 
\frac{\partial P}{\partial t_{\rm vir}}(t_{\rm vir},M,t)~.
\end{equation}
%$\rho_0$ is a fixed density that essentially defines our unit of
%comoving volume. 

\subsection{Entropy production}
\label{sec-entropy}

As explained in Sec.~\ref{sec-observers}, we model the rate of
observation either by the rate star formation plus a time delay, or by
the rate of entropy production by stars.  A time delay is trivial to
implement: we simply shift the star formation rate by 5 or 10 Gyr.
Here we will discuss how to estimate the rate of entropy production
per unit comoving volume and unit time.  The entropy production rate
at the time $t$ is given by the total luminosity of the stars shining
at that time, divided by the effective temperature at which this power
is ultimately dissipated.  A significant fraction of starlight is
absorbed by interstellar dust grains and re-emitted in the infrared.
This process converts one optical photon into many infrared photons,
so it dominates entropy production~\cite{BouHar07}.  Hence, the
appropriate temperature is the interstellar dust temperature.

We will follow Ref.~\cite{BouHar07} for computing the total luminosity
at time $t$ from the star formation rate at earlier times.  We will
follow Ref.~\cite{BouHarTA} for estimating the temperature of dust at
time $t$ in a galaxy that formed at time $t_{\rm f}$.  This doubly
time-dependent treatment of the dust temperature is a refinement over
Ref.~\cite{BouHar07}, where the temperature was held fixed.

We take the luminosity of an individual star to be related to its mass
by $L_\star \propto m^{3.5}$.  The mass distribution of newly formed
stars is assumed to be described by the Salpeter initial mass
function, independently of time and of the parameters $(\Lambda, Q,
\Delta N)$:
\begin{equation}
\xi_{\rm IMF}(m)\equiv \frac{dN_\star}{dm}=\begin{cases}
am^{-2.35} & \text{if $m>0.5 \, m_\odot$}\\
bm^{-1.5} & \text{if $m\leq 0.5 \, m_\odot$}
\end{cases}
\end{equation}
No stars form ($\xi_{\rm IMF}(m)=0$) outside the range $0.08\,m_\odot
<m<100\,m_\odot$. Here $a$ and $b$ are constants chosen so that the
function is continuous and integrates to one over the allowed mass
range.

The lifetime of a star is also controlled by its mass; smaller stars
live longer.  It is convenient to work with the inverted
relation~\cite{BouHar07}:
\begin{equation}
  m_{\rm max}(\Delta t) = \begin{cases}
    100\,m_\odot & \text{if $\Delta t<10^{-4}$ Gyr}~;\\
    \left(\frac{10\,{\rm Gyr}}{\Delta t}\right)^{2/5}\,m_\odot
    & \text{if $\Delta t\geq 10^{-4}$ Gyr}~,
\end{cases}
\end{equation}
where $m_{\rm max}(\Delta t)$ is the mass of the largest survivors in
an ensemble of stars created a time $\Delta t$ ago.  Setting $m_{\rm
  max} = 0.08\,m_\odot$ corresponds to $\Delta t = 8^{-5/2}10^6 \,{\rm
  Gyr}\approx 5500 \, {\rm Gyr}$, the lifetime of the longest-lived
stars.

Now consider an ensemble of stars of total mass $dm_\star$ that formed
at the time $t_f$.  Their combined luminosity at the time
$t=t_f+\Delta t$ is independent of $t_f$ and is given by
\begin{equation}
  \frac{dL}{dm_\star}(\Delta t)=
  \frac{1}{\left<m\right>}
  \int_{0.08\,m_\odot}^{m_{\rm max}(\Delta t)}dm\,\xi_{\rm IMF}(m)\,
  L_\odot\left(\frac{m}{m_\odot}\right)^{3.5}~.
\end{equation}
The mass and luminosity of the sun, $m_\odot$ and $L_\odot$, and the
average initial mass, $\left<m\right>$, are constant and drop out in
all normalized probabilities.  We will continue to display them for
clarity.

Next, we turn to estimating the temperature of interstellar dust, at
which this luminosity is ultimately dissipated.  The dust temperature
will depend on the mass density of the host galaxy (higher density
means higher temperature), and on the CMB temperature.  The CMB
temperature is borderline negligible in our universe.  But in a large
portion of our parameter space (specifically, for $Q>Q_0$),
significant star formation occurs earlier than in our universe.  Then
the CMB temperature can be larger than the temperature the dust would
reach just from stellar heating, and so it can effectively control the
dust temperature.  This is an important effect mitigating the
preference for larger values of $Q$~\cite{BouHarTA}.

Ref.~\cite{Andriesse} [see Eq.~170 therein] models how the temperature
of the interstellar dust scales with the distance to a
star:\footnote{The powers of temperature in this relation arise
  because the dust neither absorbs nor emits as a blackbody.  However,
  our results would not change much if we adopted the blackbody ansatz
  $T^4= \left(\frac{R_\star}{R}\right)^2 T_\star^4 + T_{\rm CMB}^4$.
  This is reassuring since the regime of validity of
  Eq.~(\ref{eq-DustTemp}) depends on material properties which are
  difficult to estimate.}
\begin{equation}\label{eq-DustTemp}
  T(t_{\rm vir},t)^6 \propto \left(\frac{R_\star}{R(t_{\rm vir})}\right)^2 
  T_\star^5 + T_{\rm CMB}(t)^5~,
\end{equation}
where we have included the last term to account for the heating of
dust by the CMB.  Here, $R$ is a typical distance between stars, and
$T_\star$ is a typical stellar temperature (we use $6000\,{\rm K}$).
We are explicitly dropping an overall dimensionful factor because we
are only interested in normalized probabilities.  One expects the
interstellar distance to scale inversely with the density of the host
galaxy, which in turn is proportional to the virial density of halo:
\begin{equation}
  R^3 \propto \rho_{\rm vir}^{-1}~.
\label{eq-r3}
\end{equation}
We normalize to the value $(R_\star/R)^2=3.5\times 10^{-14}$ for our
galaxy~\cite{Andriesse}, which we assume formed with a virial density
typical of halos forming at $t_{\rm vir}=3.7\,{\rm Gyr}$.  Then
Eq.~(\ref{eq-r3}) determines the relevant $R$ for other galaxies that
form at different times in our universe or in others.  Note that the
virial density is set by the time $t_{\rm vir}$ of the last major
merger, whereas the CMB temperature appearing in
Eq.~(\ref{eq-DustTemp}) must be evaluated at the time of emission.  In our model, stars do not form for a long time after virialization,
\begin{equation}
\frac{t_f-t_{\rm vir}}{t_{\rm vir}}\lesssim 1~,
\end{equation}
Thus we can approximate $t_{\rm vir} \approx t_f$ for the purposes of
estimating the relevant dust temperature, leaving us with one fewer
time variable to keep track of.

To compute the entropy production rate, at the time $t=t_f+\Delta t$,
per unit stellar mass formed at the time $t_f$, we divide the
luminosity by the temperature:
\begin{eqnarray} 
  \frac{d^2S}{dm_\star dt}(t,t_{\rm f}) & =  &
  \frac{1}{T(t_f,t_f+\Delta t)}~
  \frac{dL}{dm_\star}(\Delta t)\\ & = &
  \frac{1}{T(t_{\rm f},t)}~\frac{L_\odot}{\left<m\right>}
  \int_{0.08 m_\odot}^{m_{\rm max}(t-t_f)}dm\,\xi_{\rm IMF}(m)\,
  \left(\frac{m}{m_\odot}\right)^{3.5}~.
\end{eqnarray} 
From this rate we can obtain the total entropy production rate at the
time $t$, by integrating over $t_f$ and using the SFR to account for
the amount of stellar mass produced per comoving volume at the time
$t_f$:
\begin{equation}
  \frac{d^2 S}{dV_{\rm c}dt}(t) = \int_0^t dt_{\rm f}~
  \frac{d^2S}{dm_\star dt}(t,t_{\rm f})~\dot{\rho}_\star(t_{\rm f})~.
\end{equation}
By Eq.~(\ref{eq-3mod}), this becomes the integrand in
Eq.~(\ref{eq-nobs}) in the case where observers are modeled by entropy
production.\footnote{Numerically, the resulting double integral is
  most efficiently evaluated by exchanging the order of integration:
\begin{align}
  n_{\rm obs} &\propto \int_0^\infty dt\,
  V_{\rm c}(t)~\frac{d^2 S}{dV_{\rm c}dt}(t) \\
  &= \int_0^\infty dt \int_0^t dt_{\rm f}~ V_{\rm c}(t)~\frac{d^2
    S}{dm_\star dt}(t,t_{\rm f})~
  \dot{\rho}_\star(t_{\rm f})\\
  &= \int_0^\infty dt_{\rm f} \int_{t_{\rm f}}^\infty dt~ V_{\rm
    c}(t)~\frac{d^2S}{dm_\star dt}(t,t_{\rm f})~
  \dot{\rho}_\star(t_{\rm f})\\
  &=\int_0^\infty dt_{\rm f} \left[\int_0^\infty d(\Delta t)~ V_{\rm
      c}(t_{\rm f}+\Delta t)~\frac{d^2S}{dm_\star dt}(t_{\rm f}+\Delta
    t,t_{\rm f})\right]\dot{\rho}_\star(t_{\rm f})~.
\end{align}
The inner integral represents the entropy that will eventually be
produced inside the causal patch or diamond by the stars created at
time $t_{\rm f}$.  Because it does not depend on $Q$, it is more
efficient to compute this integral separately as a function of
$(t_{\rm f};\Lambda,\Delta N)$, before multiplying by
$\dot{\rho}_\star(t_{\rm f};\Lambda,Q,\Delta N)$ and computing the
outer integral.}

\section{Results}
\label{sec-results}

This section contains the probability distributions we have computed
over the parameters $\Lambda, Q, \Delta N$.  

\paragraph{Ordering} There are six subsections, corresponding to the
six models described in Sec.~\ref{sec-framework} (two different
measures, and three different ways of modeling observers).  Each
subsection contains three pages of plots.  On the first page,
$\Lambda$ runs only over positive values; on the second, $\Lambda<0$.
This division is useful since some of the interesting features we find
depend on the sign of the cosmological constant.  Moreover, visually
the clearest way to display the probability distribution over
$\Lambda$ is as a function of $\log_{10}|\Lambda|$, which makes it
difficult to include $\Lambda=0$.  On the third page of each
subsection, we display some distributions over all values of
$\Lambda$.

We have computed a full trivariate probability distribution for each
case, which cannot be displayed in a single plot.  For this reason,
the first page ($\Lambda>0$) and second page ($\Lambda<0$) of every
subsection contain 12 plots each.  The first six plots (the top two
rows of plots) are single variable distributions over $\Lambda$, over
$Q$, and over $\Delta N$.  In the first three, the remaining two
variables are held fixed.  This corresponds to asking about the
probability distribution over a single parameter in the portion of the
landscape in which the two other parameters take the observed values.
In the second group of three plots, the remaining two variables are
marginalized (i.e., integrated out).  This corresponds to asking about
the probability distribution over a single parameter in the entire
three-parameter landscape we consider.

The remaining six plots on each page are bivariate probability
distributions.  Of these, the first three are distributions over two
parameters with the third held fixed.  This corresponds to asking
about a bivariate probability distribution in the portion of the
landscape in which the remaining parameter takes its observed value.
In the other three bivariate plots, the remaining variable is
integrated out.  This corresponds to asking about the probability
distribution over some pair of parameters in the entire
three-parameter landscape we consider.

The third page of each subsection shows distributions in which
$\Lambda$ takes both positive and negative values, either explicitly
or by marginalization.  The three plots in which $\Lambda$ is fixed to
the observed value would be identical to the corresponding plots shown
on the $\Lambda>0$ page.  Moreover, we do not display any plots
corresponding to a parameter combination that led to a pathological
probability distribution for either $\Lambda>0$ or $\Lambda<0$, when
the inclusion of both signs can only worsen the problem.  (This case
arises for the causal patch only.)

\paragraph{Confidence regions} In most plots (see the discussion
below), we show the one-sigma (68\% confidence) and two-sigma (95\%
confidence) parameter regions.  The one-sigma region is unshaded, the
two-sigma region is lightly shaded, and the remaining region is shaded
dark.  In the one-parameter plots, confidence regions are centered on
the median of the probability distribution.  In the two parameter
plots, they are centered on the maximum probability density and
bounded by contour lines of constant probability.  Additional contours
are included for better visualization of the probability distribution.
They not drawn at any special values of the probability density or of
its integral.

The displayed confidence regions are strictly based on the probability
distribution over the portion of the landscape for which we have
computed probabilities:
\begin{eqnarray}
  10^{-3}\Lambda_0< & \abs{\Lambda} & <10^5\Lambda_0\\
  10^{-1}Q_0< & Q & <10^2Q_0\\
  -3.5 < & \Delta N & <\infty ~,
  \label{eq-range2}
\end{eqnarray}
with all other physical parameters held fixed.  In other words, we are
setting the probability to zero outside the above range, and for
universes in which other parameters differ from ours.  As we noted in
the introduction, this is legitimate: we are asking whether or not our
observations are typical among those made by observers in this portion
of the landscape, described by a particular range of three particular
parameters.  If they are highly atypical, then there is a problem.

In certain plots involving $\Lambda<0$ with the causal patch measure,
the probability density increases very rapidly towards the boundary of
our parameter range. Because of this runaway behavior, the 1 and
$2\sigma$ regions would depend sensitively on the precise value of the
parameter boundary. In these cases, we do not display confidence
intervals in single-variable plots; in bivariate plots, we display
only the contours of constant probability density, along with an arrow
indicating the direction of increasing probability density. Other
runaways are less strong; in this case we do display confidence
intervals based on the above parameter range.  Finally, not every
probability distribution that increases monotonically towards a
boundary is indicative of a runaway, because it might be cut off by a
change of regime at finite distance beyond the boundary: For any value
of $Q$ and $\Lambda$, sufficiently large curvature will disrupt
structure formation.  And for any $\Delta N$ and $\Lambda$,
sufficiently large $Q$ (of order the upper bound we consider, $10^2
Q_0$) leads to a change of regime.  We will note in the captions which
plots have a true runaway direction.

\paragraph{Display range and data point} We display the entire range
of parameters for which we have computed the probability density,
Eq.~(\ref{eq-range2}), except, of course, for $\Delta N$, where we cut
off the display at $\Delta N=1$.  For larger values of $\Delta N$,
curvature is too weak to affect either the dynamics of structure
formation or the geometry of the causal patch or diamond.  In this
regime, the probability distribution over $\Delta N$ is proportional
to the prior distribution, Eq.~(\ref{eq-nprior}).  All contour
intervals take this undisplayed portion of the probability
distribution into account.  Also, when we marginalize over $\Delta N$,
the undisplayed portion is included in the range of the integral.

The display variables are not $\Lambda$, $Q$, and $\Delta N$, but
$\log_{10} (|\Lambda|/\Lambda_0)$, $\log_{10} (Q/Q_0)$, and $\Delta
N$.  Therefore, the observed values correspond to $0$ on every axis.
To guide the eye, the vertical axis intersects the horizontal axis at
$0$ in all single-parameter plots, so the observed value is where the
vertical axis is.  In the two-parameter plots, the data point $(0,0)$
is shown by a green triangle.

There are two subtleties: First, in the figures that display only the
negative range of $\Lambda$, the observed (positive) value cannot be
shown.  We find it useful to show as ``data'' our ``evil twin''
universe, with $\Lambda=-\Lambda_0$, $Q=Q_0$, and $\Delta N\geq0$, in
these plots.  Secondly, spatial curvature has not been detected, only
constrained.  Thus, $\Delta N=0$ corresponds to a lower bound, and not
to the actual value of $\Delta N$ in our universe.  The reader should
keep in mind, therefore, that in single-variable plots over $\Delta N$
the entire region to the right of the vertical axis is compatible with
observation.  In two-parameter plots involving $\Delta N$, the
observed universe lies somewhere on a semi-infinite line starting at
the triangle and running upward towards larger $\Delta N$.  As long as
some part of the $\Delta N\geq 0$ range is not very improbable, there
would be no conflict with experiment, even if the point $\Delta N=0$
were highly improbable.

\paragraph{Comparing the probability distributions to observation} 

Because life is short, we reject scientific theories with finite
confidence only, taking a gamble that freak chance might be leading us
astray.  Often, we can increase our confidence by repeating
experiments.  In cosmology, we sometimes cannot.  This limitation has
nothing to do with the multiverse, but stems from the finiteness of
the observable universe.  (Because $\Lambda_0\neq 0$, the observable
universe will not grow indefinitely, so this is not merely an accident
of our present era.)  For example, cosmic variance does not prevent us
from drawing conclusions from the CMB, but it does prevent us from
sharpening them when they do not meet a confidence standard we are
comfortable with, as may be the case for the low quadrupole.

There is one data point for each parameter considered in this paper,
the value observed in our universe (or, in the case of $\Delta N$, the
range not ruled out).  If this data point happened to have very small
probability (e.g., if it lay well beyond $6\sigma$, if this were our
desired level of confidence), then our observations would be extremely
unexpected given the theory from which the probability distribution
was computed.  In other words, the theory would conflict with
experiment at that level of confidence.  Since the theory consists of
a combination of a prior distribution in the landscape (including the
possible range of parameters), a choice of measure, and a choice of
observer model, at least one of these elements is excluded.

\paragraph{Main conclusions}

Our conclusions are more fully described in the introduction.  The
causal diamond measure is remarkably successful.  The observed values
of parameters lie in the $2\sigma$ confidence region of all but one or
two out of thirty-three probability distributions we show for each
observer model (where they lie within $3\sigma$).

The causal patch is problematic, independently of the observer model.
In the absence of curvature, results are similar to the causal
diamond, if not quite as successful quantitatively.  In the presence
of curvature, we find significant conflicts between prediction and
observation.  They are sharpest in distributions that include negative
values of $\Lambda$ (see, e.g., Fig.~\ref{posnegPatch5Gyr}), where we
find a strong runaway towards small $|\Lambda|$ and strong pressure
towards the large-curvature boundary on structure formation.  If we
restrict to positive $\Lambda$, there is still a weak runaway to small
positive values of $\Lambda$, though this is barely visible in the
parameter range we display.  As discussed in the introduction, these
runaways imply that the causal patch measure is incorrect, or that the
prior probabilties differ significantly from those we have assumed, or
that the finite size of the landscape effectively provides a cut-off
on how small $\Lambda$ can be, and that we find ourselves near this
cut-off.  The level of confidence at which the measure is excluded by
the data depends sensitively on this cut-off; with no cut-off, it is
infinite.

\subsection{Weighting by entropy production in the causal diamond}

\begin{figure}[h!]
 \begin{center}
 $\begin{array}{c@{\hspace{.1in}}c@{\hspace{.1in}}c}
 \includegraphics[width=1.8in]{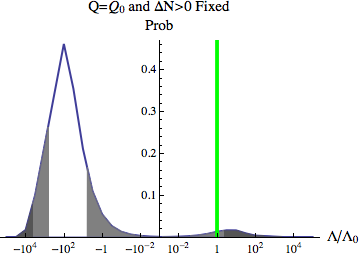} & & \\
 \includegraphics[width=1.8in]{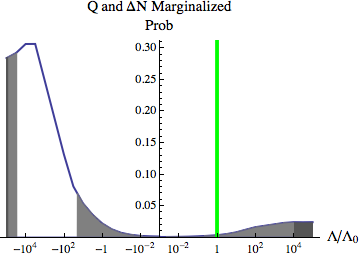} &
 \includegraphics[width=1.8in]{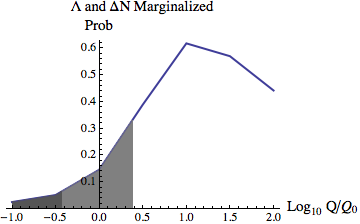} &
 \includegraphics[width=1.7in]{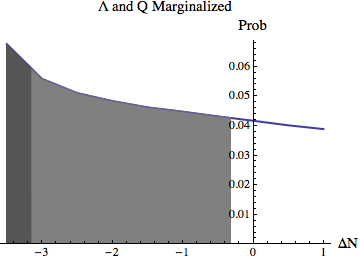} \\
 \end{array}$
 $\begin{array}{c@{\hspace{.1in}}c@{\hspace{.1in}}c}
 &
 \includegraphics[width=1.7in]{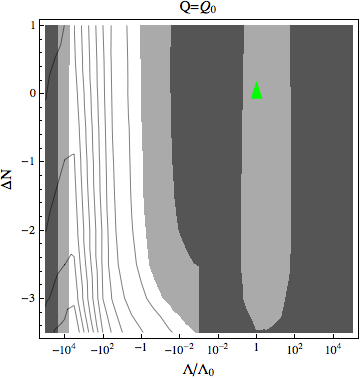} &
 \includegraphics[width=1.8in]{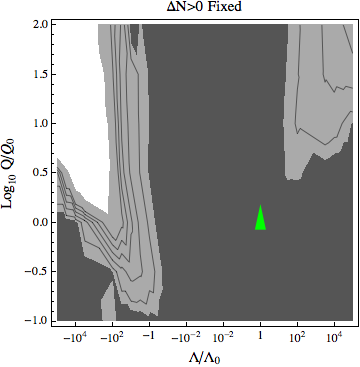} \\
 \includegraphics[width=1.7in]{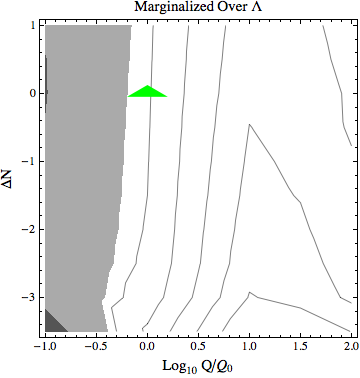} &
 \includegraphics[width=1.7in]{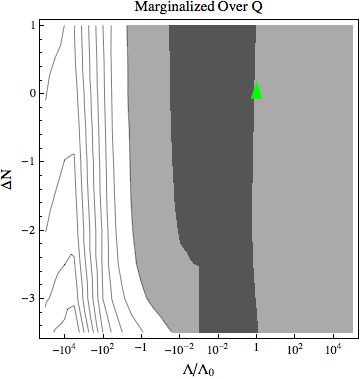} &
 \includegraphics[width=1.8in]{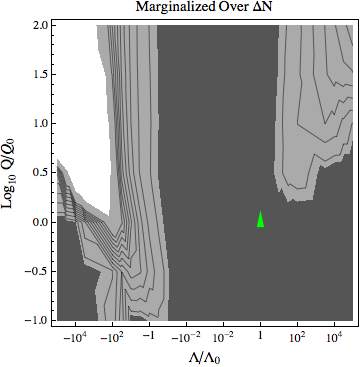} \\
 \end{array}$
 \caption{All values of $\Lambda$, Causal Diamond, Entropy Production.
   $\Lambda<0$ is preferred over $\Lambda>0$ by a factor of 25 (11) in
   the 1st (2nd) plot.  The observed values of $(\Lambda,Q)$ lie
   outside $2\sigma$, but within $3\sigma$, in the 6th and 9th plot.}
 \label{posnegDia}
 \end{center}
 \end{figure}

With this observer model and measure, the observed parameters fall
within the central $2\sigma$ of almost all probability distributions
we compute.  The only exceptions are the bivariate distributions over
$Q$ and the full range of $\Lambda$ (both positive and negative
$\Lambda$), where we find ourselves outside of $2\sigma$ but within
$3\sigma$ (6th and 9th plot, Fig.~\ref{posnegDia}).  The total
probability for $\Lambda<0$ is $25$ times larger than that for
$\Lambda>0$ when $Q=Q_0$ and $\Delta N>0$ are held fixed, and $11$
times larger when $Q$ and $\Delta N$ are marginalized (1st and 2nd
plot, Fig.~\ref{posnegDia}).

\begin{figure}[h!]
\begin{center}
$\begin{array}{c@{\hspace{.1in}}c@{\hspace{.1in}}c}
 \includegraphics[width=1.8in]{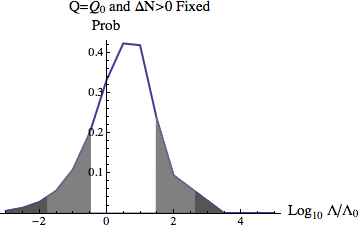} &
 \includegraphics[width=1.8in]{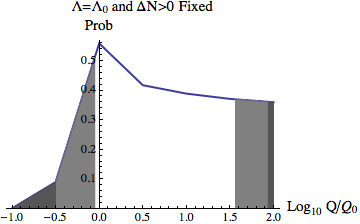} &
 \includegraphics[width=1.7in]{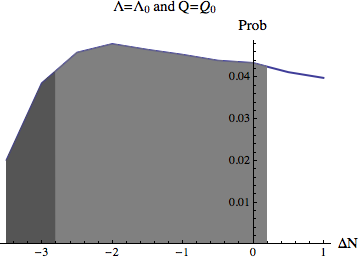} \\
 \includegraphics[width=1.8in]{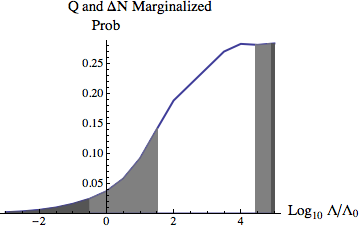} &
 \includegraphics[width=1.8in]{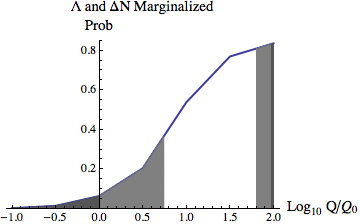} &
 \includegraphics[width=1.7in]{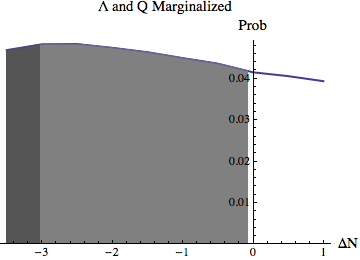} \\
 \end{array}$
 $\begin{array}{c@{\hspace{.1in}}c@{\hspace{.1in}}c}
 \includegraphics[width=1.7in]{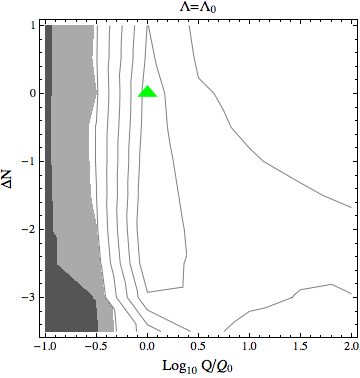} &
 \includegraphics[width=1.7in]{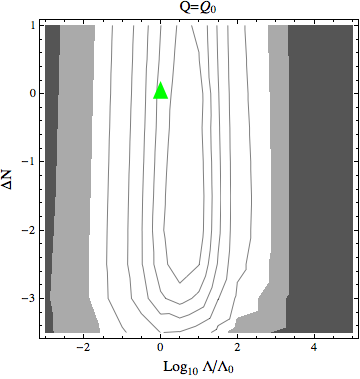} &
 \includegraphics[width=1.8in]{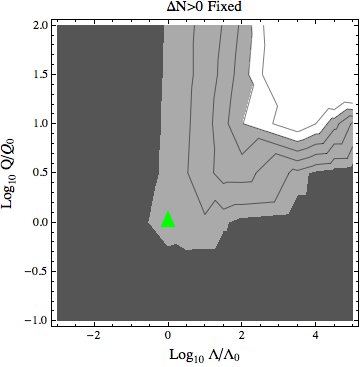} \\
 \includegraphics[width=1.7in]{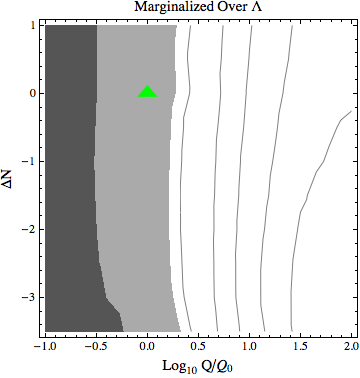} &
 \includegraphics[width=1.7in]{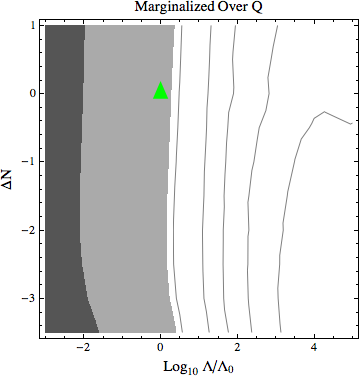} &
 \includegraphics[width=1.8in]{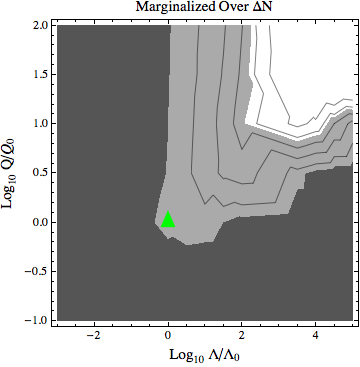} \\
 \end{array}$
 \caption{$\Lambda>0$, Causal Diamond, Entropy Production.  This is
   the only model studied which produces a peak in the $Q$
   distribution of the 2nd plot.}
 \label{posDia}
 \end{center}
 \end{figure}

As explained in Sec.~\ref{sec-qonly}, the distribution over $Q$, with
$\Lambda=\Lambda_0$ and $\Delta N>0$, has a maximum centered on the
observed value (2nd plot, Fig.~\ref{posDia}).  This feature is not
reproduced with any other choice of measure or observer model.
However, this does not in itself lend particular weight to this
measure and observer model.  Any model in which the observed value
falls, say, within $2\sigma$ should be regarded as unproblematic.

As explained in Sec.~\ref{sec-qlambdaonly}, the most likely
value of $Q$ grows with $\Lambda$ (9th and 12th plot,
Fig.~\ref{posDia}); therefore, the observed value of $Q$ is no
longer the most likely after $\Lambda$ is integrated out (5th plot,
Fig.~\ref{posDia}), though it is well within $2\sigma$.

\begin{figure}[h!]
 \begin{center}
 $\begin{array}{c@{\hspace{.1in}}c@{\hspace{.1in}}c}
 \includegraphics[width=1.8in]{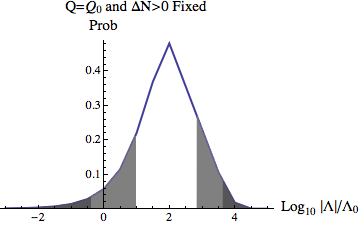} &
 \includegraphics[width=1.8in]{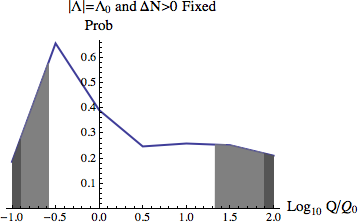} &
 \includegraphics[width=1.7in]{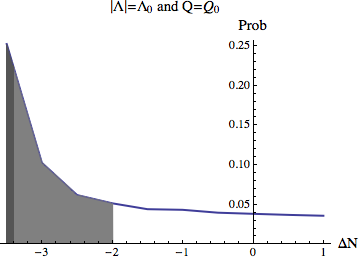} \\
 \includegraphics[width=1.8in]{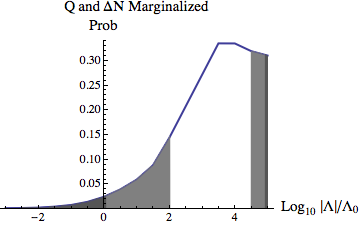} &
 \includegraphics[width=1.8in]{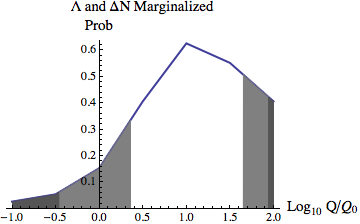} &
 \includegraphics[width=1.7in]{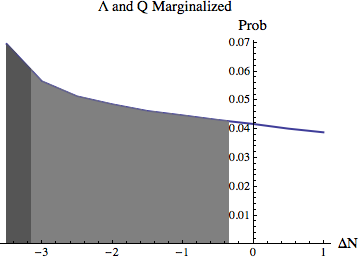} \\
 \end{array}$
 $\begin{array}{c@{\hspace{.1in}}c@{\hspace{.1in}}c}
 \includegraphics[width=1.7in]{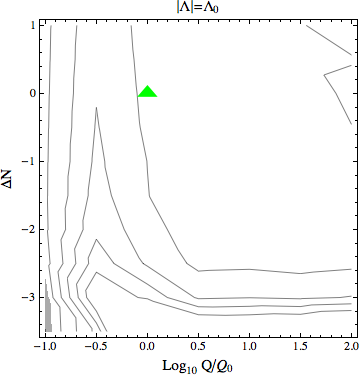} &
 \includegraphics[width=1.7in]{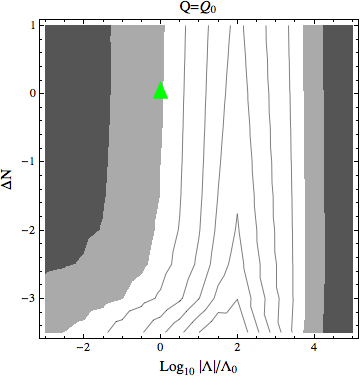} &
 \includegraphics[width=1.8in]{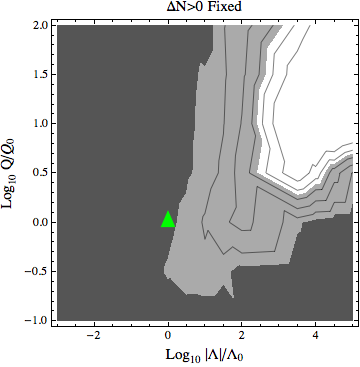} \\
 \includegraphics[width=1.7in]{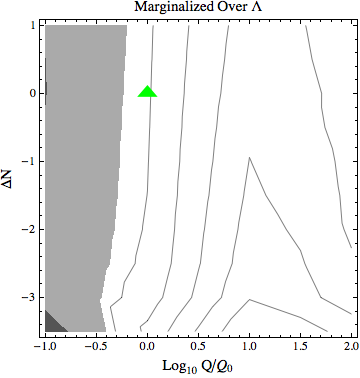} &
 \includegraphics[width=1.7in]{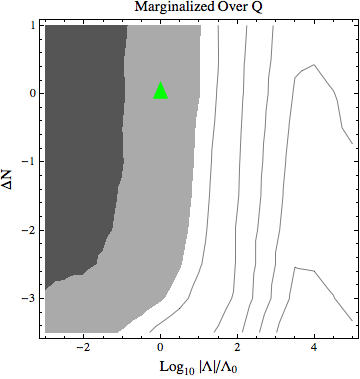} &
 \includegraphics[width=1.8in]{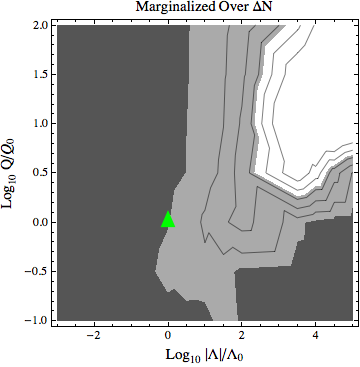} \\
 \end{array}$
 \caption{$\Lambda<0$, Causal Diamond, Entropy Production.  Unlike for
   $\Lambda>0$, the probability density increases monotonically with
   curvature (3rd and 6th plot).  However, the range allowed by
   observation ($\Delta N>0$) overlaps with the central $1\sigma$
   region (unshaded).  This good agreement is not an artifact of our
   lower cut-off, $\Delta N>3.5$, because the probability is about to
   be suppressed by the disruption of structure formation for slightly
   smaller $\Delta N$.  This can be seen by studying the distribution
   over $\Delta N$ at $Q=10^{-0.5}Q_0$ in the 7th and 10th plots.}
 \label{negDia}
 \end{center}
 \end{figure}

For $\Lambda=\Lambda_0$, the 3rd plot in Fig.~\ref{posDia} shows that
too much curvature suppresses structure formation: the probability
distribution increases towards smaller $\Delta N$ (governed mainly by
the prior distribution), but turns around near $\Delta N=-2$.  For
negative cosmological constant (Fig.~\ref{negDia}), the analogous plot
does not show this feature because the turnaround occurs just outside
our display range.  The reason for this difference is discussed in
Sec.~\ref{sec-lambdacurvonly}.

\newpage

\newpage

 \newpage

\newpage

\subsection{Weighting by star formation + 5 Gyr in the causal diamond}

\begin{figure}[h!]
 \begin{center}
 $\begin{array}{c@{\hspace{.1in}}c@{\hspace{.1in}}c}
 \includegraphics[width=1.8in]{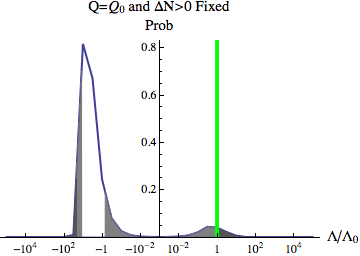} & & \\
 \includegraphics[width=1.8in]{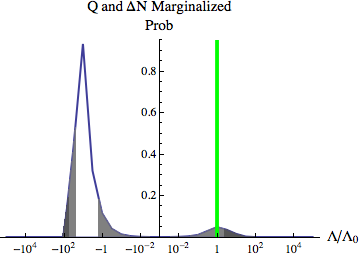} &
 \includegraphics[width=1.8in]{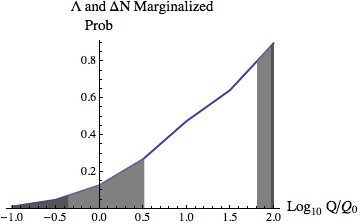} &
 \includegraphics[width=1.7in]{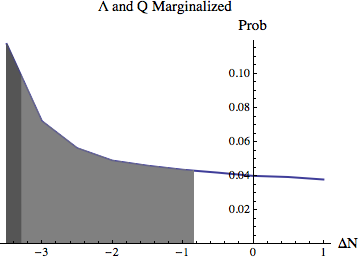} \\
 \end{array}$
 $\begin{array}{c@{\hspace{.1in}}c@{\hspace{.1in}}c}
 &
 \includegraphics[width=1.7in]{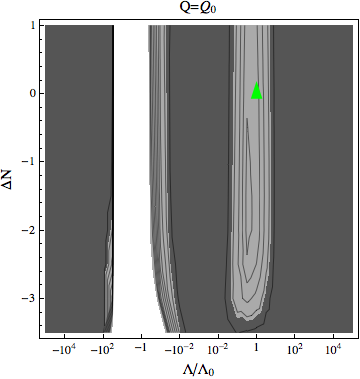} &
 \includegraphics[width=1.8in]{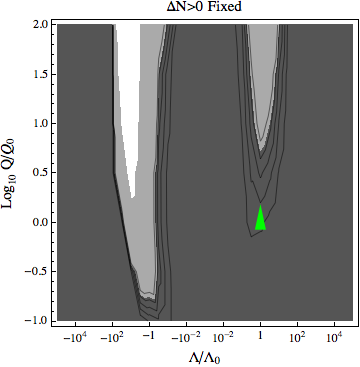} \\
 \includegraphics[width=1.7in]{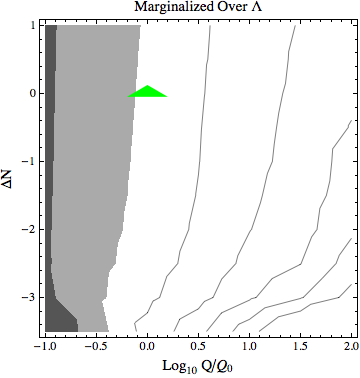} &
 \includegraphics[width=1.7in]{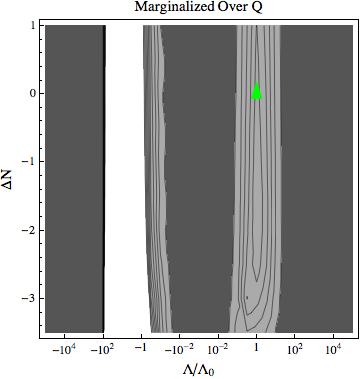} &
 \includegraphics[width=1.8in]{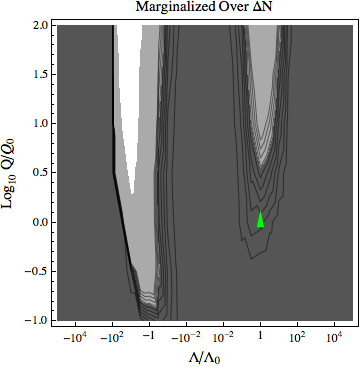} \\
 \end{array}$
 \caption{All values of $\Lambda$, Causal Diamond, 5 Gyr delay time.
   As with entropy production, $\Lambda<0$ is preferred over
   $\Lambda>0$ (here, by a factor of 13).  The observed values of
   $(\Lambda,Q)$ lie outside $2\sigma$, but within $3\sigma$, in the
   6th and 9th plot.}
 \label{posnegDia5Gyr}
 \end{center}
 \end{figure}

For this choice of measure and observer model, we find that the
observed values of parameters lie within or at the $2\sigma$ contours
of all plots with a single exception: in the bivariate distribution
over $Q$ and the full range of $\Lambda$ (both positive and negative
$\Lambda$) with $\Delta N$ marginalized (9th plot,
Fig.~\ref{posnegDia5Gyr}), we find ourselves within $3\sigma$.
The total probability for $\Lambda<0$ is $13$ times larger than that
for $\Lambda>0$ when $Q$ and $\Delta N$ are held fixed, and $12.6$
times larger when $Q$ and $\Delta N$ are marginalized.

\begin{figure}[h!]
 \begin{center}
 $\begin{array}{c@{\hspace{.1in}}c@{\hspace{.1in}}c}
 \includegraphics[width=1.8in]{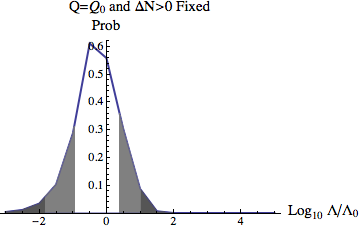} &
 \includegraphics[width=1.8in]{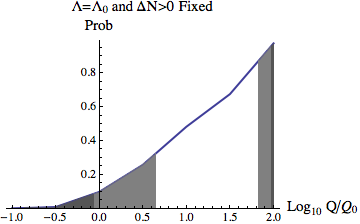} &
 \includegraphics[width=1.7in]{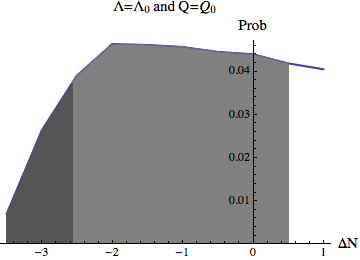} \\
 \includegraphics[width=1.8in]{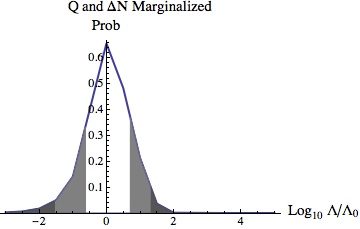} &
 \includegraphics[width=1.8in]{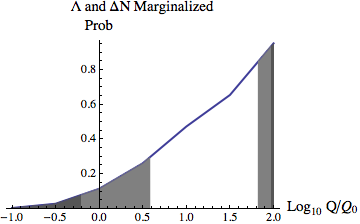} &
 \includegraphics[width=1.7in]{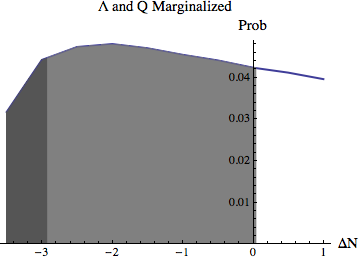} \\
 \end{array}$
 $\begin{array}{c@{\hspace{.1in}}c@{\hspace{.1in}}c}
 \includegraphics[width=1.7in]{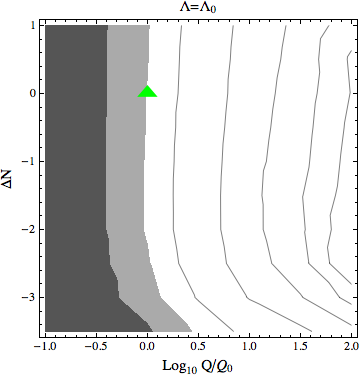} &
 \includegraphics[width=1.7in]{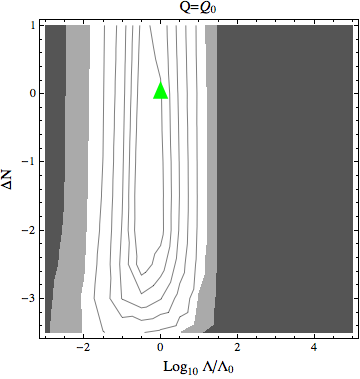} &
 \includegraphics[width=1.8in]{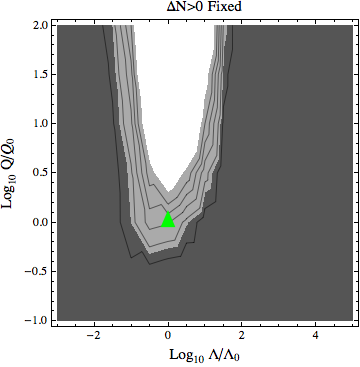} \\
 \includegraphics[width=1.7in]{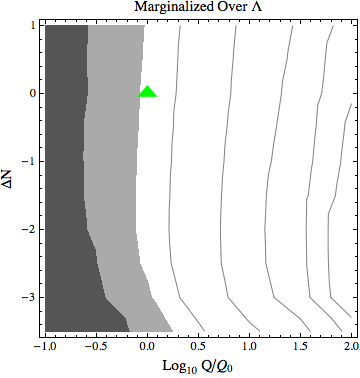} &
 \includegraphics[width=1.7in]{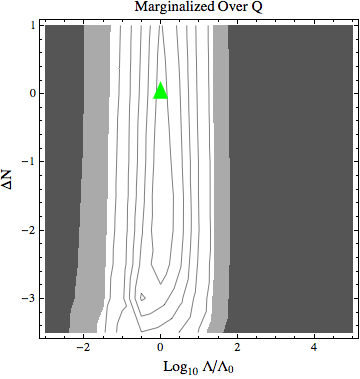} &
 \includegraphics[width=1.8in]{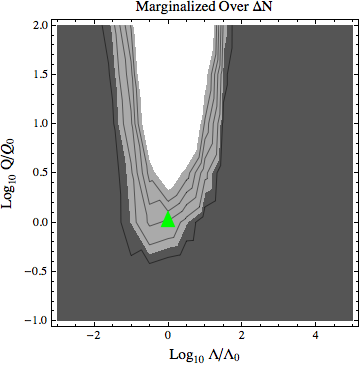} \\
 \end{array}$
 \caption{$\Lambda>0$, Causal Diamond, 5 Gyr delay time.  The
   preferred value of $\Lambda$ is independent of $Q$, as seen in the
   9th and 12th plots.}
 \label{posDia5Gyr}
 \end{center}
 \end{figure}

Unlike in the entropy production model shown previously, the
distribution grows monotonically towards large $Q$, but not rapidly
enough to render the observed value unlikely.  The preferred positive
value of $\Lambda$ is virtually independent of $Q$, as explained in
Sec.~\ref{sec-qlambdaonly}, and is roughly equal to the observed
value, as seen in the 9th and 12th plot in Fig.~\ref{posDia5Gyr}.

\newpage

 \newpage

 \begin{figure}[h!]
 \begin{center}
 $\begin{array}{c@{\hspace{.1in}}c@{\hspace{.1in}}c}
 \includegraphics[width=1.8in]{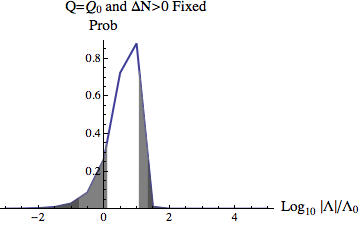} &
 \includegraphics[width=1.8in]{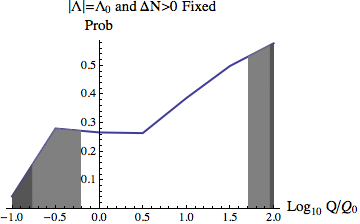} &
 \includegraphics[width=1.7in]{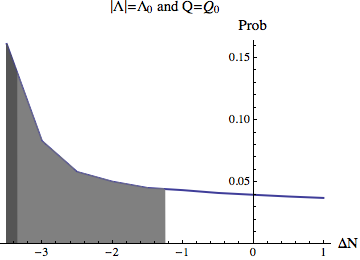} \\
 \includegraphics[width=1.8in]{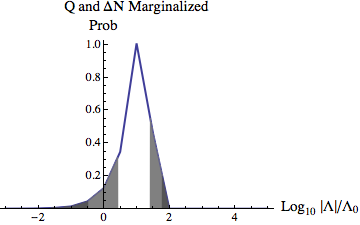} &
 \includegraphics[width=1.8in]{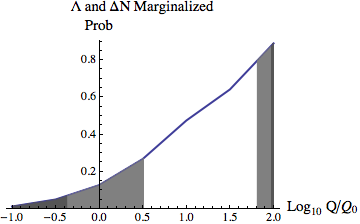} &
 \includegraphics[width=1.7in]{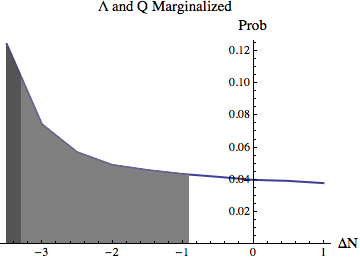} \\
 \end{array}$
 $\begin{array}{c@{\hspace{.1in}}c@{\hspace{.1in}}c}
 \includegraphics[width=1.7in]{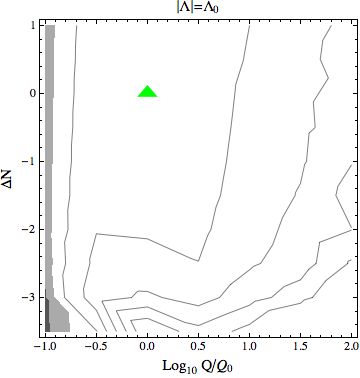} &
 \includegraphics[width=1.7in]{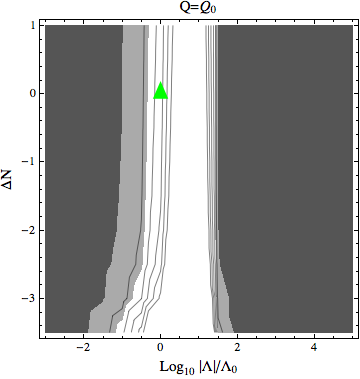} &
 \includegraphics[width=1.8in]{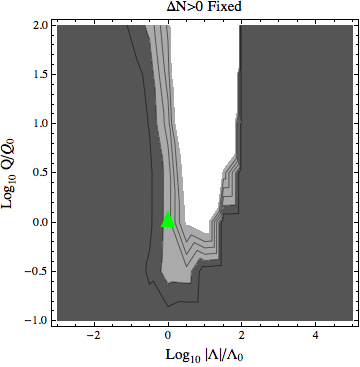} \\
 \includegraphics[width=1.7in]{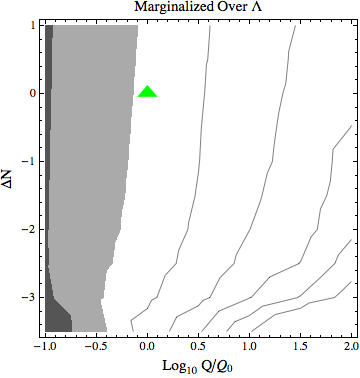} &
 \includegraphics[width=1.7in]{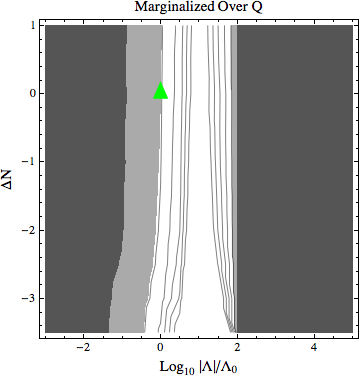} &
 \includegraphics[width=1.8in]{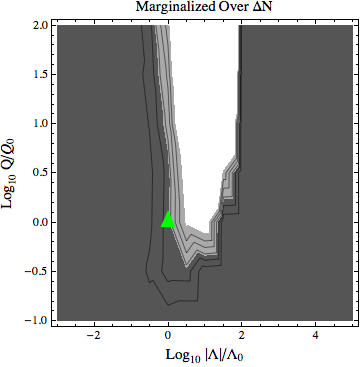} \\
 \end{array}$
 \caption{$\Lambda<0$, Causal Diamond, 5 Gyr delay time.  Note the
   peculiar feature around $Q=Q_0$ in the 2nd plot.  (It can also be
   seen in the 7th plot.)  The increasing probability towards small
   $\Delta N$ in the 3rd and 6th plot is not a runaway; see the
   caption of Fig.~\protect\ref{negDia}.}
 \label{negDia5Gyr}
 \end{center}
 \end{figure}

 \newpage

 \newpage

\subsection{Weighting by star formation + 10 Gyr in the causal diamond}

The $10\,{\rm Gyr}$ delay model shares many features with the $5\,{\rm
  Gyr}$ delay model, with only small numerical differences.  The
preferred value of $\Lambda$ is again nearly independent of $Q$ but
slightly smaller than with the $5\,{\rm Gyr}$ delay.  This feature is
explained in Sec.~\ref{sec-qlambdaonly}.  

The observed values of parameters lie within or at the $2\sigma$
contours of almost all plots. They are within $3\sigma$ of some of the
distributions that range over both positive and negative values of
$\Lambda$.  The total probability for negative $\Lambda$ is $12.4$
times larger than that for positive $\Lambda$ when $Q$ and $\Delta N>0$ are fixed, and $12.8$ times larger when $Q$ and $\Delta N$ are
marginalized.

\newpage

 \begin{figure}[h!]
 \begin{center}
 $\begin{array}{c@{\hspace{.1in}}c@{\hspace{.1in}}c}
 \includegraphics[width=1.8in]{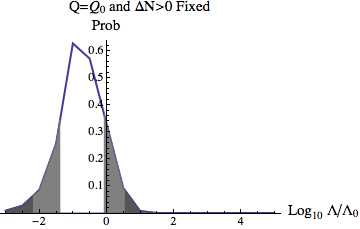} &
 \includegraphics[width=1.8in]{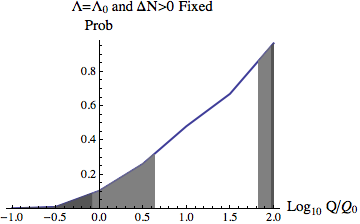} &
 \includegraphics[width=1.7in]{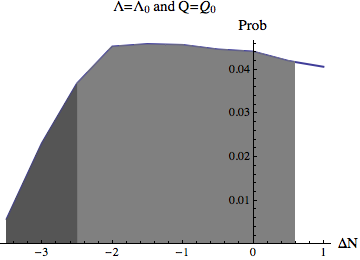} \\
 \includegraphics[width=1.8in]{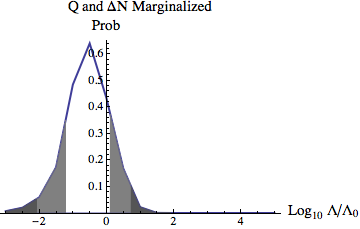} &
 \includegraphics[width=1.8in]{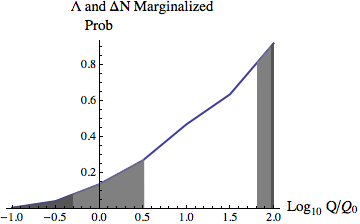} &
 \includegraphics[width=1.7in]{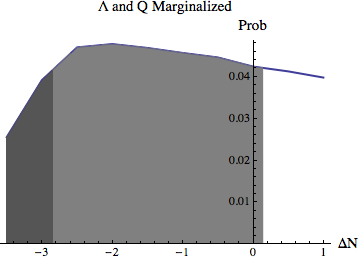} \\
 \end{array}$
 $\begin{array}{c@{\hspace{.1in}}c@{\hspace{.1in}}c}
 \includegraphics[width=1.7in]{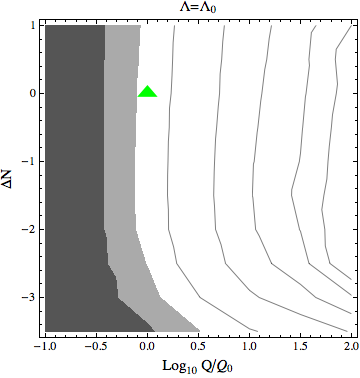} &
 \includegraphics[width=1.7in]{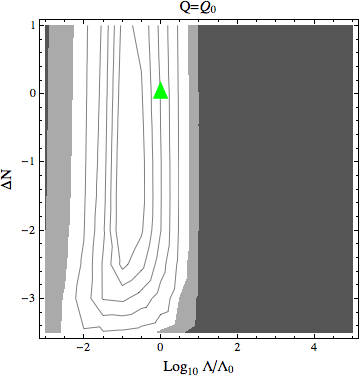} &
 \includegraphics[width=1.8in]{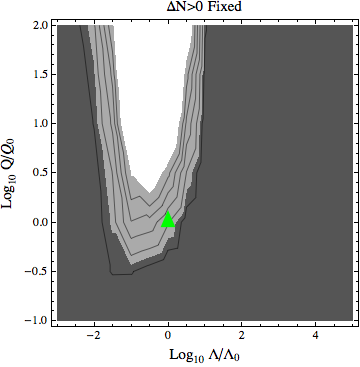} \\
 \includegraphics[width=1.7in]{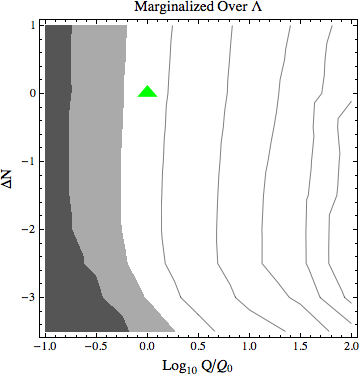} &
 \includegraphics[width=1.7in]{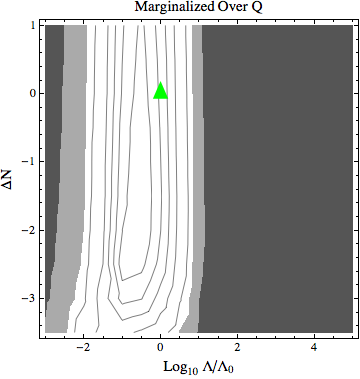} &
 \includegraphics[width=1.8in]{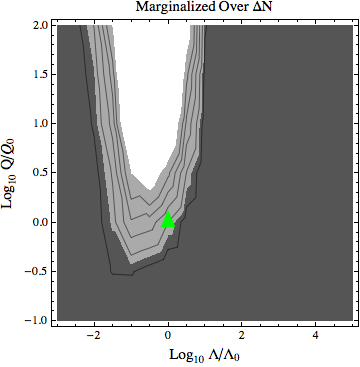} \\
 \end{array}$
 \caption{$\Lambda>0$, Causal Diamond, 10 Gyr delay time.  As with the
   5 Gyr delay, we find a monotonic distribution over $Q$ (2nd and 5th
   plot).  The preferred value of $\Lambda$ is independent of $Q$, and
   is somewhat smaller than with the 5 Gyr delay.}
 \label{posDia10Gyr}
 \end{center}
 \end{figure}

 \newpage

 \begin{figure}[h!]
 \begin{center}
 $\begin{array}{c@{\hspace{.1in}}c@{\hspace{.1in}}c}
 \includegraphics[width=1.8in]{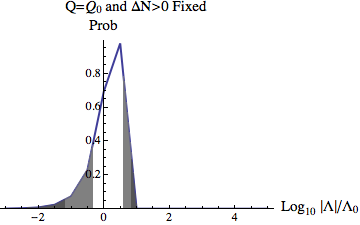} &
 \includegraphics[width=1.8in]{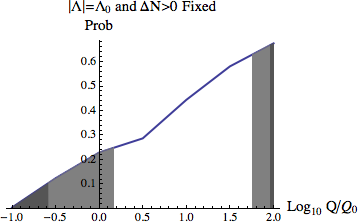} &
 \includegraphics[width=1.7in]{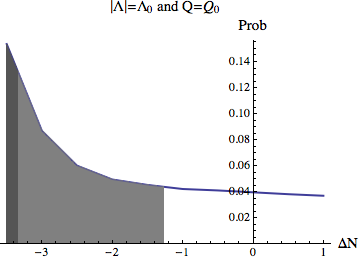} \\
 \includegraphics[width=1.8in]{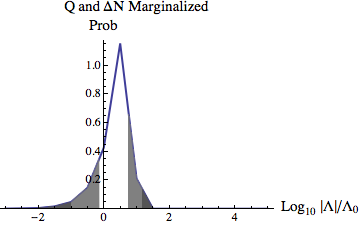} &
 \includegraphics[width=1.8in]{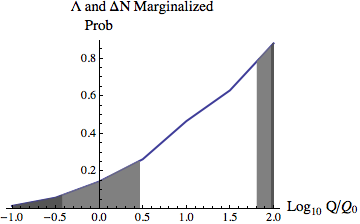} &
 \includegraphics[width=1.7in]{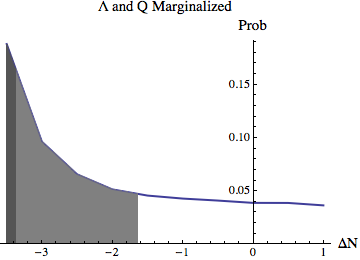} \\
 \end{array}$
 $\begin{array}{c@{\hspace{.1in}}c@{\hspace{.1in}}c}
 \includegraphics[width=1.7in]{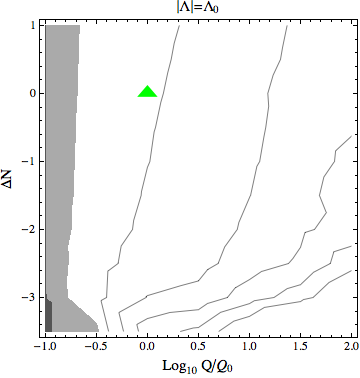} &
 \includegraphics[width=1.7in]{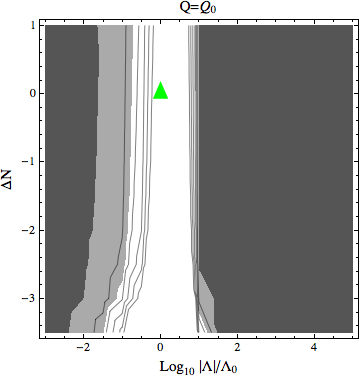} &
 \includegraphics[width=1.8in]{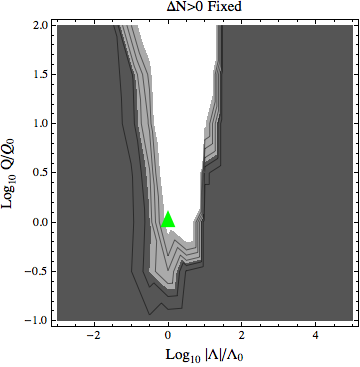} \\
 \includegraphics[width=1.7in]{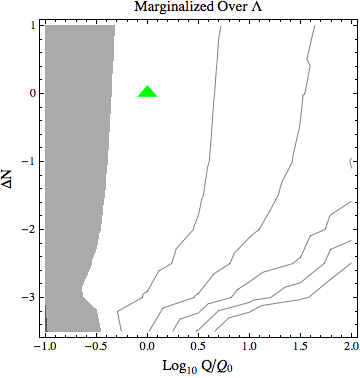} &
 \includegraphics[width=1.7in]{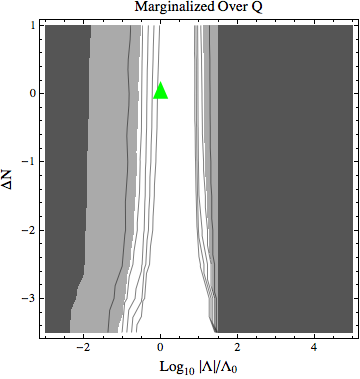} &
 \includegraphics[width=1.8in]{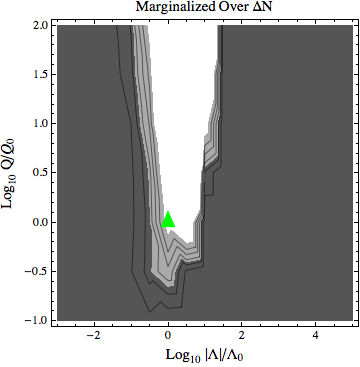} \\
 \end{array}$
 \caption{$\Lambda<0$, Causal Diamond, 10 Gyr delay time.  The 3rd
   and 6th plots show an increase in probability toward small
   $\Delta N$.  This is not a runaway; see the caption of
   Fig.~\protect\ref{negDia}.}
 \label{negDia10Gyr}
 \end{center}
 \end{figure}

 \newpage

 \begin{figure}[h!]
 \begin{center}
 $\begin{array}{c@{\hspace{.1in}}c@{\hspace{.1in}}c}
 \includegraphics[width=1.8in]{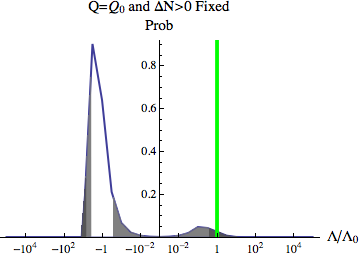} & & \\
 \includegraphics[width=1.8in]{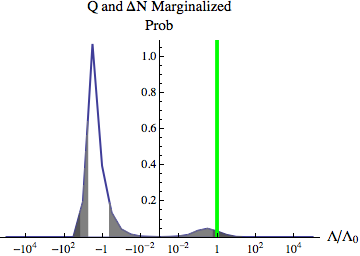} &
 \includegraphics[width=1.8in]{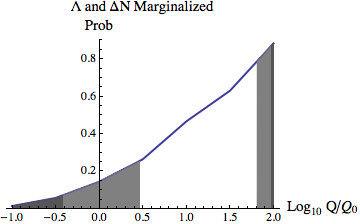} &
 \includegraphics[width=1.8in]{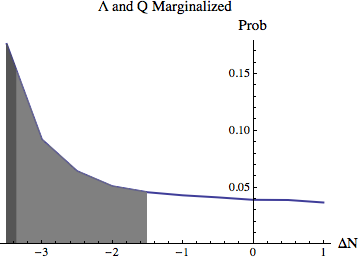} \\
 \end{array}$
 $\begin{array}{c@{\hspace{.1in}}c@{\hspace{.1in}}c}
 &
 \includegraphics[width=1.8in]{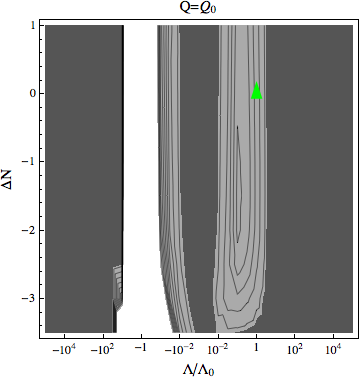} &
 \includegraphics[width=1.8in]{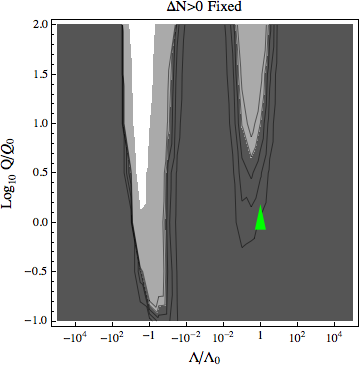} \\
 \includegraphics[width=1.8in]{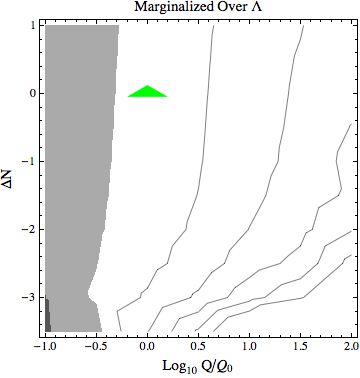} &
 \includegraphics[width=1.8in]{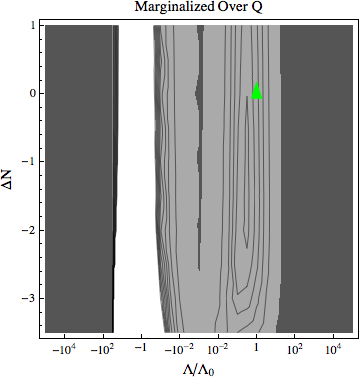} &
 \includegraphics[width=1.8in]{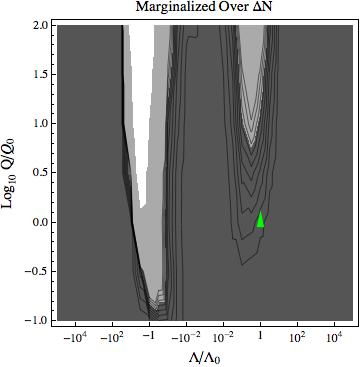} \\
 \end{array}$
 \caption{All values of $\Lambda$, Causal Diamond, 10 Gyr delay time.
   As usual, $\Lambda<0$ is more likely than $\Lambda>0$.  In the 9th
   and 12th plot, the observed values of $(\Lambda,Q)$ lie outside
   $2\sigma$ but within $3\sigma$.}
 \label{posnegDia10Gyr}
 \end{center}
 \end{figure}

\newpage

\subsection{Weighting by entropy production in the causal patch}
\label{sec-entpatch}

\begin{figure}[h!]
 \begin{center}
 $\begin{array}{c@{\hspace{.1in}}c@{\hspace{.1in}}c}
 \includegraphics[width=1.8in]{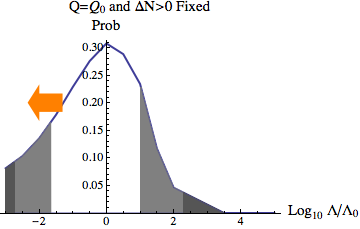} &
 \includegraphics[width=1.8in]{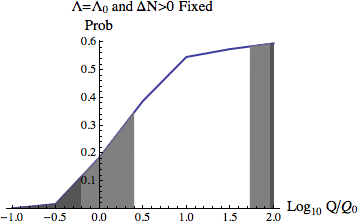} &
 \includegraphics[width=1.7in]{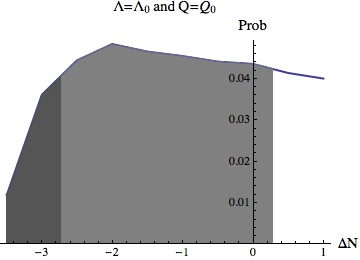} \\
 \includegraphics[width=1.8in]{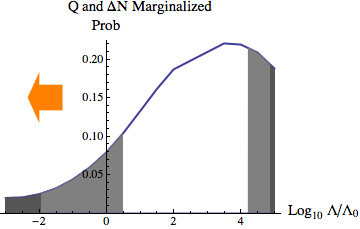} &
 \includegraphics[width=1.8in]{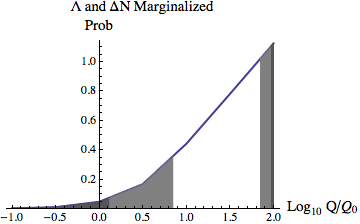} &
 \includegraphics[width=1.7in]{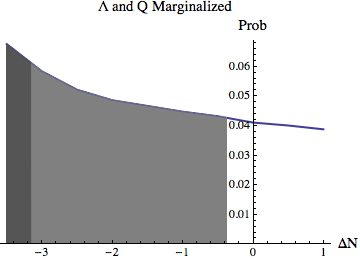} \\
 \end{array}$
 $\begin{array}{c@{\hspace{.1in}}c@{\hspace{.1in}}c}
 \includegraphics[width=1.7in]{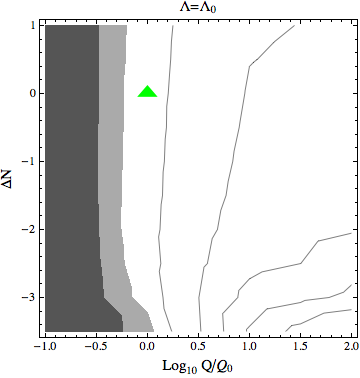} &
 \includegraphics[width=1.7in]{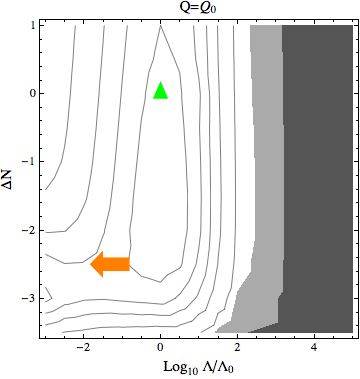} &
 \includegraphics[width=1.8in]{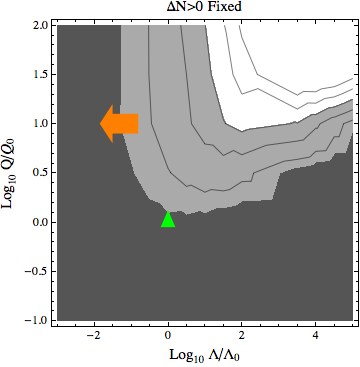} \\
 \includegraphics[width=1.7in]{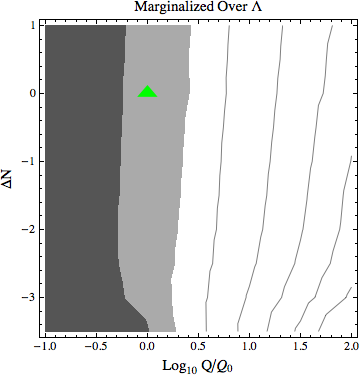} &
 \includegraphics[width=1.7in]{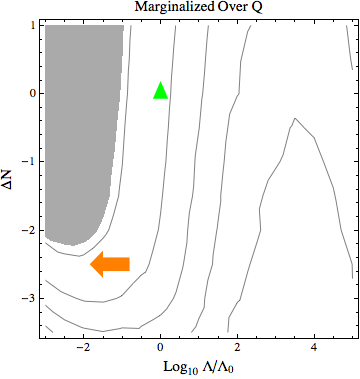} &
 \includegraphics[width=1.8in]{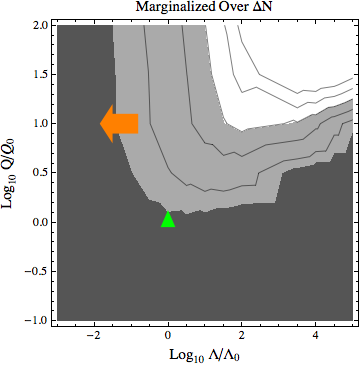} \\
 \end{array}$
 \caption{$\Lambda>0$, Causal Patch, Entropy Production.  The
   probability density goes to a constant for $\Lambda\ll t_{\rm
     c}^{-2}$.  This runaway is indicated by arrows.  With $\Delta N>
   0$, curvature dominates so late that the runaway is not evident
   even at the smallest values of $\Lambda$ in the first plot.  In the
   8th and 11th plot, however, the distribution can be seen to flatten
   out towards small $\Lambda$ at $\Delta N=-3$.}
 \label{posPatch}
 \end{center}
 \end{figure}

\begin{figure}[h!]
 \begin{center}
 $\begin{array}{c@{\hspace{.1in}}c@{\hspace{.1in}}c}
 \includegraphics[width=1.8in]{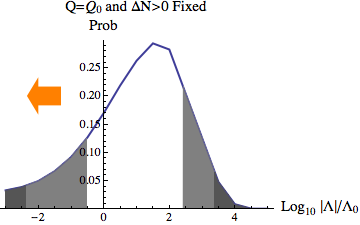} &
 \includegraphics[width=1.8in]{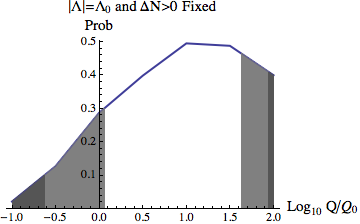} &
 \includegraphics[width=1.7in]{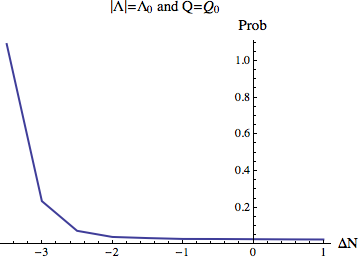} \\
 \includegraphics[width=1.8in]{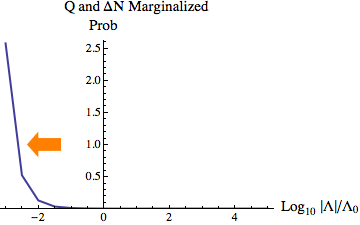} &
 \includegraphics[width=1.8in]{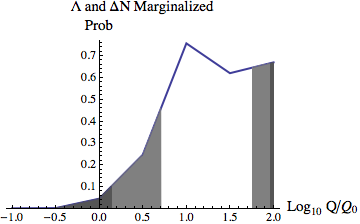} &
 \includegraphics[width=1.7in]{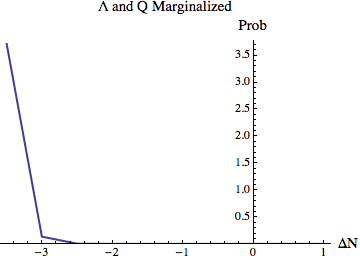} \\
 \end{array}$
 $\begin{array}{c@{\hspace{.1in}}c@{\hspace{.1in}}c}
 \includegraphics[width=1.7in]{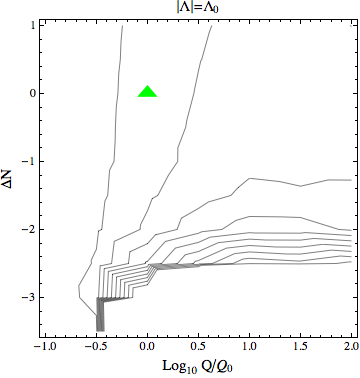} &
 \includegraphics[width=1.7in]{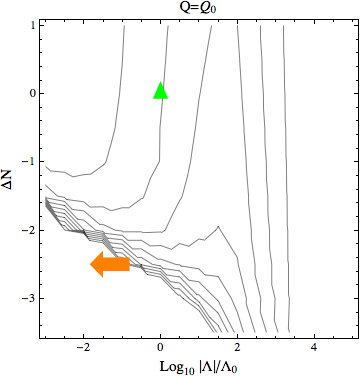} &
 \includegraphics[width=1.8in]{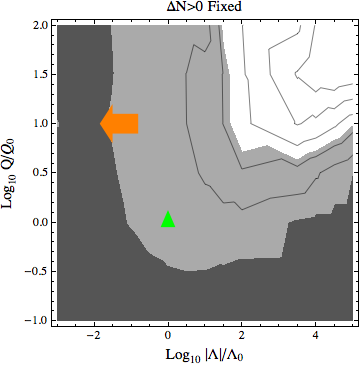} \\
 \includegraphics[width=1.7in]{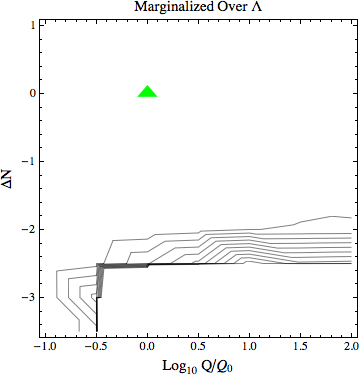} &
 \includegraphics[width=1.7in]{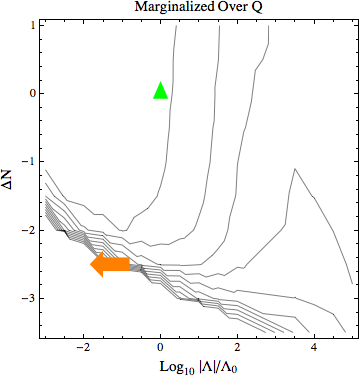} &
 \includegraphics[width=1.8in]{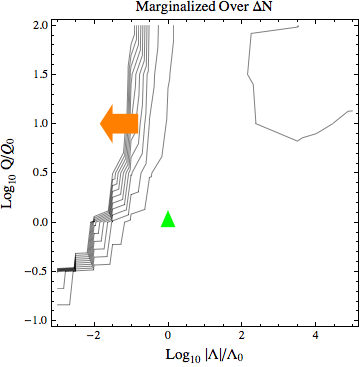} \\
 \end{array}$
 \caption{$\Lambda<0$, Causal Patch, Entropy Production.  The
   probability density grows like $|\Lambda|^{-1}$ for $\Lambda\ll
   t_{\rm c}^{-2}$.  This runaway is indicated by arrows.  At fixed
   $\Lambda$, small $\Delta N$ is strongly preferred (3rd and 7th
   plots).}
 \label{negPatch}
 \end{center}
 \end{figure}

\begin{figure}[h!]
\begin{center}
 $\begin{array}{c@{\hspace{.1in}}c@{\hspace{.1in}}c}
 \includegraphics[width=1.8in]{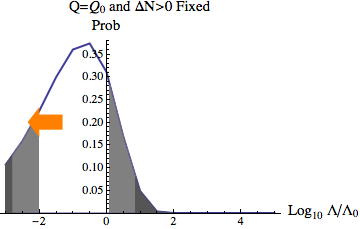} &
 \includegraphics[width=1.8in]{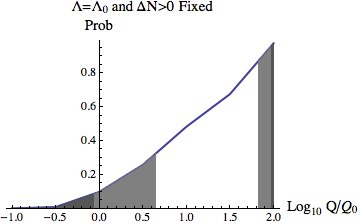} &
 \includegraphics[width=1.7in]{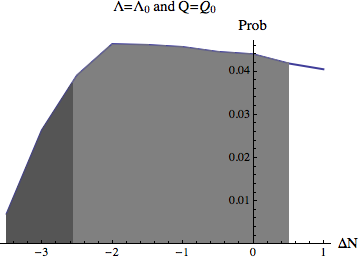} \\
 \includegraphics[width=1.8in]{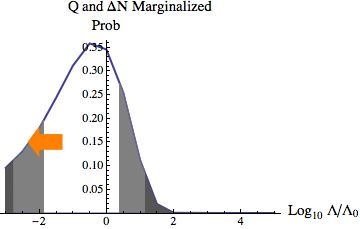} &
 \includegraphics[width=1.8in]{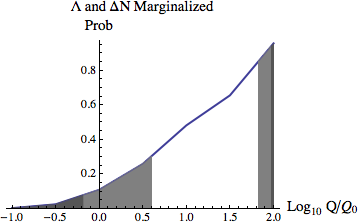} &
 \includegraphics[width=1.7in]{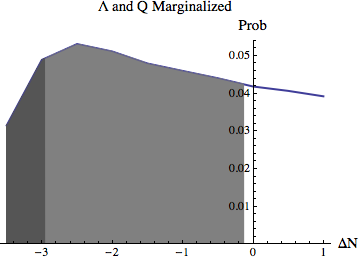} \\
 \end{array}$
 $\begin{array}{c@{\hspace{.1in}}c@{\hspace{.1in}}c}
 \includegraphics[width=1.7in]{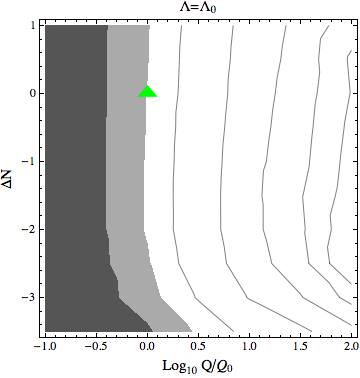} &
 \includegraphics[width=1.7in]{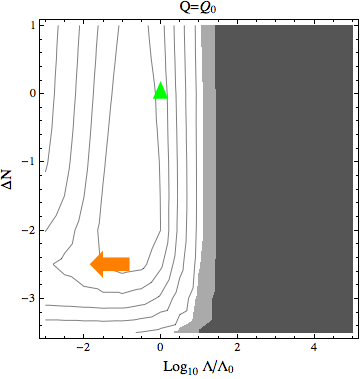} &
 \includegraphics[width=1.8in]{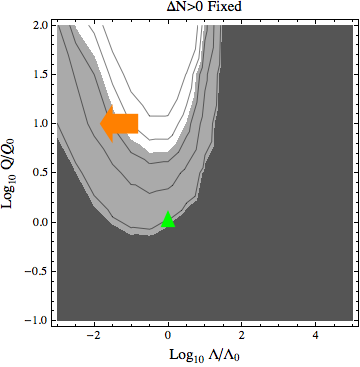} \\
 \includegraphics[width=1.7in]{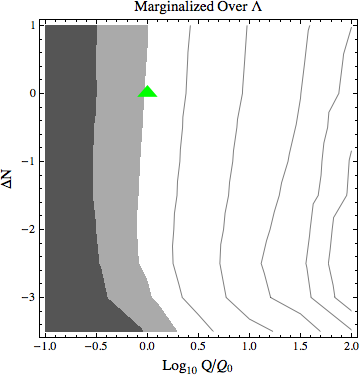} &
 \includegraphics[width=1.7in]{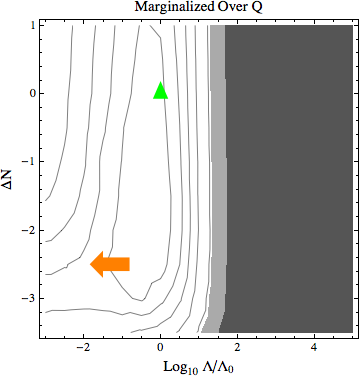} &
 \includegraphics[width=1.8in]{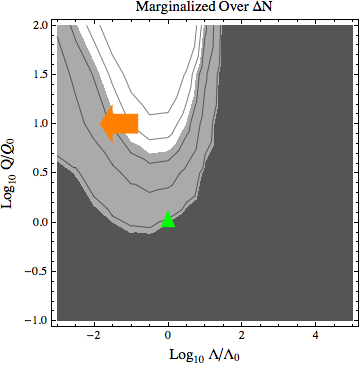} \\
 \end{array}$
 \caption{$\Lambda>0$, Causal Patch, 5 Gyr delay time.  In the absence
   of curvature, the preferred value of $\Lambda$ is independent of
   $Q$, as seen in the 9th plot.  For other comments, see the caption
   of Fig.~\protect\ref{posPatch}.}
 \label{posPatch5Gyr}
 \end{center}
 \end{figure}

\begin{figure}[h!]
 \begin{center}
 $\begin{array}{c@{\hspace{.1in}}c@{\hspace{.1in}}c}
 \includegraphics[width=1.8in]{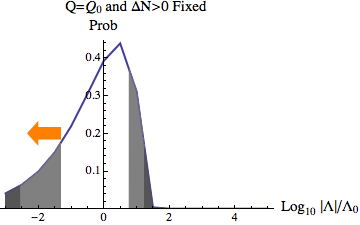} &
 \includegraphics[width=1.8in]{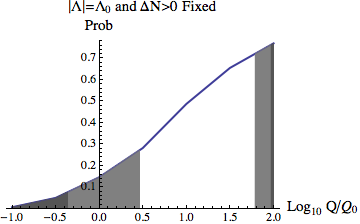} &
 \includegraphics[width=1.7in]{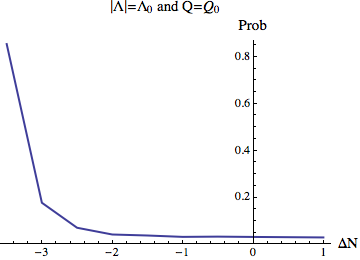} \\
 \includegraphics[width=1.8in]{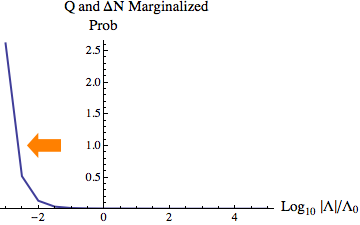} &
 \includegraphics[width=1.8in]{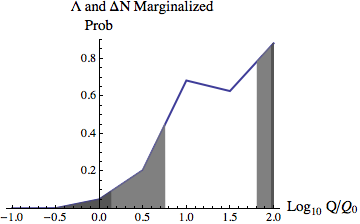} &
 \includegraphics[width=1.7in]{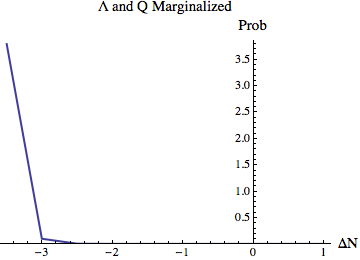} \\
 \end{array}$
 $\begin{array}{c@{\hspace{.1in}}c@{\hspace{.1in}}c}
 \includegraphics[width=1.7in]{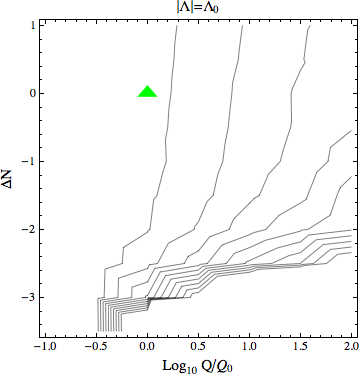} &
 \includegraphics[width=1.7in]{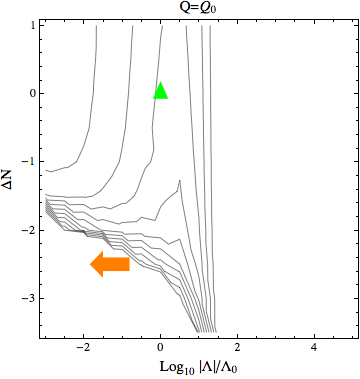} &
 \includegraphics[width=1.8in]{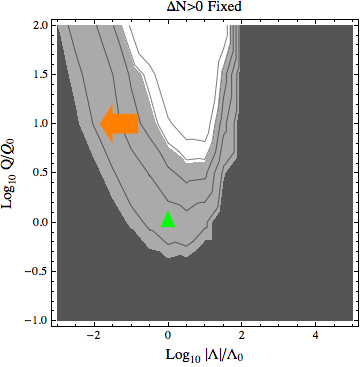} \\
 \includegraphics[width=1.7in]{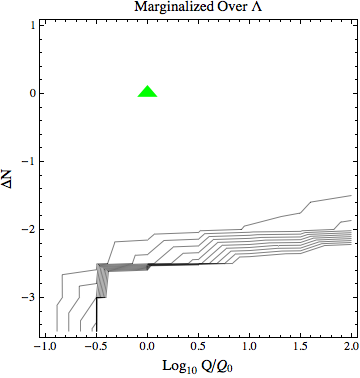} &
 \includegraphics[width=1.7in]{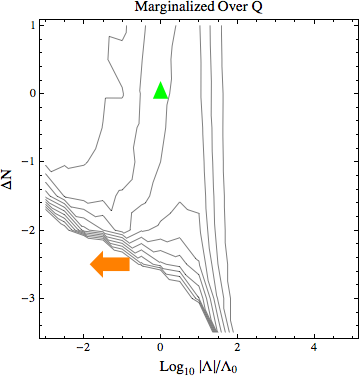} &
 \includegraphics[width=1.8in]{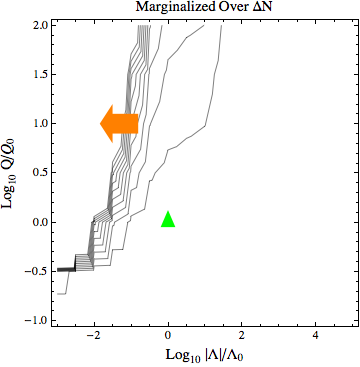} \\
 \end{array}$
 \caption{$\Lambda<0$, Causal Patch, 5 Gyr delay time.  In the absence
   of curvature, the preferred value of $\Lambda$ is independent of
   $Q$, as seen in the 9th plot.  For other comments, see the caption
   of Fig.~\protect\ref{negPatch}.}
 \label{negPatch5Gyr}
 \end{center}
 \end{figure}

We now turn to the causal patch measure.  Independently of the
observer model, for $\Lambda<0$, the rapid divergence as $|\Lambda|\to
0$ (Sec.~\ref{sec-lambdacurvonly}) prevents us from displaying
confidence regions in the affected plots in Figs.~\ref{negPatch},
\ref{negPatch5Gyr}, and~\ref{negPatch10Gyr}.  For $\Lambda>0$, the
probability density over $\log_{10}(|\Lambda|/\Lambda_0)$ goes to a
constant for $\Lambda\ll t_{\rm c}^{-2}$.  This runaway is mild enough
that we display confidence regions (based, as always, on an assumed
cut-off at the end of our parameter range) in the relevant plots in
Figs.~\ref{posPatch}, \ref{posPatch5Gyr}, and~\ref{posPatch10Gyr}.
The runaway is not always readily apparent but we have indicated all
runaway directions with arrows.  For $\Lambda<0$, the growth towards
small $\Delta N$ (also described in Sec.~\ref{sec-lambdacurvonly}) is
so rapid at $\Delta N=-3.5$ (the lower end of our parameter space)
that confidence regions are very sensitive to the exact position of
this cut-off and are not displayed.

In this subsection, we begin with the case where observers are
modelled by entropy production.  For $\Lambda>0$, we find that the
observed values are at or within $2\sigma$, except in those
distributions which are susceptible to the above runaway towards small
$\Lambda$.  For $\Lambda<0$, the only acceptable results obtain when
curvature is fixed and negligible.  Even in the absence of curvature,
with $\Delta N$ and $Q$ fixed, negative values of the cosmological
constant are more probable than positive values, by a factor of order
10.  (As explained in Sec.~\ref{sec-lambdaonly}, the distribution in
$\Lambda$ is wider than in the causal diamond case and so has
significant support at the boundary of our parameter space.  Thus,
computing a more precise value of the relative probability for
different signs of $\Lambda$, from our data alone, would not be very
informative.)

Unlike the case of the causal diamond with the same observer model,
the distribution in $Q$ is monotonically increasing for
$\Lambda=\Lambda_0$ and $\Delta N>0$.  This difference is
explained in Sec.~\ref{sec-qonly}.  Like in the case of the causal
diamond, the preferred value of $\Lambda$ grows with $Q$ (see
Sec.~\ref{sec-qlambdaonly}).

\newpage

\newpage

\newpage

 \begin{figure}[h!]
 \begin{center}
 $\begin{array}{c@{\hspace{.1in}}c@{\hspace{.1in}}c}
 \includegraphics[width=1.8in]{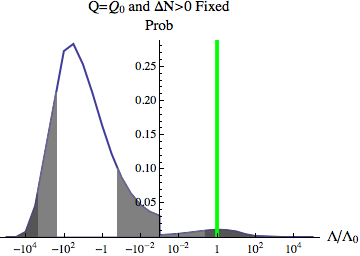} & & \\
\hspace{1.8in}&
 \includegraphics[width=1.8in]{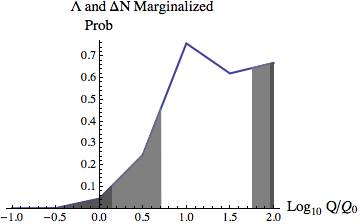} &
\hspace{1.8in}\\
 \end{array}$
 $\begin{array}{c@{\hspace{.1in}}c@{\hspace{.1in}}c}
 \hspace{1.8in}&
 \hspace{1.8in}&
 \includegraphics[width=1.8in]{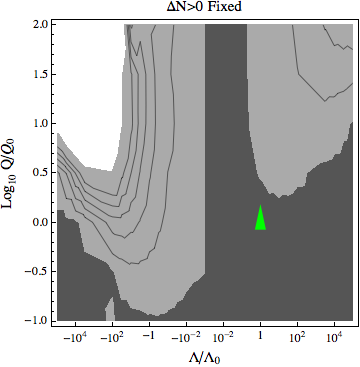} \\
\hspace{1.8in}&
 \hspace{1.8in}&
 \hspace{1.8 in} \\
 \end{array}$
 \caption{All values of $\Lambda$, Causal Patch, Entropy Production.
   In the presence of curvature, $\Lambda<0$ is preferred over
   $\Lambda>0$ by an arbitrarily large amount, depending on the
   cut-off in $\Lambda$.  In the 5th plot, the observed value of
   $(\Lambda, Q)$ lies outside $2\sigma$, but within $3\sigma$.}
 \label{posnegPatch}
 \end{center}
 \end{figure}

\newpage

\subsection{Weighting by star formation + 5 Gyr in the causal patch}

This case is very similar to the previous subsection, and we refer the
reader to its text and captions.  \newpage

 \newpage

 \newpage

 \begin{figure}[h!]
 \begin{center}
 $\begin{array}{c@{\hspace{.1in}}c@{\hspace{.1in}}c}
 \includegraphics[width=1.8in]{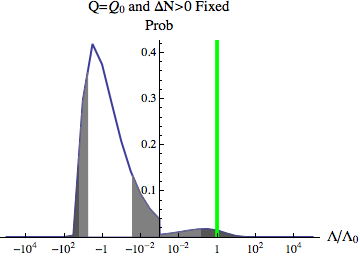} &
 &
  \\
\hspace{1.8in} &
 \includegraphics[width=1.8in]{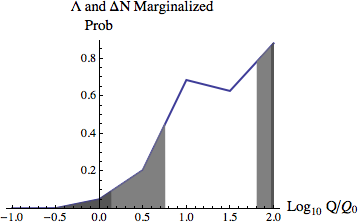} &
\hspace{1.8in} \\
 \end{array}$
 $\begin{array}{c@{\hspace{.1in}}c@{\hspace{.1in}}c}
 \hspace{1.8in}&
 \hspace{1.8in}&
 \includegraphics[width=1.8in]{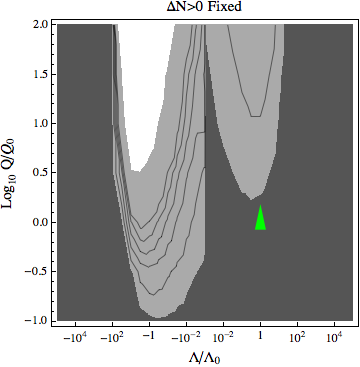} \\
\hspace{1.8in}&
 \hspace{1.8in}&
 \hspace{1.8in}\\
 \end{array}$
 \caption{All Values of $\Lambda$, Causal Patch, 5 Gyr delay time.
   Negative $\Lambda$ is preferred here by a factor of order $10$ when
   curvature is absent.  In the 5th plot, the data point is within
   $3\sigma$.  For other comments, see the caption
   of Fig.~\protect\ref{posnegPatch}.}
 \label{posnegPatch5Gyr}
 \end{center}
 \end{figure}

\newpage

\subsection{Weighting by star formation + 10 Gyr in the causal patch}

This case is very similar to the previous subsection, and we refer the
reader to its text and captions. 

\newpage

 \begin{figure}[h!]
 \begin{center}
 $\begin{array}{c@{\hspace{.1in}}c@{\hspace{.1in}}c}
 \includegraphics[width=1.8in]{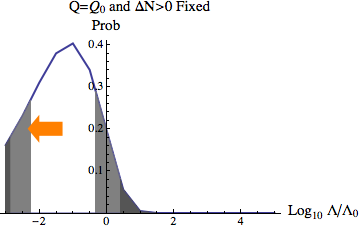} &
 \includegraphics[width=1.8in]{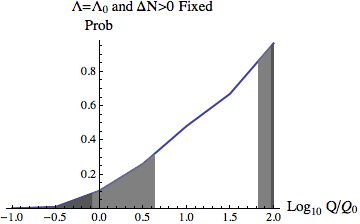} &
 \includegraphics[width=1.7in]{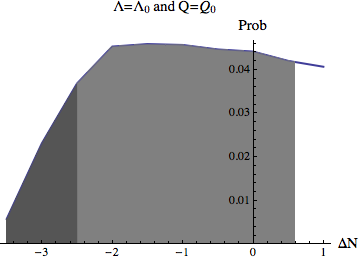} \\
 \includegraphics[width=1.8in]{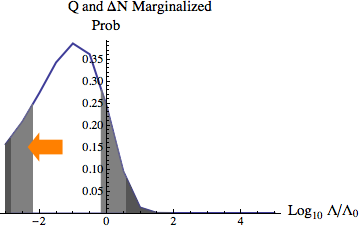} &
 \includegraphics[width=1.8in]{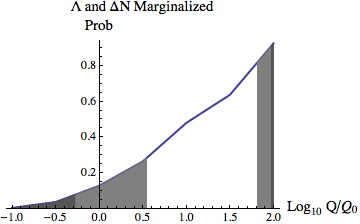} &
 \includegraphics[width=1.7in]{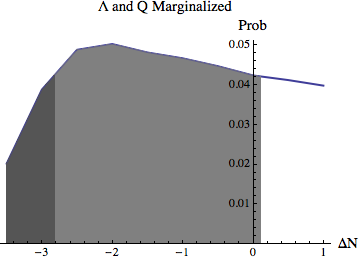} \\
 \end{array}$
 $\begin{array}{c@{\hspace{.1in}}c@{\hspace{.1in}}c}
 \includegraphics[width=1.7in]{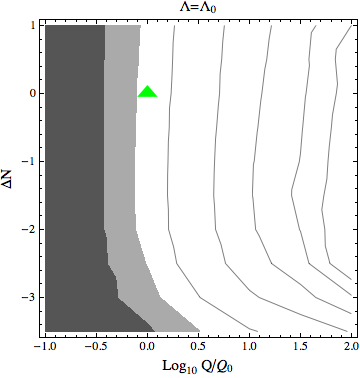} &
 \includegraphics[width=1.7in]{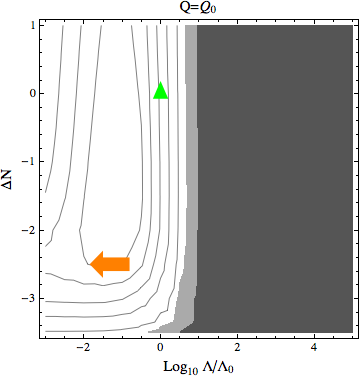} &
 \includegraphics[width=1.8in]{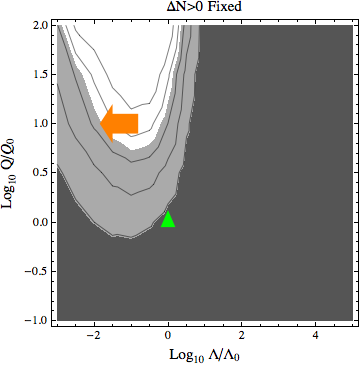} \\
 \includegraphics[width=1.7in]{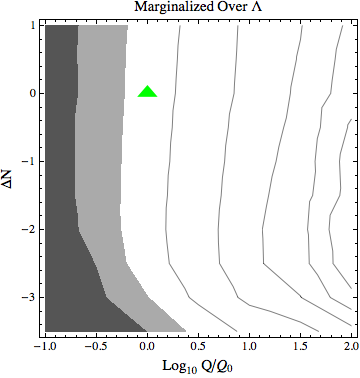} &
 \includegraphics[width=1.7in]{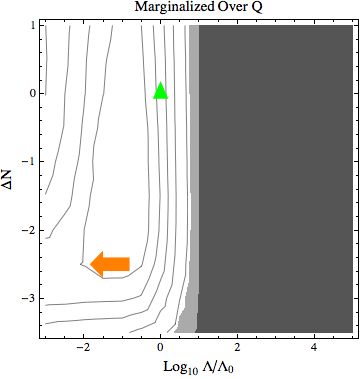} &
 \includegraphics[width=1.8in]{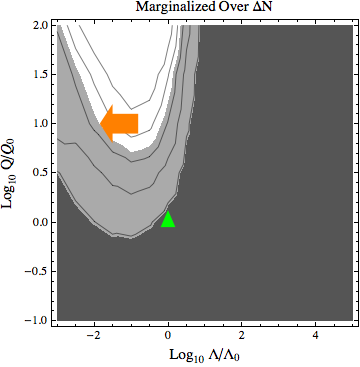} \\
 \end{array}$
 \caption{$\Lambda>0$, Causal Patch, 10 Gyr delay time.  See caption of Fig.~\protect\ref{posPatch5Gyr}.}
 \label{posPatch10Gyr}
 \end{center}
 \end{figure}

 \newpage

 \begin{figure}[h!]
 \begin{center}
 $\begin{array}{c@{\hspace{.1in}}c@{\hspace{.1in}}c}
 \includegraphics[width=1.8in]{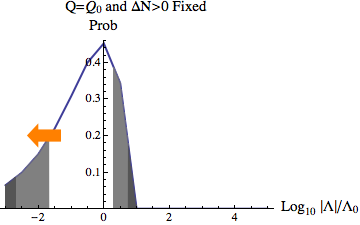} &
 \includegraphics[width=1.8in]{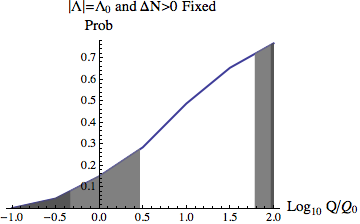} &
 \includegraphics[width=1.7in]{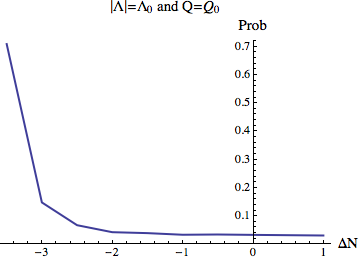} \\
 \includegraphics[width=1.8in]{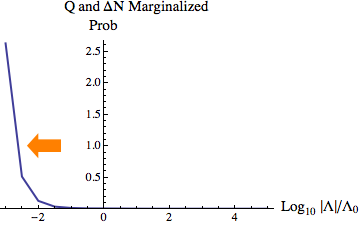} &
 \includegraphics[width=1.8in]{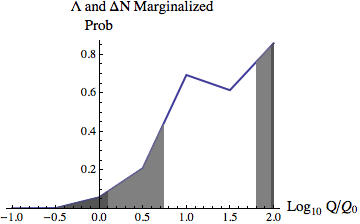} &
 \includegraphics[width=1.7in]{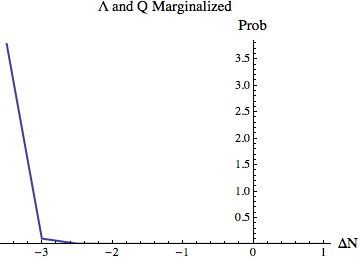} \\
 \end{array}$
 $\begin{array}{c@{\hspace{.1in}}c@{\hspace{.1in}}c}
 \includegraphics[width=1.7in]{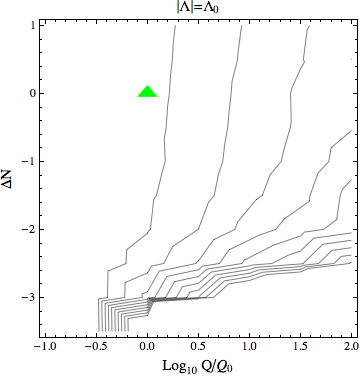} &
 \includegraphics[width=1.7in]{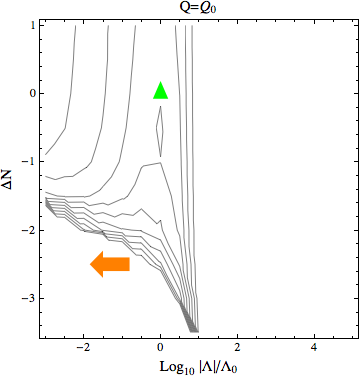} &
 \includegraphics[width=1.8in]{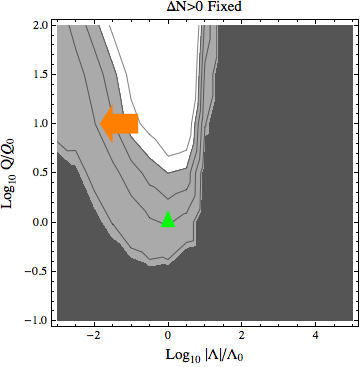} \\
 \includegraphics[width=1.7in]{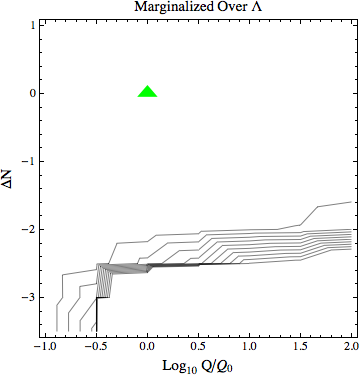} &
 \includegraphics[width=1.7in]{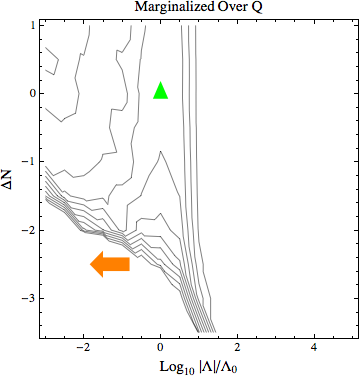} &
 \includegraphics[width=1.8in]{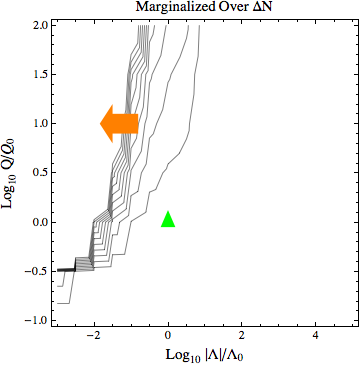} \\
 \end{array}$
 \caption{$\Lambda<0$, Causal Patch, 10 Gyr delay time.  See caption
   of Fig.~\protect\ref{negPatch5Gyr}. }
 \label{negPatch10Gyr}
 \end{center}
 \end{figure}

 \newpage

 \begin{figure}[h!]
 \begin{center}
 $\begin{array}{c@{\hspace{.1in}}c@{\hspace{.1in}}c}
 \includegraphics[width=1.8in]{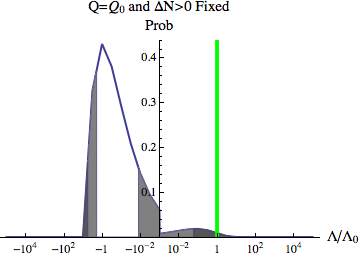} &
  &
  \\
\hspace{1.8in}&
 \includegraphics[width=1.8in]{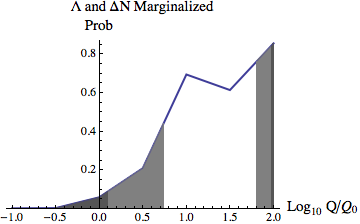} &
\hspace{1.8in} \\
 \end{array}$
 $\begin{array}{c@{\hspace{.1in}}c@{\hspace{.1in}}c}
 \hspace{1.8in} &
 \hspace{1.8 in}&
 \includegraphics[width=1.8in]{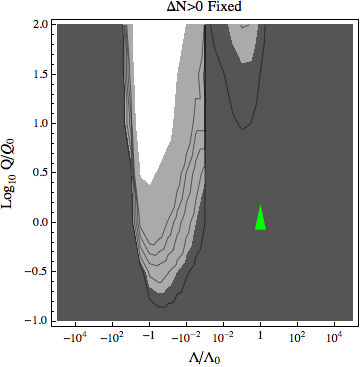} \\
\hspace{1.8in}&
 \hspace{1.8in}&
 \hspace{1.8in} \\
 \end{array}$
 \caption{All Values of $\Lambda$, Causal Patch, 10 Gyr delay time.
   See caption of Fig.~\protect\ref{posnegPatch5Gyr}.}
 \label{posnegPatch10Gyr}
 \end{center}
 \end{figure}

 \newpage

\section{Discussion}
\label{sec-discussion}

The main purpose of this section is to provide an intuitive
qualitative understanding of the most important features in the plots
of Sec.~\ref{sec-results}.  We will also supply some additional
quantitative results that are not immediately apparent in the plots.
(In particular, we will compute the probability that nonvanishing
spatial curvature will be detected in the near future, at the end of
subsection \ref{sec-curvonly}.)

What are the physical consequences of varying the parameters
$(\Lambda$, $Q$, $\Delta N)$?  Varying $\Lambda$ or $\Delta N$ changes
both the dynamical evolution, and the geometry of the cut-off region.
Dynamically, these parameters affect the rates of star formation and
entropy production.  Geometrically, they affect the amount of comoving
volume contained within the causal patch or the causal diamond.  In
several probability distributions, the geometric effect is
quantitatively more important than the dynamical effect.  

The parameter $Q$, on the other hand, enters the probability
distribution only dynamically, through its effects on the rate at
which observers form.  In an approximately homogeneous universe, the
initial strength of density perturbations does not have an important
effect on the comoving volume within the patch or diamond.
(Ref.~\cite{PhiAlb09} discusses possible effects of the breakdown of
the homogeneous approximation, which are not modeled here.)

\subsection{Varying the cosmological constant only}
\label{sec-lambdaonly}

Let us begin by discussing the probability distribution over the
cosmological constant alone, with both remaining parameters fixed to
their observed values. Its effects on star formation and on the
geometry of the causal patch or diamond depend strongly on the sign of
the cosmological constant. We begin by discussing the case
$\Lambda>0$; our treatment follows Ref.~\cite{BouHar07}.

With $\Lambda>0$, structure formation, and thus star formation, halts
after a time of order
\begin{equation}
t_\Lambda\equiv (3/\Lambda)^{1/2}~.
\end{equation}
However, this effect becomes important only for relatively large
values of $\Lambda$ of order $100 \Lambda_0$. As we shall now see,
these values are already suppressed by geometric effects. We will
self-consistently treat the rate of observation, $\dot n_{\rm
  obs}(t)$, as fixed. (We emphasize that this approximation, like all
others in this section, is made for the sole purpose of elucidating
our plots, which are always obtained using a full numerical
calculation of both the geometry and the observation rate.)

For $\Lambda>0$, a period of matter domination is succeeded, around
the time $t_\Lambda$, by an infinite period of vacuum domination. In
the sudden transition approximation, the scale factor is given by
\begin{equation} a(t) \propto \begin{cases}
    t^{2/3} & \text{if $t<\frac{2}{3}t_\Lambda$}~, \\
    \left(\frac{2}{3}t_\Lambda\right)^{2/3}\exp(t/t_\Lambda-2/3) &
    \text{if $t\geq \frac{2}{3}t_\Lambda$}~.
\end{cases} 
\end{equation}

Matching both the scale factor and its derivative requires cumbersome
shift terms, or order-one corrections to the matching time, like the
$2/3$ appearing in the above equation.  Our goal is only to understand
the rough features of our plots, not to keep track of all factors of
order one.  For this purpose, it suffices to match the value of the
scale factor but not its derivative at the transition time.  We will
do so in all formulas below.  For the present case, the simplified
version of the scale factor is
\begin{equation} a(t) \propto \begin{cases}
    t^{2/3} & \text{if $t<t_\Lambda$}~, \\
     t_\Lambda^{2/3}\exp(t/t_\Lambda-1) & \text{if $t\geq t_\Lambda$}~.
  \end{cases} \end{equation}
By Eq.~(\ref{eq-patch}), the comoving radius of the causal patch is
given by
%\begin{equation}
%\chi_{\rm patch}(t) \propto \begin{cases}
%    \left(\frac{3}{2}\right)^{2/3}t_\Lambda^{1/3}+3\left(\frac{2}{3}t_\Lambda\right)^{1/3} - 3 t^{1/3} & \text{if $t<\frac{2}{3}t_\Lambda$}~, \\
 %    \left(\frac{3}{2}\right)^{2/3}t_{\Lambda}^{1/3}\exp(-t/t_\Lambda+2/3)  & \text{if $t\geq \frac{2}{3}t_\Lambda$}~.
  %  \end{cases}
%\end{equation}
\begin{equation}
\chi_{\rm patch}(t) \propto \begin{cases}
    4t_{\Lambda}^{1/3} - 3 t^{1/3} & \text{if $t<t_\Lambda$}~, \\
    t_{\Lambda}^{1/3}\exp(-t/t_\Lambda+1)  & \text{if $t\geq t_\Lambda$}~.
    \end{cases}
\end{equation}
By Eq.~(\ref{eq-diamond}), the comoving radius of the causal diamond
is given by
%\begin{equation}
%\chi_{\rm dia}(t) \propto \begin{cases}
%3t^{1/3} & \text{if $t\ll t_\Lambda$}~,\\
%\left(\frac{3}{2}\right)^{2/3}t_{\Lambda}^{1/3}\exp(-t/t_\Lambda+2/3)& \text{if $t\gg t_\Lambda$}~.
%\end{cases}
%\end{equation}
\begin{equation}
\chi_{\rm dia}(t) \propto \begin{cases}
3t^{1/3} & \text{if $t\ll t_\Lambda$}~,\\
t_{\Lambda}^{1/3}\exp(-t/t_\Lambda+1) & \text{if $t\gg t_\Lambda$}~.
\end{cases}
\label{eq-lambdadia}
\end{equation}
The ``edge'' of the causal diamond, where the future light-cone from
the reheating surface meets intersects the boundary of the causal
patch, occurs at the time $0.23 t_\Lambda$~\cite{BouHar07}.  Since
this is approximately the same time at which the scale factor changes
from power law to exponential growth, there is no need for another
case distinction in Eq.~(\ref{eq-lambdadia}) at this level of
approximation.

Since in this section we are assuming negligible spatial curvature,
the comoving volume for the patch is
%\begin{equation}
%V_{\rm c}^{\rm patch}(t) \propto \begin{cases}
    % \frac{4\pi}{3}\left(\left(\frac{3}{2}\right)^{2/3}t_\Lambda^{1/3}+3\left(\frac{2}{3}t_\Lambda\right)^{1/3} - 3 t^{1/3}\right)^3 & \text{if $t<\frac{2}{3}t_\Lambda$}~, \\
 %  3\pi t_{\Lambda}\exp(-3t/t_\Lambda+2)  & \text{if $t\geq \frac{2}{3}t_\Lambda$}~.
    %\end{cases}
%\end{equation}
\begin{equation}
  V_{\rm c}^{\rm patch}(t) \propto \begin{cases}
    \frac{4\pi}{3}(4 t_{\Lambda}^{1/3} - 3 t^{1/3})^3 & \text{if $t<t_\Lambda$}~, \\
    \frac{4\pi}{3} t_{\Lambda}\exp(-3t/t_\Lambda+3)  & \text{if $t\geq t_\Lambda$}~.
    \end{cases}
    \label{eq-patchlambdavol}
\end{equation}
while that for the diamond is
%\begin{equation}
%V_{\rm c}^{\rm dia}(t) \propto \begin{cases}
%36\pi t & \text{if $t\ll t_\Lambda$}~,\\
%3\pi t_{\Lambda}\exp(-3t/t_\Lambda+2) & \text{if $t\gg t_\Lambda$}~.
%\end{cases}
%\end{equation}
\begin{equation}
V_{\rm c}^{\rm dia}(t) \propto \begin{cases}
36\pi t & \text{if $t\ll t_\Lambda$}~,\\
\frac{4\pi}{3}t_{\Lambda}\exp(-3t/t_\Lambda+3) & \text{if $t\gg t_\Lambda$}~.
\end{cases}
\label{eq-dialambdavol}
\end{equation}
Now we are in a position to derive a probability distribution for
$\Lambda$ by counting observers:
\begin{equation}
  \frac{dp}{d\log_{10} \Lambda} \propto \frac{d\tilde p}{d\log_{10} \Lambda}
  \times \int dt\, \dot{n}_{\rm obs}(t) V_{\rm c}(t)
\end{equation}
Recall from section~\ref{sec-prior} that we assume $d \tilde
p/d\log_{10}\Lambda \propto \Lambda$, and that here we are assuming
$\dot{n}_{\rm obs}(t)$ is independent of $\Lambda$.  

Beginning with the {\em causal diamond}, we see that
\begin{equation}
  \frac{dp}{d\log_{10} \Lambda} \propto \Lambda \left[ 36\pi \int_0^{t_\Lambda} dt\, \dot{n}_{\rm obs}(t) t + \frac{4\pi}{3}t_\Lambda \int_{t_\Lambda}^\infty dt\, \dot{n}_{\rm obs}(t) \exp(-3t/t_\Lambda+3)\right]~.
\label{eq-lambdadiaprob}
\end{equation}
If $\dot{n}_{\rm obs}(t)$ were constant in time, the bracketed terms
would be proportional to $t_\Lambda^2\propto \Lambda^{-1}$ and we
would obtain a distribution flat in $\log_{10}\Lambda$.  In reality,
$\dot{n}_{\rm obs}(t)$ has a peak at a time $t_{\rm peak}$ of order
Gyr with a width that is also of order Gyr.  This will serve to select
a preferred value of $\Lambda$ as follows.  If $t_\Lambda \ll t_{\rm
  peak}$, then only the second integral in Eq.~(\ref{eq-lambdadiaprob})
contributes, and this integral features an exponential suppression due
to the rapid emptying of the de Sitter horizon during
$\Lambda$-domination.  If $t_\Lambda \gg t_{\rm peak}$, then only the
first integral contributes, and its contribution will be independent
of $\Lambda$.  But there are more vacua with larger $\Lambda$ than
smaller $\Lambda$; this is encoded in the overall factor of $\Lambda$
coming from the prior, which tend to push the probability toward
larger values of $\Lambda$.  Thus we conclude that the most favorable
value of $\Lambda$ is one where $t_{\rm peak} \approx t_\Lambda$ (more
precisely we conclude that $t_{\rm peak}\approx t_{\rm edge}$, but
these are the same up to order-one factors).  Similarly, the width of
our distribution depends on the width of $\dot{n}_{\rm obs}(t)$: if
$t_{\rm on}$ and $t_{\rm off}$ are the times when $\dot{n}_{\rm
  obs}(t)$ is at half-maximum, then the corresponding values of
$\Lambda$ will give the approximate $1\sigma$ boundaries in the the
$\log_{10}\Lambda$ distribution~\cite{BouHar07}.

The same analysis holds for the {\em causal patch\/}, with one
modification.  In Eq.~(\ref{eq-patchlambdavol}), we see that for
$t<t_\Lambda$ the patch volume has some residual $\Lambda$ dependence.
So when $t_{\rm peak}\ll t_\Lambda$, the factor of $\Lambda$ from the
prior is partially cancelled by the factor of $t_\Lambda\sim
\Lambda^{-1/2}$ in the comoving volume.  The result is that the
probability distributions using the patch are more tolerant of small
values of $\Lambda$ than those using the diamond.

These estimates are confirmed by our plots for the probability
distribution over $\Lambda>0$, with $Q=Q_0$ and $\Delta N>0$ fixed
(the first plot in each figure).  Fig.~\ref{posDia} shows the result
for entropy production, where the most likely value of $\Lambda$ is
about $10\Lambda_0$.  This corresponds to $t_{\rm peak} \approx 2~{\rm
  Gyr}$, so we expect that in the $5~{\rm Gyr}$ delay time model of
Fig.~\ref{posDia5Gyr} the most likely value of $\Lambda$ would be
smaller by a factor of $(2/7)^2\approx.08$, and indeed we can see in
that plot that the most likely value is now slightly smaller than
$\Lambda_0$.  With a $10~{\rm Gyr}$ time delay, our estimate says that
most likely value of $\Lambda$ should be be $(2/12)^2\approx.03$ times
that for entropy production, and in Fig.~\ref{posDia10Gyr} we see that
the most likely value is down to nearly $\Lambda_0/10$.  Also, for the
patch, in Figs.~\ref{posPatch}, \ref{posPatch5Gyr}, and~\ref{posPatch10Gyr}
we see a braodening of the distributions compared to those of the
diamond.

The $\Lambda<0$ case has several important differences from the
$\Lambda>0$ case. Instead of halting structure formation, a collapsing
universe leads to a period of enhanced structure growth.  However, the
structures which grow during the collapse phase are much larger than
the structures which grow in the early universe. The largest of these
cannot cool and do not form stars.  As a result, enhanced structure
growth is actually a subdominant effect in explaining the difference
between the probability distributions.  Far more important is the
difference in geometry for a $\Lambda<0$ universe, which we will now
discuss.

The scale factor or $\Lambda<0$ (and negligible curvature) is
\begin{equation}
  a(t) = \left[\frac{2}{3}t_\Lambda
    \sin\left(\frac{3t}{2t_\Lambda}\right)\right]^{2/3}~.
\end{equation}
The comoving volume can be expressed in terms of hypergeometric
functions.  Once again the behavior of the {\em causal diamond\/} is
easier to estimate.  A reasonable approximation is
\begin{equation}
V_{\rm c}^{\rm dia}(t) \propto \begin{cases}
36\pi t & \text{if $t< \pi t_\Lambda/3$}~,\\
36\pi \left(\frac{2\pi}{3}t_\Lambda-t\right) & \text{if $t\geq \pi t_\Lambda/3$}~.
\end{cases}
\end{equation}
For $\Lambda<0$, the ``edge'' time for the causal diamond is $t_{\rm
  edge} = \pi t_\Lambda/3$ (coinciding with the turnaround time for
the scale factor).  Since $\pi/3> 0.23$, the probability distribution
for $\Lambda<0$ peaks at a value of $\abs{\Lambda}$ which is higher by
a factor of $(\pi/(3\times 0.23))^{2}\approx 21$ than that for
$\Lambda>0$ with the same observer model.  This evident when
comparing, for example, the first plot of Fig.~\ref{posDia} with that
of Fig.~\ref{negDia}.  It is also manifest in the distributions over
both positive and negative values of $\Lambda$, such as the first plot
in Fig.~\ref{posnegDia}.

According to our approximate equation, the diamond's volume at
$t=t_{\rm edge}$ should be larger for $\Lambda<0$, by the same factor
of $21$, for equal $|\Lambda|$.  Indeed, the first plot of
Fig.~\ref{posnegDia} shows that height of the peak for $\Lambda<0$ is
larger than for $\Lambda>0$ by about this amount.  We can also compare
the integrated probabilities for each sign of $\Lambda$,
$p(\Lambda<0)/p(\Lambda>0)$.  For the entropy production model, this
ratio is $25$, which we can attribute to the enhancement of diamond
volume for those universes with $\Lambda<0$ where $t_{edge}$ coincides
with the peak of entropy production.  In the $5~{\rm Gyr}$ and
$10~{\rm Gyr}$ time delay models the ratio is $13$ and $12.4$,
respectively.  The ratio is smaller than in the entropy model, because
about half of the range of $\log_{10}\Lambda$ has zero probability in
the time delay models because those universes crunch before $t=t_{\rm
  delay}$.

Similar conclusions hold for the {\em causal patch\/} when
$\Lambda<0$.  As for $\Lambda>0$, small values of $\abs{\Lambda}$ are
less suppressed by the causal patch than by the diamond.  This
broadens the probability distribution so much that our range of
$\log_{10}\abs{\Lambda}$ misses some of the probability for small
$\abs{\Lambda}$, as one can see in the first plot of
Figs.~\ref{negPatch}, \ref{negPatch5Gyr}, and~\ref{negPatch10Gyr}.  This
means we cannot accurately compute certain quantities, such as the
ratio $p(\Lambda<0)/p(\Lambda>0)$, for the patch.  We can make some
qualitative statements, though.  For the entropy production model, we
expect that the ratio should be of the same order for the patch as for
the diamond.  For the time delay models, the ratio should decrease for
the same reason that it decreased in the diamond: the universe
crunches too early in part of the parameter space.  However, the
decrease should be smaller in the patch than it was in the diamond.
That is because the universes which are crunching ``too early'' are
the ones with large $\abs{\Lambda}$, but as discussed above this is
precisely the region of parameter space which is relatively
de-emphasized in the patch as compared with the diamond.

\subsection{Varying spatial curvature only}
\label{sec-curvonly}

Now we will consider the implications of varying spatial curvature,
while keeping $\Lambda$ (and $Q$) fixed at its observed value (or at
minus this value; see below).  How is structure formation and observer
formation affected by period of curvature domination following matter
domination, beginning at $t\sim t_{\rm c}$.  In the sudden transition
approximation, the scale factor is
%\begin{equation}
%a(t) = \begin{cases}
%\frac{3}{2}t^{2/3}t_{\rm c}^{1/3} & \text{if $t<t_{\rm c}$}~, \\
%t+\frac{1}{2}t_{\rm c}& \text{if $t_{\rm c}\leq t <  t_\Lambda-\frac{1}{2}t_{\rm c}$}~, \\
%t_\Lambda \exp(t/t_\Lambda+t_{\rm c}/{2t_\Lambda}-1) & \text{if $t_\Lambda-\frac{1}{2}t_{\rm c}\leq t$}~.
%\end{cases}
%\end{equation}
\begin{equation}
a(t) = \begin{cases}
t^{2/3}t_{\rm c}^{1/3} & \text{if $t<t_{\rm c}$}~, \\
t& \text{if $t_{\rm c}\leq t <  t_\Lambda$}~, \\
t_\Lambda \exp(t/t_\Lambda-1) & \text{if $t_\Lambda\leq t$}~.
\end{cases}
\label{eq-curvscalefactor}
\end{equation}
By Eq.~(\ref{eq-patch}), the comoving radius of the causal patch is
given by
%\begin{equation}
%\chi_{\rm patch}(t) \propto \begin{cases}
  %  3- 2 (t/t_{\rm c})^{1/3} +\ln\frac{2t_\Lambda}{3t_{\rm c}}& \text{if $t<t_{\rm c}$}~, \\
  % 1+\ln\frac{2t_\Lambda}{2t+t_{\rm c}}& \text{if $t_{\rm c}\leq t < t_\Lambda-\frac{1}{2}t_{\rm c}$} ~,\\
%   \exp(-t/t_\Lambda-t_{\rm c}/2t_\Lambda+1)  & \text{if $ t_\Lambda-\frac{1}{2}t_{\rm c} \leq t$}~.
   % \end{cases}
%\end{equation}
\begin{equation}
\chi_{\rm patch}(t) \propto \begin{cases}
    4- 3 (t/t_{\rm c})^{1/3} +\ln\frac{t_\Lambda}{t_{\rm c}}& \text{if $t<t_{\rm c}$}~, \\
   1+\ln\frac{t_\Lambda}{t}& \text{if $t_{\rm c}\leq t < t_\Lambda$} ~,\\
   \exp(-t/t_\Lambda+1)  & \text{if $ t_\Lambda \leq t$}~.
    \end{cases}
\end{equation}
%\begin{equation}
%\chi_{\rm dia}(t) \propto \begin{cases}
%2(t/t_{\rm c})^{1/3} & \text{if $t\ll t_{\rm c}$}~,\\
%   2+\ln\frac{2t+t_{\rm c}}{3t_{\rm c}}& \text{if $t_{\rm c}\ll t \ll t_\Lambda$} ~,\\
 %\exp(-t/t_\Lambda-t_{\rm c}/2t_\Lambda+1)& \text{if $t_\Lambda \ll t$}~.
%\end{cases}
%\end{equation}
To compute the radius of the diamond, we need to first find the edge time $t_{\rm edge}$ where the the forward and backward lightcones meet.
We will assume (and later check for consistency), that this time is during the period of curvature domination.
During curvature domination, the radius of the forward lightcone is
\begin{equation}
\chi_{\rm forward}(t) =  3+\ln\frac{t}{t_{\rm c}}
\end{equation}
while that of the backward lightcone is given by $\chi_{\rm patch}$.
Setting the two equal, we find that
\begin{equation}\label{eq-tedge}
t_{\rm edge} = e^{-1}\sqrt{t_{\rm c}t_\Lambda}~.
\end{equation}
One can easily see that $t_{\rm c}<t_{\rm edge}<t_\Lambda$ when $t_\Lambda \gg t_{\rm c}$, which is the limit we are most interested in.
So we have, by Eq.~(\ref{eq-diamond}),
\begin{equation}
\chi_{\rm dia}(t) \propto \begin{cases}
3(t/t_{\rm c})^{1/3} & \text{if $t< t_{\rm c}$}~,\\
   3+\ln\frac{t}{t_{\rm c}}& \text{if $t_{\rm c}\leq t < t_{\rm edge}$} ~,\\
      1+\ln\frac{t_\Lambda}{t}& \text{if $t_{\rm edge}\leq t < t_\Lambda$} ~,\\
\exp(-t/t_\Lambda+1)& \text{if $t_\Lambda \leq t$}~.
\end{cases}
\label{eq-lambdadia}
\end{equation}

When computing the comoving volume, we must keep in mind that space is
curved; the comoving volume scales like $\pi\sinh(2\chi) -2\pi\chi$.
For large $\chi$, this reduces to $\pi \exp(2\chi) / 2$, while for
small $\chi$ we recover $4\pi \chi^3 /3$.  As an approximation, we
will use the exponential form during the period of curvature
domination for both the patch and the diamond, as well as during
matter domination for the patch.  We will use the flat form during
matter domination for the diamond, and during $\Lambda$ domination for
both the patch and the diamond.  This is a good approximation in the
regime where the scales are widely separated.  We find
%\begin{equation}
%V_{\rm c}^{\rm patch}(t) \propto \begin{cases}
   % \frac{\pi}{2}\left(\frac{2t_\Lambda}{3t_{\rm c}}\right)^2\exp\left(6- 4 (t/t_{\rm c})^{1/3}\right)& \text{if $t\ll t_{\rm c}$}~, \\
 %  \frac{\pi e^2}{2}\left(\frac{2t_\Lambda}{2t_+t_{\rm c}}\right)^2 & \text{if $t_{\rm c}\ll t \ll t_\Lambda$} ~,\\
 % \frac{4\pi}{3} \exp(-3t/t_\Lambda-3t_{\rm c}/2t_\Lambda+3)  & \text{if $t_\Lambda \ll t$}~.
    %\end{cases}
%\end{equation}
\begin{equation}
V_{\rm c}^{\rm patch}(t) \propto \begin{cases}
    \frac{\pi}{2}\left(\frac{t_\Lambda}{t_{\rm c}}\right)^2\exp\left(8- 6 (t/t_{\rm c})^{1/3}\right)& \text{if $t\ll t_{\rm c}$}~, \\
   \frac{\pi e^2}{2}\left(\frac{t_\Lambda}{t}\right)^2 & \text{if $t_{\rm c}\ll t \ll t_\Lambda$} ~,\\
  \frac{4\pi}{3} \exp(-3t/t_\Lambda+3)  & \text{if $t_\Lambda \ll t$}~.
    \end{cases}
    \label{eq-patchvolcurv}
\end{equation}
for the patch and
%\begin{equation}
%V_{\rm c}^{\rm dia}(t) \propto \begin{cases}
%\frac{32\pi}{3}(t/t_{\rm c}) & \text{if $t\ll t_{\rm c}$}~,\\
  % \frac{\pi e^4}{2}\left(\frac{2t+t_{\rm c}}{3t_{\rm c}}\right)^2& \text{if $t_{\rm c}\ll t \ll t_\Lambda$} ~,\\
%\frac{4\pi}{3}\exp(-3t/t_\Lambda-3t_{\rm c}/2t_\Lambda+3)& \text{if $t_\Lambda \ll t$}~.
%\end{cases}
%\end{equation}
\begin{equation}
V_{\rm c}^{\rm dia}(t) \propto \begin{cases}
36\pi(t/t_{\rm c}) & \text{if $t\ll t_{\rm c}$}~,\\
   \frac{\pi e^6}{2}\left(\frac{t}{t_{\rm c}}\right)^2& \text{if $t_{\rm c}\ll t < t_{\rm edge}$} ~,\\
      \frac{\pi e^2}{2}\left(\frac{t_\Lambda}{t}\right)^2& \text{if $t_{\rm edge}\leq t \ll t_\Lambda$} ~,\\
\frac{4\pi}{3}\exp(-3t/t_\Lambda+3) & \text{if $t_\Lambda \ll t$}~.
\end{cases}
\label{eq-diavolcurv}
\end{equation}
for the diamond.

We can count observers in the causal patch or diamond by integrating
the comoving volume against $\dot{n}_{\rm obs}(t)$, the rate at which
observations are made per unit time and comoving volume.  Note that
the scale factor in Eq.~(\ref{eq-curvscalefactor}) contains an explicit
dependence on $t_{\rm c}$ during matter domination.  This means that
the comoving observer density also depends on $t_{\rm c}$ through a
trivial overall multiplicative factor.  But the physical observer
density is independent of $t_{\rm c}$ during matter domination.  It
will be clearer to work with a variable that makes this explicit:
\begin{equation}
\dot{\tilde n}_{\rm obs}\equiv \dot{n}_{\rm obs}(t)/ t_{\rm c}~.
\label{eq-tilden}
\end{equation}

For sufficiently large $\Delta N$ at fixed $\Lambda$, curvature never
plays a dynamical role, because $t_\Lambda < t_{\rm c}$.  In this
regime the number of observers is independent of $\Delta N$, and the
probability distribution over $\Delta N$ is given by the prior:
\begin{equation}
\frac{dp}{d\Delta N} \propto \frac{1}{(60+\Delta N)^4}~.
\label{eq-largenprob}
\end{equation}
This can be seen in the 3rd plot of
Figs.~\ref{posDia}, \ref{posDia5Gyr}, \ref{posDia10Gyr}, \ref{posPatch}, \ref{posPatch5Gyr}, and~\ref{posPatch10Gyr}.
In all figures, we have used this analytic formula to continue our
results beyond the displayed upper limit of $\Delta N=1$ when we
calculate probabilities and confidence intervals. The (prior)
suppression of large $\Delta N$ means that there is no runaway in that
direction.

Let us turn to the more interesting case where $t_\Lambda>t_{\rm c}$.
(In general, there would be two subcases, $t_\Lambda>t_{\rm peak} >
t_{\rm c}$ and $t_\Lambda> t_{\rm c}>t_{\rm peak}$. However, in this
subsection we are discussing only the variation of curvature, in a
universe otherwise like ours. For all our observer models, $t_{\rm
  peak}$ is comparabale to $ t_\Lambda $ in our universe, so the
second subcase does not arise.) For the {\em
  causal diamond\/}, we find
\begin{equation}
\frac{dp}{d\Delta N} \propto \frac{1}{(60+\Delta N)^4}
\int_{t_{\rm c}}^{t_\Lambda}dt~ \dot{\tilde n}_{\rm obs}(t) \frac{t^2}{t_{\rm c}}~.
\label{eq-pdn}
\end{equation}
The geometric factor of $t_{\rm c}^{-1}\propto \exp(-3\Delta N)$,
along with the prior distribution, favors curvature.

However, for sufficiently large curvature, dynamical effects become
important and cut off the probability distribution. With all observer
models we consider, the number of observations will decrease if
structure formation is suppressed. This effect becomes severe if not
even the earliest structures can form, i.e., if the density contrast,
$\sigma(M,t)$, never reaches unity even for the smallest mass scales
that can cool efficiently, $M_{\rm min}$. Let $t_{\rm vir}$ denote the
time when these structures would have formed in a universe without
curvature. By Eq.~(\ref{eq-PSFraction}), for $t_{\rm c}\ll t_{\rm vir}$
these structures will be suppressed like 
\begin{equation}
\exp\left[-\left(\frac{1.68}{\sqrt{2}\sigma(M,t)}\right)^2\right] = \exp\left[-B\left(\frac{t_{\rm vir}}{t_{\rm c}}\right)^{4/3} \right]~.
\label{eq-curvsuppression}
\end{equation}
(Here $B$ is some order-one coefficient and $t_{\rm vir}$ depends weakly on the mass scale.)
This corresponds to a doubly-exponential suppression of the probability
distribution over $\Delta N$, Eq.~(\ref{eq-pdn}). In our universe, the
value of $t_c$ corresponding to $\Delta N=0$ is somewhat larger than
$t_{\rm vir}$, and the suppressed regime is reached close to the lower
end of our parameter space, $\Delta N=-3.5$. This can be seen in the
3rd plot of Figs.~\ref{posDia}, \ref{posDia5Gyr}, and~\ref{posDia10Gyr}.

Now let us consider the same regime,  $t_\Lambda > t_{\rm c}$,
in the {\em causal patch\/}. Using Eq.~(\ref{eq-patchvolcurv}) we find
\begin{equation}
\frac{dp}{d\Delta N} \propto \frac{\pi e^2/2}{(60+\Delta N)^4}\int_{t_{\rm c}}^{t_\Lambda}dt~ \dot{\tilde n}_{\rm obs}(t) \frac{t_{\rm c}t_\Lambda^2}{t^2}~.
\end{equation}
This time, a geometric factor of $t_c$ appears in the numerator,
suppressing curvature. Below $\Delta N\approx -3.5$, the stronger,
dynamical suppression discussed for the diamond sets in: the failure
of structure formation. This behavior is reflected in the 3rd plot
of Figs.~\ref{posPatch}, \ref{posPatch5Gyr}, and~\ref{posPatch10Gyr}.

In all plots and all observer models, our failure thus far to detect
curvature is not surprising; it is perfectly compatible with the
predicted probability distribution. However, upcoming experiments such
as Planck will be more sensitive to small amounts of spatial
curvature. In the spirit of Ref.~\cite{FreKle05}, let us use our
probability distribution to calculate the probability that curvature
will be detected in the future.

The current $1\sigma $ bound on curvature from
WMAP5+BAO+SN~\cite{WMAP5} is $\Omega_k=-0.0050^{+0.0061}_{-0.0060}$.
This corresponds roughly to the Gaussian distribution
\begin{equation}
  \frac{dp_{\rm exp}}{d\Omega_k} \propto \exp\left(-\frac{(\Omega_k +
 .0050)^2}{2(.0061)^2}\right)~.
\end{equation}
Our convention for $\Delta N=0$ is the upper $1\sigma$ bound,
$\Omega_k = .0011$. Since $\Omega_k\propto \exp(-2\Delta N)$, we can
convert our probability distribution for $\Delta N$ into one for
$\Omega_k$, which in the regime $\Delta N\gtrsim -1$ looks like
\begin{equation}
\frac{dp}{d\Omega_k} \propto 
\frac{1}{\Omega_k}\frac{1}{\left(60+\frac{1}{2}\ln\frac{.0011}{\Omega_k}\right)^4}
~,
\end{equation}
Because we are assuming an open universe, the probability for $\Omega_k<0$ vanishes.
The current experimental bound is so strong that we do not need a more detailed form.
If future experiments reach a sensitivity level of $\Delta \Omega_k$, then we will be able to detect the openness of the universe if $\Omega_k \geq \Delta \Omega_k$.
The probability for this to occur is
\begin{eqnarray}
p(\Omega_k \geq \Delta \Omega_k) &=&\int_{\Delta \Omega_k}^\infty \frac{dp_{\rm expt}}{d\Omega_k}(\Omega_k)\frac{dp}{d\Omega_k}(\Omega_k)~  d\Omega_k \\
&\propto& \int_{\Delta \Omega_k}^\infty \frac{d\Omega_k}{\Omega_k}\frac{\exp\left(-\frac{(\Omega_k + .0050)^2}{2(.0061)^2}\right)}{\left(60+\frac{1}{2}\ln\frac{.0011}{\Omega_k}\right)^4}~,
\end{eqnarray}
which is normalized so that $p(\Omega_k \geq 0) = 1$. Then we find
$p(\Omega_k \geq 10^{-3}) \approx 0.033$, which might be realized in
the near future, and $p(\Omega_k \geq 10^{-4}) \approx 0.088$, which
is roughly the limit of achievable sensitivity.

\subsection{Varying both the cosmological constant and curvature}
\label{sec-lambdacurvonly}

When both the curvature and the cosmological constant vary, we can use
a similar analysis to obtain a qualitative understanding of the
probability distributions. Again, we will distinguish different
regimes determined by the relative sizes of $t_\Lambda$, $t_{\rm c}$,
and $t_{\rm peak}$. We will now have to consider both positive and
negative values of the cosmological constant. 

The cases with $t_\Lambda\ll t_{\rm c}$ correspond to negligible
curvature. In this regime, the joint probability distribution is the
direct product of the probability distribution over $\Lambda$ (with
negligible curvature, see Sec.~\ref{sec-lambdaonly}) and the prior
probability distribution over $\Delta N$ (see Eq.~(\ref{eq-nprior}).
which we have already examined. We can also immediately dispense with
the cases $t_\Lambda \ll t_{\rm peak}$: there will always be pressure
toward smaller $\abs{\Lambda}$ either because of suppressed structure
formation and a small causal diamond/patch ($\Lambda>0$) or because
the universe has already collapsed ($\Lambda<0$). 

The case $t_{\rm c}\ll t_{\rm peak} \ll t_\Lambda$, for $\Lambda>0$,
was essentially discussed in the previous subsection, except that
$\Lambda$ was held fixed there. But since $t_{\rm peak}$ is
essentially independent of $\Lambda$ and $\Delta N$, $\Lambda$ can
vary only in the direction of smaller values while preserving the
above inequality. Therefore, in this regime, the joint probability
distribution is the direct product of the probability distribution
over Lambda given in Sec.~\ref{sec-lambdaonly} and the distribution
over $\Delta N$ derived in Sec.~\ref{sec-curvonly}. Moreover, the
doubly-exponential suppression of structure formation by curvature is
unaffected by the sign of $\Lambda$. Therefore, the joint distribution
over $\Delta N$ and negative values of $\Lambda$ is given by the
direct product of the $\Delta N$ distribution from
Sec.~\ref{sec-curvonly} and the negative $\Lambda$ distribution from
Sec.~\ref{sec-lambdaonly}.

There is one subtlety, which is apparent when comparing the 3rd plot
of Fig.~\ref{posDia} with that of Fig.~\ref{negDia}.  In
Fig.~\ref{negDia} the probability is {\em increasing\/} toward the
boundary of the plot at $\Delta N=-3.5$, whereas in Fig.~\ref{posDia}
the suppression of structure formation has caused the probability
distribution to {\em decrease\/} at the same value of $\Delta N$.  We
have already argued above that structure suppression due to large
curvature works the same way for $\Lambda<0$ as for $\Lambda>0$, so we
must reconcile the apparent discrepancy.  First, we should note that
the probability does not increase indefinitely in toward small $\Delta
N$, our plot range is merely inadequate to show the eventual peak and
subsequent decrease.  We must explain why the suppression does not
happen at the same value of $\Delta N$ in the positive and negative
cases, for equal values of $\abs{\Lambda}$.

The answer lies in the geometry of the causal diamond, specifically in
the difference in edge times, $t_{\rm edge}$, for positive and
negative $\Lambda$.  As we saw in Eq.~(\ref{eq-tedge}), $t_{\rm edge}$
will actually decrease as curvature is increased while $\Lambda$
remains constant.  It turns out that $t_{\rm edge} \approx t_{\rm
  peak}$ for $\Lambda=\Lambda_0$, $Q=Q_0$, and $\Delta N = -3.5$.
However, $t_{\rm edge}$ for $\Lambda<0$ is always of order $t_\Lambda$
(to be precise, it is equal to the time of maximum expansion).  In
particular, $t_{\rm edge}\gg t_{\rm peak}$ for $Q=Q_0$,
$\Lambda=-\Lambda_0$ and all values of $\Delta N$.  The entropy
production curves are nearly identical when the sign of $\Lambda$ is
flipped (since we are safely in the limit $t_\Lambda \gg t_{\rm
  peak}$), but in the $\Lambda<0$ case the tail of the entropy
production lies entirely within the growing phase of the causal
diamond.  The extra boost in probability granted by this effect means
that more curvature is required to suppress the {\em probability} for
$\Lambda<0$, even though {\em structure formation\/} is suppressed by
an amount independent of the sign of $\Lambda$.

Let us turn to the case 
\begin{equation}
t_{\rm peak} \ll t_{\rm c} \ll t_\Lambda; ~~\Lambda>0~.
\label{eq-hjk}
\end{equation}
Structure formation is uninhibited in this regime, so
only prior probabilities and geometric effects control the joint
probability distribution. The comoving volumes of the patch and
diamond are given in Eqs.~\ref{eq-patchvolcurv} and
\ref{eq-diavolcurv}. Combining this with the prior distribution, we
find for the {\em causal diamond}:
\begin{equation}
\frac{d^2p}{d\log_{10}\Lambda\, d\Delta N}\propto 
\frac{\Lambda}{(60+\Delta N)^4}\int_0^{t_{\rm c}}dt~\dot{\tilde n}(t)~t~.
\end{equation}
The integral is independent of $\Lambda$ and $\Delta N$, so the
probability distribution is governed by the prefactor. The pressure is
toward larger $\Lambda$ and smaller $\Delta N$ suppresses any
hierarchy between the timescales appearing in the above double
inequality. Indeed, in our universe, the inequality is either violated
or approximately saturated.

In the same regime, Eq.~(\ref{eq-hjk}), the {\em causal patch\/} introduces the geometric factor $t_\Lambda^2/t_c$.  This leads to a somewhat different result:
\begin{align}
\frac{d^2p}{d\log_{10}\Lambda\, d\Delta N} &\propto 
\frac{\Lambda}{(60+\Delta N)^4}\int_0^{t_{\rm c}}dt~
\dot{\tilde n}(t) \frac{t_\Lambda^2}{t_{\rm c}}
\exp\left(8-6(t/t_{\rm c})^{1/3}\right)\\
&\propto \frac{1}{(60+\Delta N)^4}\exp\left(-3\Delta N\right)~.
\end{align}
There is a pressure toward smaller $\Delta N$, suppressing a large
hierarchy $t_{\rm peak}\ll t_c$. This is qualitatively the same as for
the diamond, though the pressure is stronger. But unlike in the
diamond case, small values of the cosmological constant are
unsuppressed: at fixed $\Delta N$, the probability is flat in
$\log_{10}\Lambda$. This is a runaway problem: Without a lower bound
on $\log_{10} \Lambda$, the probability distribution is not integrable
in the direction of small $\log_{10} \Lambda$. Such a lower bound may
well be supplied by the finiteness of the landscape. It need not be
particularly close to the observed value, since the probability is
flat. (We are about to discover, however, that a more severe runaway
problem arises for the causal patch in the analogous case with
negative $\Lambda$.)

Now we analyze the same case, Eq.~(\ref{eq-hjk}), but with $\Lambda<0$.
Our conclusions for the {\em causal diamond\/} remain the same, with $\Lambda$ replaced by $\abs{\Lambda}$, since neither $\dot{n}_{\rm obs}(t)$ nor $V_{\rm c}^{\rm dia}(t)$ depend on the sign of $\Lambda$ when $t\ll t_{\rm c}\ll t_\Lambda$.
Turning to the {\em causal patch}, we note that the scale factor can be approximated by
\begin{equation}
a(t) = \begin{cases}
t^{2/3} t_{\rm c}^{1/3}& \text{if $t<t_{\rm c}$}~,\\
t_{\Lambda}\sin\left( \frac{t+t_1}{t_\Lambda}\right) & \text{if $t_{\rm c} \leq t<t_2 - t_{\rm c}$}~,\\\
\left(t_2-t\right)^{2/3} t_{\rm c}^{1/3} & \text{if $t_2- t_{\rm c} \leq t<t_2$}~.
\end{cases}
\end{equation}
in the simplified sudden transition approximation, where
\begin{equation}
t_1 = t_\Lambda \sin^{-1}\left(\frac{t_{\rm c}}{t_\Lambda}\right)-t_{\rm c}
\end{equation}
and 
\begin{equation}
t_2 = \pi t_\Lambda - 2t_1~.
\end{equation}
Since $t_{\rm peak} \ll t_c$, we need only compute the comoving volume in the regime $t\ll t_{\rm c}$.  For the comoving radius, we find
\begin{eqnarray}
&\chi_{\rm patch}(t) &= \int_t^{t_2} \frac{dt}{a(t)}\\
&  &=\int_t^{t_{\rm c}}\frac{1}{t^{2/3}t_{\rm c}^{1/3}}+ \int_{t_{\rm c}}^{t_2 - t_{\rm c}} \frac{dt}{t_{\rm \Lambda}\sin\left(\frac{ t+t_1}{t_\Lambda}\right)}+\int_{t_2-t_{\rm c}}^{t_2}\frac{1}{\left(t_2-t\right)^{2/3}t_{\rm c}^{1/3}}\\
& &= 6 - 3\left(\frac{t}{t_{\rm c}}\right)^{1/3} + \ln\frac{\tan(t_2/2t_\Lambda-t_{\rm c}/2t_\Lambda+t_1/2t_\Lambda)}{\tan(t_{\rm_c}/2t_\Lambda+t_1/2t_\Lambda)}\\
& &\approx 6 + 2\ln \frac{2t_\Lambda}{t_{\rm c}}~.
\end{eqnarray}
(In the final line we used $t \ll t_{\rm c} \ll t_\Lambda$.)
Using $V(\chi)\propto \exp(2\chi)$, the comoving volume is approximately
\begin{equation}
V^{\rm patch}_{\rm c}(t)\propto \left( \frac{t_\Lambda}{t_{\rm c}}\right)^4~,
\end{equation}
yielding the probability distribution
\begin{equation}
\frac{d^2p}{d\log_{10}\abs{\Lambda}\, d\Delta N}\propto 
\frac{\exp\left(-9\Delta N\right)}{\abs{\Lambda}\,(60+\Delta N)^4}~.
\end{equation}

Again we find the causal patch leads to runaway towards small values
of $|\Lambda|$ in the presence of curvature.  The runaway is stronger,
like $|\Lambda|^{-1}$ for fixed $\Delta N$, than the flat distribution
over $\log_{10}|\Lambda|$ that we found in the case of positive
$\Lambda$.  The preference for small $\abs{\Lambda}$ is evident in the
8th plot of
Figs.~\ref{negPatch}, \ref{negPatch5Gyr}, and~\ref{negPatch10Gyr}, where the
probability is concentrated in the small $\Delta N$ and small
$\Lambda$ corner of the plots.

This runaway implies one of two things.  Either the causal patch is
the wrong measure, at least in regions with nonpositive cosmological
constant.\footnote{These regions were defined more sharply in
  Ref.~\cite{Bou09,BouYan09} (``hat domains'' and ``singular
  domains'').  If the causal patch is viewed as arising from an
  analogy with the AdS/CFT correspondence via a global/local duality,
  it is not implausible that it may apply only in the ``eternal
  domain'' (roughly, regions with positive $\Lambda$).} Or our vacuum
has the smallest cosmological constant of all (anthropic) vacua in the
landscape.  The latter possibility is quite interesting, since it
offers a potential first-principles explanation of the actual
magnitude of $|\Lambda|$ (and by correlation, many other unexplained
hierarchies) in terms of a fundamental property of the underlying
theory: the size of the landscape.  (A variant of this possibility is
that the landscape does contain vacua with smaller $\Lambda$ but the
prior probability distribution for $\Delta N$ is weighted towards
large values, so that curvature is negligible in all pocket universes.)

\subsection{Varying the density contrast only}
\label{sec-qonly}

Unlike $\Lambda$ and $\Delta N$, $Q$ has no geometric effects (at
least in the homogeneous approximation we employ).  It does have
dynamical implications, however.  The effects of $Q$ on the observer
production rate $\dot{n}_{\rm obs}$ differ somewhat depending on the
observer model.  Since all three models involve star formation, let us
begin by considering how it is affected by $Q$.  $Q$ enters the
structure formation equations in the combination $QG(t)$, and
$G(t)\propto t^{2/3}$ during matter domination.  Thus, the time of
structure formation (and thus, star formation) scales as $t_{\rm vir}
\sim Q^{-3/2}$.  

Let us discuss first the effects of increasing $Q$. Star formation
that happens earlier also happens faster. The star formation rate for
a single halo, Eq.~(\ref{eq-SingleHaloSFR}), scales like the inverse of
the gravitational timescale of the halo. Since $t_{\rm grav}$ is
proportional to the virialization time of the halo, the star formation
rate scales as $\dot{\rho}_\star(t_{\rm peak})\propto Q^{3/2}$.  By
similar arguments, the width of the star formation rate also scales
like $\Delta t\propto Q^{-3/2}$.  Thus, one might expect that the
total integrated star formation rate stays constant as one varies $Q$.
However, there is an effect which is not seen by this rough analysis.
As $Q$ increases, the range of halo masses which are able to cool and
form stars also increases.  The lower mass limit, which comes from the
requirement that halos be hotter than $10^4\, K$ to start, is smaller
at earlier times.  And the upper mass barrier, which arises from
radiative cooling failure, completely disappears at sufficiently early
times (prior to about $1$ Gyr), because Compton cooling becomes
efficient~\cite{TegRee97}.\footnote{Dominant Compton cooling, however,
  corresponds to a drastic change of regime and may be catastrophic.
  This possibility is explored in Ref.~\cite{BouHal09}.} Thus, a
larger mass fraction forms stars. Numerically, we find that these two
effects combine to make the integrated star formation grow roughly
logarithmically with $Q$:
\begin{equation} 
  \int dt \dot{\rho}_\star(t)
  \propto \log_{10}Q~.
\label{eq-logq}
\end{equation}
In the entire analysis, so far, we have assumed that structure
formation is not disrupted by curvature or vacuum energy; this
assumption is certainly justified in the case at hand, where $\Lambda$
and $\Delta N$ are held fixed and set to the observed values.

There is a limit to how far we can increase $Q$ given the model of
star formation outlined in Sec.~\ref{sec-stars}: for large enough $Q$,
structure formation happens before recombination.  Even though dark
matter halos can form prior to recombination, baryons cannot collapse
into galaxies.  One expects that if there are large dark matter halos
already formed at the time of recombination, then there will be a huge
surge in star formation as the baryons fall into them after decoupling
from the photons.  Our star formation code takes this into account in
a very simplistic way: star formation which would have happened prior
to recombination is delayed until after recombination, and at that
time there is a large and essentially instantaneous spike in the star
formation rate, after which it drops back down to normal levels.  (The
code knows nothing of the actual interactions between baryons and
photons.)  There may be a phenomenological reason why such
instantaneous star formation is not hospitable for observers, but at
the very least this is a change in regime for the star formation rate.
At $Q=10^2Q_0$, $32\%$ of the total star formation is created in this
spike, up from $13\%$ at $10^{1.75}Q_0$ and $0.1\%$ at $10^{1.5}Q_0$,
and this percentage will continue to rise if $Q$ is increased.  This
motivates our cut-off at $10^2Q_0$.  As seen in
Sec.~\ref{sec-results}, the behavior of the probability distribution
in $Q$ near the upper boundary is fairly mild (at worst a logarithmic
growth when $\Lambda=\Lambda_0$ and $\Delta N>0$ are fixed), so our
results should not change dramatically if the cut-off is extended.

If $Q$ is decreased compared to the observed value, then the range of
halo masses that can cool efficiently shrinks. It soon disappears
altogether, for $Q\approx 10^{-1} Q_0$. There are no stars, and no
observers, for smaller values of $Q$. This cuts off the probability
distribution over $Q$ independently of $\Delta N$, $\Lambda$, the
measure, and the observer model, so we will not discuss this case
further.

Let us now estimate the probability distribution for $Q>Q_0$. We
begin with the time delay observer models. The time delay is held fixed as $Q$ varies. (The assumption
underlying these models is that $t_{\rm delay}$ is determined, at
least in a substantial part, by dynamics unrelated to $Q$, $\Delta N$,
or $\Lambda$). In our universe, $t_{\rm vir}$ is already somewhat smaller than 5 Gyr, and for larger $Q$, $t_{\rm vir}$ will be entirely negligible compared to the time delay, as will the width of the star formation rate.  Thus, observers will live at a time given approximately by $t_{\rm delay}$.  Using a flat prior, $d\tilde p/d\log_{10} Q\sim 1$, and the logarithmic growth of the integrated SFR described, one expects
\begin{align}
\frac{dp}{d\log_{10}Q} &\propto \int dt \dot{\rho}_\star(t-t_{\rm delay}) V_{\rm c}(t_{\rm delay})\\
&\propto \log_{10}Q~.
\end{align}
This distribution holds in both the patch and the diamond and with
either a $5$ or $10\,{\rm Gyr}$ time delay, as is evident in the
2nd plot of
Figs.~\ref{posDia5Gyr}, \ref{posDia10Gyr}, \ref{posPatch5Gyr}, and~\ref{posPatch10Gyr}.

There are additional complications with entropy production, and these
complications are such that we are only able to give a very rough
qualitative account of the probability distribution; numerical
calculations are essential to finding the true shape. The first
complication is that stars burn for a long time after they are
created. This makes the entropy production rate significantly broader
than the star formation rate, and the result is that we cannot
reliably approximate all of the entropy as being created at a single
time.

A second complication is that earlier star formation means earlier
entropy production, because much of the entropy is produced by fast
burning massive stars. The peak of the entropy production rate can
happen at arbitrarily early times, unlike in the time delay models, which ensured that the peak of $\dot{n}_{\rm
  obs}(t)$ happens after the time $t_{\rm delay}$. This has an
important consequence in the comparison of the patch and the diamond.
As $Q$ is increased, more entropy production happens at earlier times,
when the diamond is small but the patch is large. Indeed, comparing
the 2nd plot of Figs.~\ref{posDia} and~\ref{posPatch}, we see that
for the {\em causal diamond}, $dp/d\log_{10} Q$ is maximal at $Q=Q_0$;
whereas for the {\em causal patch}, it increases monotonically with
$\log_{10}Q$.

The third complication is the effect of the dust temperature. The
interstellar dust temperature is a complicated function of both the
halo virialization time and the actual emission time
(Eq~\ref{eq-DustTemp}). However, we can say qualitatively that the
effect of variation in dust temperature is to suppress the entropy
production of early-time stars relative to late-time stars. This is
why, for example, in the 2nd plot of Fig.~\ref{posPatch}, which
uses the causal patch measure, the probability distribution begins to
flatten out for large $Q$ rather than continuing to increase in a
manner similar to that of the time delay model in
Figs.~\ref{posDia5Gyr} and~\ref{posDia10Gyr}.

\subsection{Varying the density contrast and the cosmological constant}
\label{sec-qlambdaonly}

When both $\Lambda$ and $Q$ are allowed to vary, we can combine the
analysis of Sec.~\ref{sec-lambdaonly} and the previous subsection to
understand the probability distribution. In Sec.~\ref{sec-lambdaonly},
we concluded that the most likely value of $\Lambda$ for fixed $Q$ was
determined by the condition $t_\Lambda \approx t_{\rm peak}$, where
$t_{\rm
  peak}$ is the peak time of the model-dependent $\dot{n}_{\rm
  obs}(t)$. In the previous subsection we found that depending on the
observer model, $t_{\rm peak}$ can depend on $Q$.

In the time delay observer model, $t_{\rm peak}\approx t_{\rm delay}$
for all relevant values of $Q$. Indeed, in the 9th plots of
Figs.~\ref{posDia5Gyr}, \ref{posDia10Gyr}, \ref{posPatch5Gyr}, and~\ref{posPatch10Gyr},
we see that the most likely value of $\Lambda$ is essentially
independent of $Q$. Additionally, the probability increases with $Q$
proportional to $\log_{10} Q$, due to the increase of total star
formation discussed in Sec.~\ref{sec-qonly}, and this effect is
visible in the same plots. The only difference between the diamond and
the patch in these models is in the broadness of the distribution in
the $\Lambda$ direction, which was also discussed in
Sec.~\ref{sec-lambdaonly}.

In the entropy production observer model, $t_{\rm peak}$ depends
strongly on $Q$: $t_{\rm peak}\propto Q^{-3/2}$. This leads to the
relation $\Lambda\propto Q^3$ for the most likely value of $\Lambda$
at a given $Q$. In this 9th plot of
Figs.~\ref{posDia}, \ref{negDia}, \ref{posPatch}, and~\ref{negPatch} we see
this trend in the slope of the contour lines toward large $\Lambda$
and large $Q$. This looks like a runaway, but there is a cut-off
($Q\sim 10^{2}Q_0$) coming from our requirement that star formation happen after recombination (see Sec.~\ref{sec-qonly}).

\subsection{Varying the density contrast and spatial curvature} 
\label{sec-qcurvonly}

If curvature is small, $t_c\gg t_{\rm vir}$, curvature has neither
dynamical nor geometric effects for $\Lambda=\Lambda_0$. Large
curvature, however, can disrupt structure formation and thus star
formation. This is encoded in Eq.~(\ref{eq-curvsuppression}), which is
valid for $t_c\ll t_{\rm vir}$ (and was derived from Eq.~(\ref{eq-PSFraction})). When $Q$ varies simultaneously with
$\Delta N$, we must take into account the $Q$-dependence of $t_{\rm
  vir}$, $t_{\rm vir} \sim Q^{-3/2}$:
\begin{equation}
\exp\left[-B\left(\frac{t_{\rm vir}}{t_{\rm c}}\right)^{4/3} \right] = \exp\left[-CQ^{-2}\left(\frac{t_{\rm eq}}{t_{\rm c}}\right)^{4/3} \right]~.
\label{eq-curvqsuppression}
\end{equation}
(Recall that $B$ is an order-one coefficient; $C$ is a different order-one coefficient that weakly depends on the mass scale.)
For increased $Q$ a smaller $t_{\rm c}$ is required to halt structure formation.
This yields the following relation between $\log_{\rm 10}Q$ and $\Delta N_{\rm crit}$, the value of $\Delta N$ at which structure formation is completely suppressed:
\begin{equation}
(2\log_{10}e)\Delta N_{\rm crit} + \log_{10}Q = {\rm const.}
\label{eq-qncrit}
\end{equation}
We found in Sec.~\ref{sec-curvonly} that $\Delta N_{\rm
  crit}\approx-3.5$ when $Q=Q_0$, which is already at the edge of the
parameter range we consider. For larger values of $Q$ the value of
$\Delta N$ necessary to significantly suppress structure formation is
outside that range. However, the highest value of $Q$ we consider,
$10^2Q_0$, shifts $\Delta N_{\rm crit}$ only by -2.3. 

The 7th plot of each figure shows how $\Delta N_{\rm crit}$ depends on $Q$ for fixed $\Lambda=\Lambda_0$.
In the time delay models, Figs.~\ref{posDia5Gyr}, \ref{posDia10Gyr}, \ref{posPatch5Gyr}, and~\ref{posPatch10Gyr}, one clearly sees the contours of constant probability following a slope given approximately by Eq.~(\ref{eq-qncrit}).
The corresponding plots in the entropy production models, Figs~\ref{posDia} and~\ref{posPatch}, look a bit different in their gross features, owing to the complications discussed in Sec.~\ref{sec-qonly}, but the same trend in $\Delta N_{\rm crit}$ is visible.

All of the above considerations hold also for $\Lambda = -\Lambda_0$ for the {\em causal diamond\/}, which can be seen in the 7th plot of Figs.~\ref{negDia}, \ref{negDia5Gyr}, and~\ref{negDia10Gyr}.
(The {\em causal patch\/}, of course, has a runaway problem which is dominant.  See Sec.~\ref{sec-lambdacurvonly}.)
Since $t_{rm edge}$ is larger for $\Lambda<0$ for equal $\abs{\Lambda}$ (see Sec.~\ref{sec-lambdaonly}), the geometry of the diamond allows $\Lambda<0$ to tolerate more curvature.
This means $\Delta N_{\rm crit}$ is smaller for $\Lambda<0$, but does not change the fact that it scales according to Eq.~(\ref{eq-qncrit}).
It is this geometric effect which also leads to the enhancement of probability for small $\Delta N$ visible in the 3rd and 6th plots of those same figures.
As our discussion has made clear, this does not indicate a new runaway problem; our displayed parameter range is merely too small to see the suppression at small $\Delta N$ from structure suppression.

\subsection{Marginalizing}

Let us combine our previous observations to understand the behavior of
the probability distributions in which one or more parameters have
been integrated out.

\paragraph{Integrating out $\Delta N$}

For the {\em causal diamond}, the only consequence of varying $\Delta
N$ is that for small $\Delta N$ structure formation is interrupted.
This only happens over a very small part of the full $\Delta N$
parameter space, so integrating out $\Delta N$ does not have a large
effect on the distributions for $Q$ and $\Lambda$, independently of
the observer model. This is evident in the similarities between the
9th and 12th plots in
Figs.~\ref{posDia}, \ref{negDia}, \ref{posDia5Gyr}, \ref{negDia5Gyr}, \ref{posDia10Gyr}, and~\ref{negDia10Gyr}.
 
For the {\em causal patch}, however, the runaway toward small
$\abs{\Lambda}$ discussed in Sec.~\ref{sec-lambdacurvonly} means that
the smallest values of $\abs{\Lambda}$ are greatly favored after
$\Delta N$ is integrated out. Restricting to $\Lambda>0$, the effect
is weaker, but can still be seen in the comparison of the 9th and
12th plots of Figs.~\ref{posPatch5Gyr} and~\ref{posPatch10Gyr}. For
$\Lambda<0$, it is stronger, as seen in comparing the 9th and
12th plots in
Figs.~\ref{negPatch}, \ref{negPatch5Gyr}, and~\ref{negPatch10Gyr}.

\paragraph{Integrating out $Q$}

Th effect of integrating out $Q$ depends on the observer model. In the
time delay models, there is very little change in the distributions of
$\Lambda$ or $\Delta N$ after integrating out $Q$, as one can see by
comparing the 8th and 11th plots in each of
Figs.~\ref{posDia5Gyr}, \ref{negDia5Gyr}, \ref{posDia10Gyr}, and~\ref{negDia10Gyr}.
For $\Lambda$, this is easily understood since most of the $Q$
parameter space has $t_{\rm peak}=t_{\rm delay}$, and thus the
analysis is largely independent of $Q$ (see Sec.~\ref{sec-qonly}). For
curvature, integrating out $Q$ should increase the relative
probability for small $\Delta N$ compared with $Q=Q_0$ due to relaxed
restrictions on structure formation.
Comparing the 3rd and 6th plots of those same figures, we see that small $\Delta N$ is relatively more favored in the 6th plot.

For entropy production, though, there are some more significant
effects. After integrating out $Q$, larger values of both
$\abs{\Lambda}$ and curvature are generally favored. There are two
reasons for this. The first is that the prior probabilities for
$\abs{\Lambda}$ and curvature want both of them to be large. With $Q$
fixed, this tendency for largeness is countered by the suppression of structure formation when $\Lambda$ or curvature
dominate before $t_{\rm peak}$. But once $Q$ is integrated out, this
restriction is relaxed. The second reason is that the large $Q$ region
of parameter space, which is where $\Lambda$ and curvature are allowed
to be large, has more star formation and hence more entropy per
comoving volume. So the large $Q$ region of parameter space
contributes more to the resulting distributions of $\Lambda$ and
$\Delta N$ than the small $Q$ region after $Q$ is integrated out.

It is difficult to find an analytical formula for the resulting
probability distributions due to the complications mentioned in
Sec.~\ref{sec-qonly}. But, qualitatively, we can see this shift toward
larger $\abs{\Lambda}$ and smaller $\Delta N$ in comparing the 8th
and 11th plots in each of
Figs.~\ref{posDia}, \ref{negDia}, and~\ref{posPatch} (Fig.~\ref{negPatch} is
not included on this list because the runaway problem toward small
$\abs{\Lambda}$ is the dominant feature for $\Lambda<0$ in the causal
patch). The same figures also show a preference for large
$\abs{\Lambda}$ in the 4th plot. The probability flattens or turns
around for the largest values of $\abs{\Lambda}$ because of the upper
limit we have imposed on $Q$.

\paragraph{Integrating out $\Lambda$}
When integrating out $\Lambda$, the most important distinction is
whether we use the diamond or the patch. For the {\em causal patch},
as discussed in Sec.~\ref{sec-lambdacurvonly}, there is a runaway
problem toward small $\Lambda$. The probability is concentrated in the
region of smallest $|\Lambda|$. It is possible that this value is not
much smaller than the observed value, so this problem is not
necessarily fatal. When computing confidence intervals requires the
assumption of a lower bound on $\abs{\Lambda}$ (which is always
highlighted in the captions), we use the lower end of the displayed
range, $10^{-3}\Lambda_0$.

This issue does not arise for the {\em causal diamond}. In the time
delay models there is almost no difference between leaving $\Lambda$
fixed and integrating it out in both the positive and negative cases.
We can see this by comparing the 7th and 10th plots of
Figs.~\ref{posDia5Gyr}, \ref{negDia5Gyr}, \ref{posDia10Gyr}, and~\ref{negDia10Gyr}.
The cases where $\Lambda$ is integrated over both positive and
negative values are covered in
Figs.~\ref{posnegDia5Gyr} and~\ref{posnegDia10Gyr}. They are almost
identical to the negative $\Lambda$ case because the $\Lambda<0$ vacua
contribute the bulk of the probability.

With entropy production there is a qualitative change in the
distribution of $Q$ when going from $\Lambda$ fixed to $\Lambda$
integrated out. This change is most clearly seen in comparing the
2nd and 5th plots of Figs.~\ref{posDia} and~\ref{negDia}. When
$\Lambda$ is fixed, the $Q$ distribution becomes flat for large $Q$,
and also has a peak at small $Q$ ($\approx Q_0$). Recall from
Sec.~\ref{sec-qonly} that the flatness for large $Q$ is attributable
to the changing dust temperature, and the shape of the diamond gives
extra suppression for large $Q$ that effectively creates the peak at
small $Q$. After integrating out $\Lambda$, both of these effects
disappear. (In the plots referenced above, it looks like large $Q$ is
actually still suppressed, but this is due to a new effect: the
finiteness of our parameter range in $\Lambda$).

The reason why these effects disappear is because at fixed $\Lambda$
and high $Q$ the diamond is sensitive to the long tail of the entropy
production rate; its maximal comoving volume is several Gyr after
$t_{\rm peak}$. Integrating over $\Lambda$ is effectively like finding
the optimal value of $\Lambda$ for each $Q$. The optimal value for
$\Lambda$ is the one where the maximal height diamond volume is
centered on the peak of the entropy production rate, which means
resulting $Q$ distribution is insensitive to the tail of the entropy
production rate.

\acknowledgments We are grateful to B.~Freivogel, R.~Harnik, and J.~Niemeyer for discussions.  This work was supported by the Berkeley Center for Theoretical Physics, by a CAREER grant (award number 0349351) of the National Science Foundation, by fqxi grant RFP2-08-06, and by the US Department of Energy under Contract DE-AC02-05CH11231.

\bibliographystyle{utphys}
\bibliography{all}

\providecommand{\href}[2]{#2}\begingroup\raggedright\begin{thebibliography}{10}

\bibitem{BP}
R.~Bousso and J.~Polchinski, ``Quantization of four-form fluxes and dynamical
  neutralization of the cosmological constant,'' {\em JHEP} {\bf 06} (2000)
  006,
\href{http://arxiv.org/abs/hep-th/0004134}{{\tt hep-th/0004134}}.
%%CITATION = JHEPA,0006,006;%%.

\bibitem{KKLT}
S.~Kachru, R.~Kallosh, A.~Linde, and S.~P. Trivedi, ``De {S}itter vacua in
  string theory,'' {\em Phys. Rev. D} {\bf 68} (2003)  046005,
\href{http://arxiv.org/abs/hep-th/0301240}{{\tt hep-th/0301240}}.
%%CITATION = HEP-TH 0301240;%%.

\bibitem{Sche06}
A.~N. Schellekens, ``{The landscape 'avant la lettre'},''
\href{http://arxiv.org/abs/physics/0604134}{{\tt arXiv:physics/0604134}}.
%%CITATION = PHYSICS/0604134;%%.

\bibitem{Wei87}
S.~Weinberg, ``Anthropic bound on the cosmological constant,''
{\em Phys. Rev. Lett.} {\bf 59} (1987)  2607.
%%CITATION = PRLTA,59,2607;%%.

\bibitem{Pol06}
J.~Polchinski, ``The cosmological constant and the string landscape,''
\href{http://arxiv.org/abs/hep-th/0603249}{{\tt hep-th/0603249}}.
%%CITATION = HEP-TH 0603249;%%.

\bibitem{Bou07}
R.~Bousso, ``{TASI Lectures on the Cosmological Constant},''
  \href{http://dx.doi.org/10.1007/s10714-007-0557-5}{{\em Gen. Rel. Grav.} {\bf
  40} (2008)  607--637},
\href{http://arxiv.org/abs/0708.4231}{{\tt arXiv:0708.4231 [hep-th]}}.
%%CITATION = 0708.4231;%%.

\bibitem{BouLei08}
R.~Bousso and S.~Leichenauer, ``Star formation in the multiverse,''
  \href{http://dx.doi.org/10.1103/PhysRevD.79.063506}{{\em Phys. Rev.} {\bf
  D79} (2009)  063506},
\href{http://arxiv.org/abs/0810.3044}{{\tt arXiv:0810.3044 [astro-ph]}}.
%%CITATION = 0810.3044;%%.

\bibitem{LinLin96}
A.~Linde, D.~Linde, and A.~Mezhlumian, ``Nonperturbative amplifications of
  inhomogeneities in a self-reproducing universe,'' {\em Phys. Rev. D} {\bf 54}
  (1996)  2504--2518,
\href{http://arxiv.org/abs/gr-qc/9601005}{{\tt gr-qc/9601005}}.
%%CITATION = PHRVA,D54,2504;%%.

\bibitem{Gut00a}
A.~H. Guth, ``Inflation and eternal inflation,'' {\em Phys. Rep.} {\bf 333}
  (1983)  555, \href{http://arxiv.org/abs/astro-ph/0002156}{{\tt
  astro-ph/0002156}}.

\bibitem{Gut00b}
A.~H. Guth, ``Inflationary models and connections to particle physics,''
  \href{http://arxiv.org/abs/astro-ph/0002188}{{\tt astro-ph/0002188}}.

\bibitem{Gut04}
A.~H. Guth, ``Inflation,'' \href{http://arxiv.org/abs/astro-ph/0404546}{{\tt
  astro-ph/0404546}}.

\bibitem{Teg05}
M.~Tegmark, ``What does inflation really predict?,'' {\em JCAP} {\bf 0504}
  (2005)  001, \href{http://arxiv.org/abs/astro-ph/0410281}{{\tt
  astro-ph/0410281}}.

\bibitem{BouFre07}
R.~Bousso, B.~Freivogel, and I.-S. Yang, ``{Boltzmann babies in the proper time
  measure},'' \href{http://dx.doi.org/10.1103/PhysRevD.77.103514}{{\em Phys.
  Rev.} {\bf D77} (2008)  103514},
\href{http://arxiv.org/abs/0712.3324}{{\tt arXiv:0712.3324 [hep-th]}}.
%%CITATION = 0712.3324;%%.

\bibitem{BouFre06b}
R.~Bousso and B.~Freivogel, ``A paradox in the global description of the
  multiverse,'' {\em JHEP} {\bf 06} (2007)  018,
\href{http://arxiv.org/abs/hep-th/0610132}{{\tt hep-th/0610132}}.
%%CITATION = HEP-TH/0610132;%%.

\bibitem{Pag06}
D.~N. Page, ``Is our universe likely to decay within 20 billion years?,''
\href{http://arxiv.org/abs/hep-th/0610079}{{\tt hep-th/0610079}}.
%%CITATION = HEP-TH/0610079;%%.

\bibitem{Pag06b}
D.~N. Page, ``Return of the {B}oltzmann brains,''
\href{http://arxiv.org/abs/hep-th/0611158}{{\tt hep-th/0611158}}.
%%CITATION = HEP-TH/0611158;%%.

\bibitem{Lin07}
A.~Linde, ``Towards a gauge invariant volume-weighted probability measure for
  eternal inflation,'' {\em JCAP} {\bf 0706} (2007)  017,
\href{http://arxiv.org/abs/arXiv:0705.1160 [hep-th]}{{\tt arXiv:0705.1160
  [hep-th]}}.
%%CITATION = ARXIV:0705.1160;%%.

\bibitem{Gut07}
A.~H. Guth, ``Eternal inflation and its implications,'' {\em J. Phys.} {\bf
  A40} (2007)  6811--6826,
\href{http://arxiv.org/abs/hep-th/0702178}{{\tt hep-th/0702178}}.
%%CITATION = HEP-TH/0702178;%%.

\bibitem{Bou09}
R.~Bousso, ``{Complementarity in the Multiverse},''
\href{http://arxiv.org/abs/0901.4806}{{\tt arXiv:0901.4806 [hep-th]}}.
%%CITATION = 0901.4806;%%.

\bibitem{BouFre08b}
R.~Bousso, B.~Freivogel, and I.-S. Yang, ``{Properties of the scale factor
  measure},''
\href{http://arxiv.org/abs/0808.3770}{{\tt arXiv:0808.3770 [hep-th]}}.
%%CITATION = 0808.3770;%%.

\bibitem{BouYan09}
R.~Bousso and I.-S. Yang, ``{Global-Local Duality in Eternal Inflation},''
\href{http://arxiv.org/abs/0904.2386}{{\tt arXiv:0904.2386 [hep-th]}}.
%%CITATION = 0904.2386;%%.

\bibitem{Bou06}
R.~Bousso, ``Holographic probabilities in eternal inflation,'' {\em Phys. Rev.
  Lett.} {\bf 97} (2006)  191302,
\href{http://arxiv.org/abs/hep-th/0605263}{{\tt hep-th/0605263}}.
%%CITATION = HEP-TH/0605263;%%.

\bibitem{BouHar07}
R.~Bousso, R.~Harnik, G.~D. Kribs, and G.~Perez, ``Predicting the cosmological
  constant from the causal entropic principle,'' {\em Phys. Rev. D} {\bf 76}
  (2007)  043513,
\href{http://arxiv.org/abs/hep-th/0702115}{{\tt hep-th/0702115}}.
%%CITATION = HEP-TH 0702115;%%.

\bibitem{Fre08}
B.~Freivogel, ``{Anthropic Explanation of the Dark Matter Abundance},''
\href{http://arxiv.org/abs/0810.0703}{{\tt arXiv:0810.0703 [hep-th]}}.
%%CITATION = 0810.0703;%%.

\bibitem{BouHal09}
R.~Bousso, L.~J. Hall, and Y.~Nomura, ``{Multiverse Understanding of
  Cosmological Coincidences},''
\href{http://arxiv.org/abs/0902.2263}{{\tt arXiv:0902.2263 [hep-th]}}.
%%CITATION = 0902.2263;%%.

\bibitem{GarVil08}
J.~Garriga and A.~Vilenkin, ``{Holographic Multiverse},''
  \href{http://dx.doi.org/10.1088/1475-7516/2009/01/021}{{\em JCAP} {\bf 0901}
  (2009)  021},
\href{http://arxiv.org/abs/0809.4257}{{\tt arXiv:0809.4257 [hep-th]}}.
%%CITATION = 0809.4257;%%.

\bibitem{BouFre06}
R.~Bousso, B.~Freivogel, and M.~Lippert, ``Probabilities in the landscape: The
  decay of nearly flat space,'' {\em Phys. Rev.} {\bf D74} (2006)  046008,
\href{http://arxiv.org/abs/hep-th/0603105}{{\tt hep-th/0603105}}.
%%CITATION = HEP-TH/0603105;%%.

\bibitem{MerAda08}
L.~Mersini-Houghton and F.~C. Adams, ``{Limitations of anthropic predictions
  for the cosmological constant $\Lambda$: Cosmic Heat Death of Anthropic
  Observers},'' {\em Class. Quant. Grav.} {\bf 25} (2008)  165002,
  \href{http://arxiv.org/abs/0810.4914}{{\tt arXiv:0810.4914}}.

\bibitem{DenDou04b}
F.~Denef and M.~R. Douglas, ``Distributions of flux vacua,'' {\em JHEP} {\bf
  05} (2004)  072,
\href{http://arxiv.org/abs/hep-th/0404116}{{\tt hep-th/0404116}}.
%%CITATION = HEP-TH 0404116;%%.

\bibitem{CliFre07}
J.~M. Cline, A.~R. Frey, and G.~Holder, ``Predictions of the causal entropic
  principle for environmental conditions of the universe,''
\href{http://arxiv.org/abs/arXiv:0709.4443 [hep-th]}{{\tt arXiv:0709.4443
  [hep-th]}}.
%%CITATION = ARXIV:0709.4443;%%.

\bibitem{BozAlb09}
B.~Bozek, A.~J. Albrecht, and D.~Phillips, ``{Curvature Constraints from the
  Causal Entropic Principle},''
\href{http://arxiv.org/abs/0902.1171}{{\tt arXiv:0902.1171 [astro-ph.CO]}}.
%%CITATION = 0902.1171;%%.

\bibitem{PhiAlb09}
D.~Phillips and A.~Albrecht, ``{Effects of Inhomogeneity on the Causal Entropic
  prediction of Lambda},''
\href{http://arxiv.org/abs/0903.1622}{{\tt arXiv:0903.1622 [gr-qc]}}.
%%CITATION = 0903.1622;%%.

\bibitem{HerSpr02}
L.~Hernquist and V.~Springel, ``{An analytical model for the history of cosmic
  star formation},''
  \href{http://dx.doi.org/10.1046/j.1365-8711.2003.06499.x}{{\em Mon. Not. Roy.
  Astron. Soc.} {\bf 341} (2003)  1253},
\href{http://arxiv.org/abs/astro-ph/0209183}{{\tt arXiv:astro-ph/0209183}}.
%%CITATION = ASTRO-PH/0209183;%%.

\bibitem{Sal09}
M.~P. Salem, ``{Negative vacuum energy densities and the causal diamond
  measure},''
\href{http://arxiv.org/abs/0902.4485}{{\tt arXiv:0902.4485 [hep-th]}}.
%%CITATION = 0902.4485;%%.

\bibitem{SusTho93}
L.~Susskind, L.~Thorlacius, and J.~Uglum, ``The stretched horizon and black
  hole complementarity,'' {\em Phys. Rev. D} {\bf 48} (1993)  3743,
\href{http://arxiv.org/abs/hep-th/9306069}{{\tt hep-th/9306069}}.
%%CITATION = PHRVA,D48,3743;%%.

\bibitem{BouFre06a}
R.~Bousso, B.~Freivogel, and I.-S. Yang, ``Eternal inflation: The inside
  story,'' {\em Phys. Rev. D} {\bf 74} (2006)  103516,
\href{http://arxiv.org/abs/hep-th/0606114}{{\tt hep-th/0606114}}.
%%CITATION = HEP-TH 0606114;%%.

\bibitem{HarSre07}
J.~B. Hartle and M.~Srednicki, ``Are we typical?,'' {\em Phys. Rev. D} {\bf 75}
  (2007)  123523,
\href{http://arxiv.org/abs/arXiv:0704.2630 [hep-th]}{{\tt arXiv:0704.2630
  [hep-th]}}.
%%CITATION = ARXIV:0704.2630;%%.

\bibitem{GarVil07}
J.~Garriga and A.~Vilenkin, ``Prediction and explanation in the multiverse,''
\href{http://arxiv.org/abs/arXiv:0711.2559 [hep-th]}{{\tt arXiv:0711.2559
  [hep-th]}}.
%%CITATION = ARXIV:0711.2559;%%.

\bibitem{Pag09}
D.~N. Page, ``{The Born Rule Dies},''
  \href{http://arxiv.org/abs/0903.4888}{{\tt arXiv:0903.4888 [hep-th]}}.

\bibitem{FreKle05}
B.~Freivogel, M.~Kleban, M.~Rodriguez~Martinez, and L.~Susskind,
  ``{Observational consequences of a landscape},'' {\em JHEP} {\bf 03} (2006)
  039,
\href{http://arxiv.org/abs/hep-th/0505232}{{\tt arXiv:hep-th/0505232}}.
%%CITATION = HEP-TH/0505232;%%.

\bibitem{SchVil06}
D.~Schwartz-Perlov and A.~Vilenkin, ``Probabilities in the
  {B}ousso-{P}olchinski multiverse,'' {\em JCAP} {\bf 0606} (2006)  010,
\href{http://arxiv.org/abs/hep-th/0601162}{{\tt hep-th/0601162}}.
%%CITATION = HEP-TH 0601162;%%.

\bibitem{Sch06}
D.~Schwartz-Perlov, ``Probabilities in the
  {A}rkani-{H}amed-{D}imopolous-{K}achru landscape,''
\href{http://arxiv.org/abs/hep-th/0611237}{{\tt hep-th/0611237}}.
%%CITATION = HEP-TH 0611237;%%.

\bibitem{OluSch07}
K.~D. Olum and D.~Schwartz-Perlov, ``Anthropic prediction in a large toy
  landscape,''
\href{http://arxiv.org/abs/arXiv:0705.2562 [hep-th]}{{\tt arXiv:0705.2562
  [hep-th]}}.
%%CITATION = ARXIV:0705.2562;%%.

\bibitem{Sch08}
D.~Schwartz-Perlov, ``{Anthropic prediction for a large multi-jump
  landscape},'' \href{http://dx.doi.org/10.1088/1475-7516/2008/10/009}{{\em
  JCAP} {\bf 0810} (2008)  009},
\href{http://arxiv.org/abs/0805.3549}{{\tt arXiv:0805.3549 [hep-th]}}.
%%CITATION = 0805.3549;%%.

\bibitem{BouHarTA}
R.~Bousso and R.~Harnik {\em (to appear)}  .

\bibitem{TARW}
M.~Tegmark, A.~Aguirre, M.~Rees, and F.~Wilczek, ``{Dimensionless constants,
  cosmology and other dark matters},''
  \href{http://dx.doi.org/10.1103/PhysRevD.73.023505}{{\em Phys. Rev.} {\bf
  D73} (2006)  023505},
\href{http://arxiv.org/abs/astro-ph/0511774}{{\tt arXiv:astro-ph/0511774}}.
%%CITATION = ASTRO-PH/0511774;%%.

\bibitem{PressSchechter}
W.~H. Press and P.~Schechter, ``{Formation of galaxies and clusters of galaxies
  by selfsimilar gravitational condensation},''
\href{http://dx.doi.org/10.1086/152650}{{\em Astrophys. J.} {\bf 187} (1974)
  425--438}.
%%CITATION = ASJOA,187,425;%%.

\bibitem{LaceyCole}
C.~G. Lacey and S.~Cole, ``{Merger rates in hierarchical models of galaxy
  formation},''
{\em Mon. Not. Roy. Astron. Soc.} {\bf 262} (1993)  627--649.
%%CITATION = MNRAA,262,627;%%.

\bibitem{Andriesse}
C.~D. {Andriesse}, ``{Radiating cosmic dust},''
  \href{http://dx.doi.org/10.1016/0083-6656(77)90017-4}{{\em Vistas in
  Astronomy} {\bf 21} (1977)  107}.

\bibitem{WMAP5}
{\bf WMAP} Collaboration, E.~Komatsu {\em et al.}, ``{Five-Year Wilkinson
  Microwave Anisotropy Probe (WMAP) Observations: Cosmological
  Interpretation},''
\href{http://arxiv.org/abs/arxiv:0803.0547 [astro-ph]}{{\tt arxiv:0803.0547
  [astro-ph]}}.
%%CITATION = 0803.0547;%%.

\bibitem{TegRee97}
M.~Tegmark and M.~J. Rees, ``Why is the {CMB} fluctuation level $10^{-5}$?,''
  {\em Astrophys. J.} {\bf 499} (1998)  526--532,
\href{http://arxiv.org/abs/astro-ph/9709058}{{\tt astro-ph/9709058}}.
%%CITATION = ASTRO-PH 9709058;%%.

\end{thebibliography}\endgroup
\end{document}